%% file: moxc.tex
\documentclass[11pt,preprint]{aastex}

\usepackage{grffile}
\usepackage{url}                                                                                  

\newcounter{column_number}
\newcommand{\numberthecolumn}{\colhead{(\arabic{column_number})}\stepcounter{column_number}}

\renewcommand{\S}{Section }
\newcommand{\tnm}[1]{\tablenotemark{#1}}

\newcommand{\anchorfoot}[2] {\anchor{#1}{#2}\footnote{\url{#1}}}
\newcommand{\anchorparen}[2]{\anchor{#1}{#2} (\url{#1})}

        \usepackage{hyperref}                        
        \renewcommand{\anchor}[2]{\href{#1}{#2}}

\newcommand{\hii}{H{\scriptsize II} }
\newcommand{\GHIIR}{GH{\scriptsize II}R }
\newcommand{\UCHIIR}{UCH{\scriptsize II}R }

\newcommand{\mystix} {MYStIX}
\newcommand{\Spitzer} {{\em Spitzer}}
\newcommand{\Herschel} {{\em Herschel}}
\newcommand{\Chandra} {{\em Chandra~}}
\newcommand{\ACIS}    {{ACIS}}
\newcommand{\CIAO}    {{\em CIAO}}

\shorttitle{MOXC}
\shortauthors{Townsley et al.} 
\slugcomment{Accepted to ApJS, March 3, 2014}

\begin{document}

\title{THE MASSIVE STAR-FORMING REGIONS OMNIBUS X-RAY CATALOG }

\author{ Leisa K. Townsley\altaffilmark{1}\altaffilmark{*}, 
Patrick S. Broos\altaffilmark{1},
Gordon P. Garmire\altaffilmark{2},
Jeroen Bouwman\altaffilmark{3},
Matthew S. Povich\altaffilmark{4},
Eric D. Feigelson\altaffilmark{1},
Konstantin V. Getman\altaffilmark{1},
Michael A. Kuhn\altaffilmark{1}
}

\altaffiltext{*}{townsley@astro.psu.edu} 

\altaffiltext{1}{Department of Astronomy \& Astrophysics, 525 Davey Laboratory, Pennsylvania State University, University Park, PA 16802, USA}
\altaffiltext{2}{Huntingdon Institute for X-ray Astronomy, LLC, 10677 Franks Road, Huntingdon, PA 16652, USA}
\altaffiltext{3}{Max Planck Institute for Astronomy, K\"{o}nigstuhl 17, D-69117 Heidelberg, Germany}
\altaffiltext{4}{California State Polytechnic University, 3801 West Temple Ave, Pomona, CA 91768, USA}

\begin{abstract}
We present the Massive Star-forming Regions (MSFRs) Omnibus X-ray Catalog (MOXC), a compendium of X-ray point sources from {\em Chandra}/ACIS observations of a selection of MSFRs across the Galaxy, plus 30~Doradus in the Large Magellanic Cloud.  MOXC consists of 20,623 X-ray point sources from 12 MSFRs with distances ranging from 1.7~kpc to 50~kpc.  Additionally, we show the morphology of the unresolved X-ray emission that remains after the catalogued X-ray point sources are excised from the ACIS data, in the context of \Spitzer\ and {\em WISE} observations that trace the bubbles, ionization fronts, and photon-dominated regions that characterize MSFRs.  In previous work, we have found that this unresolved X-ray emission is dominated by hot plasma from massive star wind shocks.  This diffuse X-ray emission is found in every MOXC MSFR, clearly demonstrating that massive star feedback (and the several-million-degree plasmas that it generates) is an integral component of MSFR physics.
\end{abstract}




\section{INTRODUCTION \label{sec:intro}}

To further our understanding of star formation, massive star feedback, and the origin and evolution of massive star-forming regions (MSFRs) in the Milky Way and beyond, we require a more complete census of massive young stellar cluster (MYSC) members and a broader multiwavelength perspective on the interstellar medium (ISM) as it is shaped by the winds and supernovae of massive stars.  To that end, we have worked for over a decade to amass the MSFR Omnibus X-ray Catalog (MOXC), a list of X-ray point sources from {\em Chandra X-ray Observatory}  ({\em Chandra}) observations of famous MSFRs that extends to the faintest statistically-significant X-ray sources.  These observations were obtained with the Advanced CCD Imaging Spectrometer (ACIS) instrument \citep{Garmire03} on \Chandra and often include multiple overlapping pointings of the ACIS camera.  MOXC collates 20,623 X-ray point sources extracted from 11 Galactic MSFRs (with distances ranging from 1.7~kpc to 7~kpc) plus 30~Doradus in the Large Magellanic Cloud (Table~\ref{targets.tbl}).  It is presented as a single large catalog because the data analysis methods were identical for all of these MSFRs, thus the resultant data products can be presented concisely using a consistent format.  This facilitates comparisons among these MSFRs and others reduced with our software and methods \citep[e.g.,][]{Broos11a,Kuhn13a}.  

The direct precursor to MOXC methodology was the 1.2~Ms \Chandra Carina Complex Project (CCCP) \citep{Townsley11a}, a 22-pointing ACIS Imaging Array (ACIS-I) mosaic of the Great Nebula in Carina that generated a catalog of $>$14,000 X-ray point sources \citep{Broos11a}.  Our experience with the CCCP left us with a good sense of the X-ray point source data products that will be interesting to ourselves and others for studying MSFRs and built the groundwork for standardizing those products.  The Carina complex is also suffused with bright, spatially-complex diffuse X-ray emission, so the CCCP also presented us with an opportunity to develop analysis techniques and software for excising X-ray point sources from ACIS data and processing the underlying diffuse emission in a systematic way \citep{Townsley11b}.  For the final paper in the CCCP ApJ Supplement Special Issue, we extended our studies of Carina's diffuse X-ray emission to several other MSFRs (M17, NGC~3576, NGC~3603, and 30~Doradus) and compared these different examples of MYSC massive star feedback in a global sense \citep{Townsley11c}.  The ACIS X-ray point source lists for those targets are included here in MOXC.

We then applied our experience with CCCP and other \Chandra observations of MSFRs to a large archival project, re-analyzing ACIS observations of MSFRs in a consistent way and combining the resulting lists of X-ray point sources with near-IR data from the Two Micron All Sky Survey \citep[2MASS,][]{Skrutskie06} and UKIRT and with mid-IR data from the {\em Spitzer Space Telescope} ({\em Spitzer}) to build a large database for MSFR comparison.  This is the {\bf M}assive {\bf Y}oung Star-Forming Complex {\bf St}udy in {\bf I}nfrared and {\bf X}-ray (MYStIX) project \citep{Feigelson13}.  The ACIS X-ray source lists for ten MYStIX targets were given in \citet{Kuhn13a}.  Those for three more MYStIX targets came from the literature:  the Orion Nebula Cluster from \citet{Getman05}, W40 from \citet{Kuhn10}, and Carina from the CCCP catalog paper \citep{Broos11a}.  X-ray source lists for the remaining seven MYStIX targets (Table~\ref{targets.tbl}) are given here as part of MOXC.  Additionally, MOXC includes ACIS source lists for five MSFRs that were not part of the MYStIX sample.  In total, MOXC lists 14,710 X-ray sources from MYStIX targets, plus 5913 X-ray sources from the additional five MSFRs.


For Table~\ref{targets.tbl}, as a rough measure of X-ray point source detection sensitivity, we used the tool  \anchorfoot{http://asc.harvard.edu/toolkit/pimms.jsp}{{\em PIMMS}} to calculate the limiting luminosity $L_{tc}$ (``{\bf t}otal'' band 0.5--8~keV, {\bf c}orrected for extinction) for detecting a 5-count source on-axis with {\em Chandra}/ACIS-I, assuming an {\em apec} thermal plasma with $kT$=2.7~keV and abundance 0.4*Z$_{\odot}$, values typical of a  pre-main sequence (pre-MS) star \citep{Preibisch2005}.  Using the relation between X-ray luminosity and stellar mass from \citet{Preibisch2005}, we then estimate $M_{50\%}$, the mass at which this limiting $L_{tc}$ captures the brighter half of the X-ray-emitting population.  For targets with very shallow ACIS observations, this mass limit moves out of the range of typical pre-MS stars, so Column~9 of Table~\ref{targets.tbl} simply notes the limit as ``bright'', meaning that only bright X-ray sources (some massive stars and extreme flaring pre-MS stars) will be detected.

\input{target_table}

\clearpage

\section{{\em CHANDRA} OBSERVATIONS AND DATA ANALYSIS \label{sec:data}}

The \Chandra observations used for MOXC are summarized in Table~\ref{tbl:obslog} and are identified by a unique Observation Identification (ObsID) number.  Sequence numbers show which ObsIDs make up a single ACIS pointing.  All of these datasets are available in the \Chandra archive.  Observations are ordered by date for each target; target names as used in this paper are given in Column 1 in bold, followed by the original name assigned by the study's principal investigator (PI, noted in Column 10).  Those observations where Gordon Garmire is listed as PI came from the ACIS Instrument Team Guaranteed Time Observations (GTO).  We note this fact because the ACIS GTO program provided many of the original seed observations for the MSFRs studied here and for other MYStIX targets.

All observations employed ACIS-I \citep{Garmire03}, a 2$\times$2 array of 1024$\times$1024-pixel CCDs covering roughly $17\arcmin \times 17\arcmin$ on the sky with 0.492$\arcsec$ pixels.  ACIS-I observations often include data from two CCDs lying far off-axis in the \ACIS\ spectroscopy array (ACIS-S); these off-axis CCDs are not included in the analysis presented here, due to the poor angular resolution of the \Chandra mirrors at large off-axis angles.\footnote {See Figure~4.12 in the \anchorparen{http://asc.harvard.edu/proposer/POG/}{\Chandra Proposers' Observatory Guide}.}  Early ACIS-I observations of 30~Doradus were omitted from this study due to calibration issues (detailed below).  Exposure maps for the MOXC ACIS-I mosaics are shown in Section~\ref{sec:targets}.

\input{observing_log.tex}

Our data reduction, point source detection, and point source extraction techniques employ several innovations beyond standard \Chandra procedures, as discussed at length by \citet{Broos10}. These techniques were standardized for MSFRs by the CCCP, as mentioned above.  Many of the CCCP data analysis steps were implemented by the \anchor{http://www.astro.psu.edu/xray/acis/acis_analysis.html}{{\em ACIS Extract}} (AE) software package\footnote{ The {\em ACIS Extract} software package and User's Guide are available at \url{http://www.astro.psu.edu/xray/acis/acis_analysis.html}. } \citep{AE2012} and are described in detail in \citet{Broos10}.  Nearly identical procedures were applied to the MYStIX \Chandra data and the other MSFRs presented here, thus we do not provide an exhaustive review of those procedures here.  A few minor improvements to the CCCP methodology that may be of interest are described below.

In earlier \Chandra reductions we used a visible-band catalog of stars, the Naval Observatory Merged Astrometric Dataset \citep[NOMAD,][]{Zacharias05}, as the astrometric reference to which we tied each \Chandra observation.  Since many of our MSFRs are embedded and/or obscured, with few visual counterparts to X-ray sources, we have now adopted 2MASS as our astrometric reference.  As before, we adjust the astrometry of each \Chandra observation using a preliminary catalog of bright X-ray sources detected in that single observation, before combining the aligned observations to search for faint sources.  We now also re-check that alignment after source extraction has been started, using source position estimates from AE, and apply a second correction to the astrometry of each \Chandra observation, as needed.

Our procedures now explicitly address the issue of a \anchorfoot{http://cxc.harvard.edu/ciao/caveats/psf_artifact.html}{hook-shaped feature in the \Chandra point-spread function} (PSF), extending ${\sim}$0.8\arcsec\ from the main peak and containing ${\sim}$5\% of the flux.  Its effects on ACIS data were demonstrated by Vinay Kashyap at the \Chandra X-ray Center \anchorfoot{http://cxc.harvard.edu/cal/Hrc/PSF/acis_psf_2010oct.html}{in a memo} dated November 2010. 
Its energy dependence (either spatially or as a fraction of the total power) remains only roughly characterized and no PSF models incorporating this feature are currently available.  Since our point source detection and extraction procedures are carefully designed to recover crowded point sources, they are susceptible to the false identification of PSF hooks as astrophysical sources.  In the CCCP, we could only point out possible ``hook sources'' after the catalog was constructed.  Now we use the recently-developed \CIAO\ tool {\it make\_psf\_asymmetry\_region} to build an {\it SAOImage DS9} region file that marks potential hook features around bright sources.  Then we examine each of these by hand, removing all candidate point sources that are consistent (both spatially and photometrically) with the PSF hook of the bright source.  Thus, we are confident that the catalog presented here does not contain a significant number of spurious ``hook sources.''

The custom background region that AE designs for each point source aperture \citep{Broos10} now accounts for the so-called ``CCD readout streaks'' produced by every bright source in the field.\footnote { The CCD readout streak phenomenon is discussed at \url{http://cxc.harvard.edu/ciao/threads/streakextract/} and \url{http://cxc.harvard.edu/ciao/why/pileup_intro.html} and \url{http://cxc.harvard.edu/ciao/download/doc/pileup_abc.pdf}. }  Thus, when a source aperture is contaminated by a readout streak from a bright source that fell in a nearby CCD column, its background region now accounts for that extra background component, improving the accuracy of the photometry.  Conversely, when a source aperture contains no streak, its background region is constructed to avoid any streaks that may be nearby.

\citet{Broos10} describe the benefits of applying different event cleaning criteria for low-count-rate and high-count-rate sources---low-count-rate sources benefit from ``aggressive'' cleaning that reduces the background as much as possible, whereas high-count-rate sources benefit from ``mild'' cleaning that omits the steps in aggressive cleaning that suffer many false positives when the event rate is high, erroneously removing true source events and damaging the photometry of bright sources.  Now, as before, our source detection and validation procedures use aggressively-cleaned data, because the goal of these procedures is to establish a conservative, accurate inventory of the X-ray point sources present in the data.  Previously, the final source event extraction had to use mildly-cleaned data for all sources, to protect the bright sources from damage.  The version of AE used here is empowered to choose between aggressively-cleaned and mildly-cleaned data based on the event rate of each source, allowing us simultaneously to achieve low background for most sources and to protect the bright sources from photometric damage.

As in previous studies, our point source detection strategy was first to propose a liberal set of candidate point sources, derived mostly from image reconstruction using local models of the {\em Chandra}/ACIS PSF, then iteratively to prune candidates found to be insignificant after extraction and careful local background estimation.  Extraction apertures are normally sized to contain 90\% of the PSF (at 1.5~keV), but are reduced when necessary to minimize overlap among crowded sources.  This ability to detect and extract closely-separated sources was particularly helpful for NGC~3603, G333.6-0.2, and NGC~6334; in those fields ${\sim}$30\% of detected sources have reduced apertures. 

The goals of completeness and validity of a source list derived from CCD imaging data are always in conflict.  Our point source detection procedure is designed to be aggressive, emphasizing sensitivity and accepting a reasonable number of possibly-spurious detections to achieve that sensitivity. In the CCCP study, \citet[][Figure~9]{Broos11a} show that when deep NIR catalogs are available, the fraction of X-ray detections without apparent NIR counterparts rises only slowly as detection significance falls; this is further evidence that our procedures do not lead to a large number of false sources. \citet[][\S6.2]{Broos11a} discuss the impracticality of quantifying the false detection rate in our full X-ray catalog, and also point out that such an estimate would be nearly useless because most science analyses select subsets of the X-ray catalog (e.g., sources with IR photometry available, or sources classified as young stars).


The \Chandra data analysis system, \CIAO\ \citep{Fruscione06}, the {\it SAOImage DS9} visualization tool \citep{Joye03}, and the \anchorfoot{http://www.ittvis.com/idl}{{\it Interactive Data Language}} (IDL) are used throughout our data analysis workflow, from data preparation through science analysis.
The \Chandra Calibration Database version used for event calibration of each ObsID is shown in the ``CALDBVER'' column of Table~\ref{tbl:obslog}.
Point source response matrices were constructed using the following \Chandra Calibration Database versions:  
4.4.6 for M16, G333.6-0.2, NGC~3576, NGC~3603, NGC~6334, W3, and W4;
4.4.7 for W51;
4.4.10 for 30~Doradus, M17, and NGC~6357 (ObsIDs 4477, 10988, 10987, 13267);
4.5.6 for G29.96-0.02;
4.5.9 for NGC~6357 (ObsIDs 13622, 13623).


\section{THE {\em CHANDRA} POINT SOURCE CATALOG \label{sec:ptsrccat}}

Table~\ref{xray_properties.tbl} defines the columns of the point source catalog that constitutes MOXC.  It is available in FITS format from the electronic edition of this article and may be available in many other formats from VizieR \citep{Ochsenbein00}.
All photometric quantities in this table are apparent (not corrected for absorption). The suffixes ``\_t'', ``\_s'', and ``\_h'' on names of photometric quantities designate the {\em total} (0.5--8~keV), {\em soft} (0.5--2~keV), and {\em hard} (2--8~keV) energy bands. 
Correction for finite extraction apertures is applied to the ancillary reference file (ARF) calibration products \citep[see][\S5.3]{Broos10}; the SrcCounts and NetCounts quantities characterize the extraction and are not aperture-corrected. 
The only calibrated quantities presented are apparent {\em photon} flux in units of photon~cm$^{-2}$~s$^{-1}$\citep[see][\S7.4]{Broos10}, and an estimate for apparent {\em energy} flux in units of erg~cm$^{-2}$~s$^{-1}$ \citep{Getman10}.
Table notes provide additional information regarding the definition of source properties.

A similar table for other MYStIX targets was presented by \citet{Kuhn13a}.  The main difference between that and the MOXC version (Table~\ref{xray_properties.tbl}) is that we have chosen to omit absorption-corrected X-ray source luminosities from the {\it XPHOT} algorithm \citep{Getman10} here, because those quantities are given in \citet{Broos13} for relevant MYStIX X-ray sources (those classified as pre-MS stars).  For beyond-MYStIX targets, we choose to postpone {\it XPHOT} calculations until the X-ray sources are classified, since {\it XPHOT} estimates are only appropriate for pre-MS stars.  \anchorfoot{http://www.astro.psu.edu/users/gkosta/XPHOT/}{The {\it XPHOT} code} is available \citep{XPHOT2012} if others wish to use it on MOXC sources.

\input{xray_column_labels.tex}




\subsection{\bf Characteristics of the X-ray Catalogs}

\begin{figure}[htb]
\centering
\includegraphics[width=0.49\textwidth]{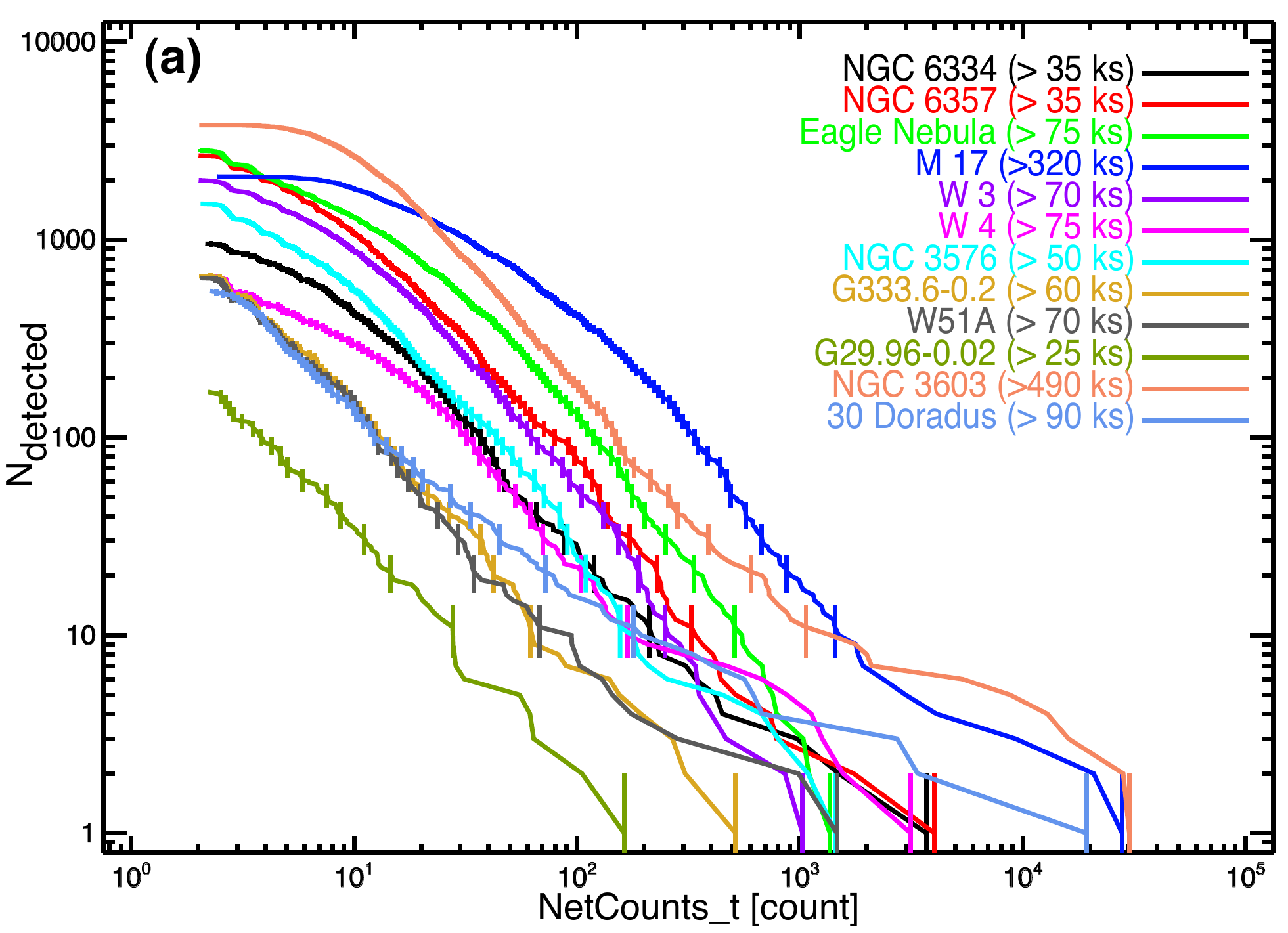} 
\hspace*{0.02in}
\includegraphics[width=0.49\textwidth]{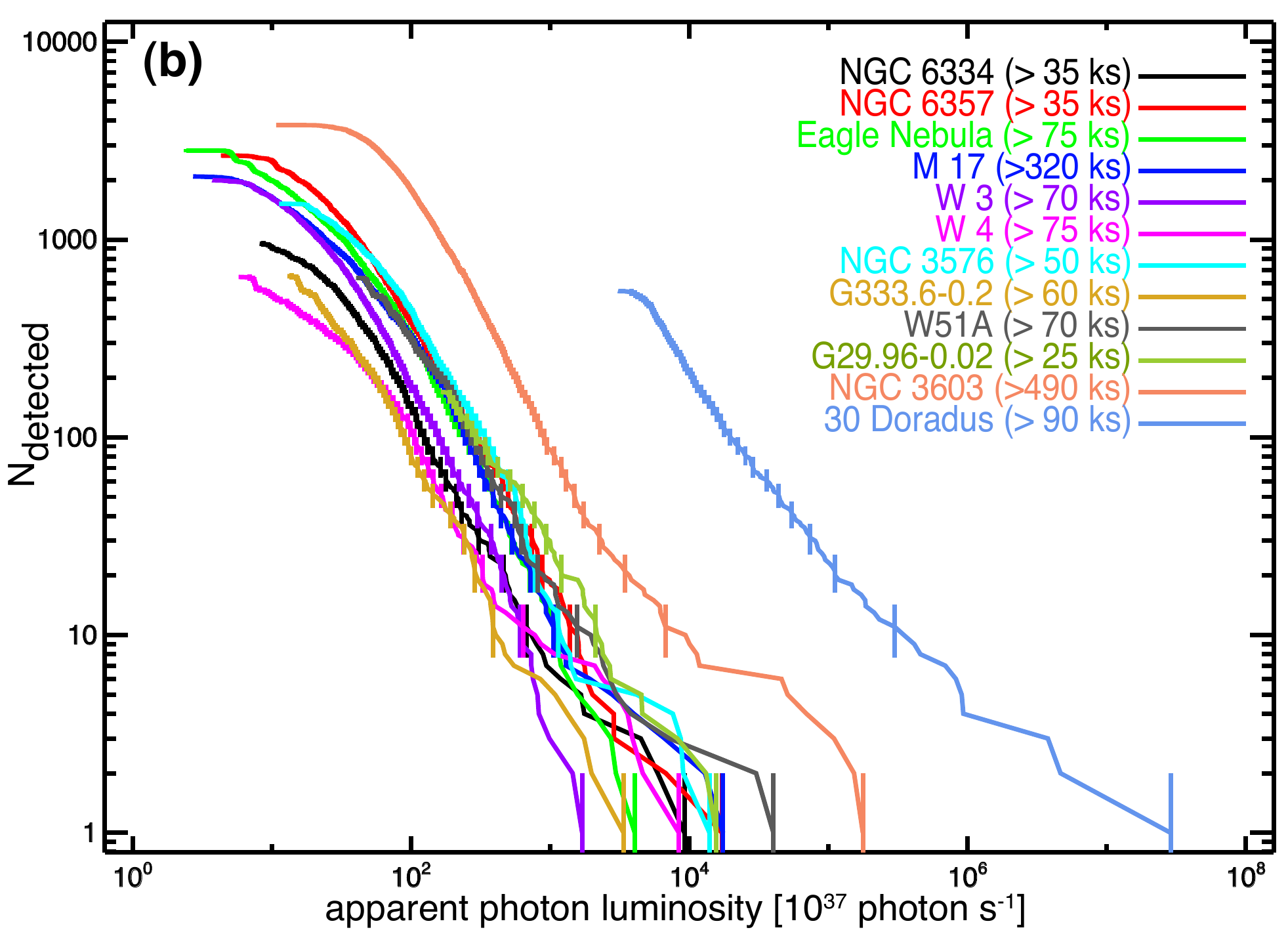}
\caption{
Cumulative number of sources detected ($N_{detected}$) above a threshold on total band (0.5--8~keV) net extracted counts (a) or on total band (0.5--8~keV) apparent photon luminosity (b), plotted against that threshold.
In both panels, sources lying in regions with lower-than-nominal exposure time (caused by misalignment among multiple ObsIDs) are excluded; those target-specific nominal exposure times are shown in the legends.
Error bars show Poisson uncertainty in 10\% of the X-axis bins.
\label{logN_vs_NetCounts_t.fig}
}
\end{figure}

{\em Chandra}'s superb PSF and the low instrumental background of the ACIS camera support detection of point sources with only a handful of net counts.
Consequently, in a typical \Chandra catalog only a small fraction of detected sources will produce the hundreds-to-thousands of counts needed for fitting complex spectral models, for time-resolved spectroscopy, or for event timing analyses.  These facts are evident in the cumulative histograms shown in Figure~\ref{logN_vs_NetCounts_t.fig}a, where the cumulative number of sources detected ($N_{detected}$) above a threshold on net extracted counts (NetCounts\_t in Table~\ref{xray_properties.tbl}) is plotted against that threshold.  Note (in the legend) that two targets, M17 and NGC~3603, have significantly longer nominal exposure times than the others.  Their cumulative histograms show that very low-count sources ($<$10 net counts) are rarely detected, consistent with the higher backgrounds produced by those longer exposures.

The net counts cumulative histograms reveal nothing about the sensitivity of these observations, because NetCounts\_t is an uncalibrated quantity.  A rough sensitivity comparision among the targets can be made by constructing cumulative histograms for the calibrated quantity ``apparent photon luminosity'', calculated from the quantity ``apparent photon flux'' (PhotonFlux\_t in Table~\ref{xray_properties.tbl}) and the distances to the targets; see Figure~\ref{logN_vs_NetCounts_t.fig}b.  Since the sensitivity goes down as the square of the distance, 30~Doradus stands out as a particularly shallow \Chandra observation.  Astrophysically useful completeness limits will not be available until sample selection decisions have been made for each MSFR, and until absorption-corrected luminosities have been estimated for each of the sources in those samples.  Sample selection may involve classification of sources to eliminate foreground and background contaminants \citep{Broos11b,Broos13}, and may involve cropping fields of view to eliminate regions with low sensitivity and/or to select sub-clusters.

\subsection{Piled Sources \label{sec:pileup}}
Photon-counting detectors, such as \ACIS, can suffer from a non-linearity known as \anchorfoot{http://cxc.harvard.edu/ciao/why/pileup_intro.html}{{\em photon pile-up}} when multiple X-ray photons arrive with a separation in time and space that is too small to allow each to be detected as a separate X-ray event.  Pile-up effects include photometric and spectroscopic mis-calibration of the observation.  Total-band photometry is underestimated, because multiple photons interact to produce only one or zero events.  The shape of the detected X-ray spectrum is hardened, because the energy assigned to a piled event will represent that from multiple photons. 

The level of pile-up in each observation of a source can vary significantly, due to variations in the off-axis angle at which the source was observed\footnote{Pile-up decreases as a source with constant flux moves off-axis, because the \Chandra PSF expands and thus the event rate per detector pixel decreases; see Figure~6.21 in the \Chandra Proposers' Observatory Guide.} and/or due to astrophysical variability in the source itself.  We screen for pile-up effects in each observation of a source by estimating the observed count rate falling on an event detection cell of size 3$\times$3 \ACIS\ pixels, centered on the source position.  The highest such rate among all obervations of the source is reported in the column RateIn3x3Cell in Table~\ref{xray_properties.tbl}.

For all extractions in which RateIn3x3Cell $> 0.05$ count/frame, we quantified the level of pile-up using a Monte Carlo forward-modeling approach that reconstructs a pile-up-free \ACIS\ spectrum from a piled \ACIS\ observation \citep{Broos11a}.  Within that process, simulated photons are propagated through a detailed model of the \ACIS\ CCD \citep{Townsley00,Townsley02a,Townsley02b}.\footnote{\citet{Townsley02a} and \citet{Townsley02b} are available in the Physics database of ADS.}  Photon pile-up effects arise naturally within the simulation from the superposition of two or more electron charge clouds within individual CCD frames.

Table~\ref{pile-up_risk.tbl} lists each extraction (source name and ObsID) that we found to suffer from significant pile-up effects.  As a metric for the level of pile-up, column (7) reports the ratio of the pile-up-free to observed (piled) count rate in the total (0.5--8~keV) energy band.\footnote{We choose not to use the phrases ``pile-up fraction'' or ``pile-up percentage'' because the \ACIS\ community has several conflicting definitions for those terms; see \S1.2 in \anchorparen{http://cxc.harvard.edu/ciao/download/doc/pileup_abc.pdf}{The \Chandra ABC Guide to Pileup}.}  For the sources in Table~\ref{pile-up_risk.tbl}, several quantities in Table~\ref{xray_properties.tbl} are expected to be biased by pile-up effects, to differing degrees depending upon the intrinsic source spectrum and the degree of pile-up.  We must emphatically warn the users of MOXC data products that all of our tabulated quantities for piled-up sources should be used with caution and with careful, informed consideration for the distortions that pile-up might be causing.

\input{piled_table}

\clearpage

\begin{figure}[htb]
\centering
\includegraphics[width=0.5\textwidth]{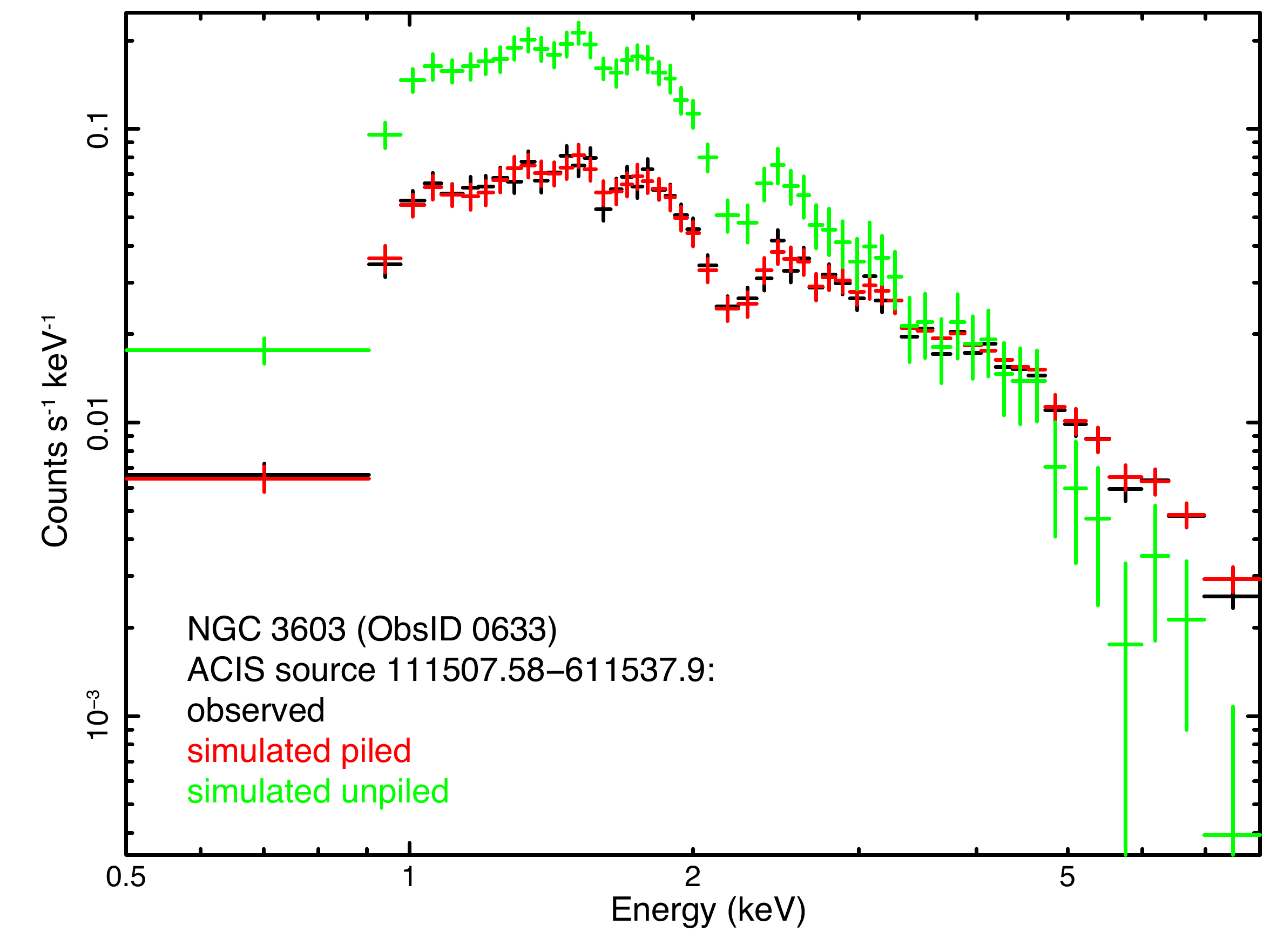}\\
\caption{An example of correcting distortion in the ACIS spectrum extracted from a piled-up point source.
 \label{pileup_example.fig}}
\end{figure}

Among the MOXC sources, the highest level of pile-up was seen in the first \Chandra observation \citep[ObsID 0633,][]{Moffat02} of the NGC~3603 source J111507.58-611537.9, the Wolf-Rayet (WR) binary NGC~3603-C.  Figure~\ref{pileup_example.fig} compares that piled spectrum to our estimate of the spectrum that \ACIS\ would have produced if pile-up effects were absent.  First, a piled simulation (red) is constructed to match the piled observed spectrum (black).  Then the simulation is run again with no pile-up (one photon per frame) to approximate the source spectrum before it was distorted by pile-up (green). The well-known effects of pile-up are clearly seen: the observed (piled) spectrum is hardened and the observed count rate is lowered.  Correcting for this more than doubles the counts below 2~keV in the reconstructed spectrum.

\subsection{MOXC Sources in Published \Chandra Catalogs \label{sec:other_analyses}}

\Chandra catalogs for the five MOXC targets in Table~\ref{tbl:other_analyses} have been published previously.  For NGC~6334 and M16, the ACIS ObsIDs used here are identical or similar to those used in the earlier studies, but MOXC uses slightly (for NGC~6334) or substantially (for M16) different analysis methodologies.  For the other targets in Table~\ref{tbl:other_analyses}, MOXC uses deeper ACIS data (or covers a wider field with more ACIS pointings) than what was available in earlier studies.
 
For the reader's convenience, we have matched those previously-published catalogs for these targets to the MOXC catalog, using a simple algorithm \citep[\S8 in][]{Broos10} in which the maximum acceptable separation between a MOXC source and a published counterpart is based on the individual source position errors, assuming Gaussian distributions, scaled so that ${\sim}$99\% of true associations should be identified as matches.  When multiple sources in the published catalog satisfy the match criterion, the closest one was adopted as the actual match.  Prior to matching, we estimated and removed small astrometric offsets between the MOXC catalogs and previously published catalog coordinate systems (Table~\ref{tbl:other_analyses}).  This step removes systematic catalog offsets; it is a very small effect for the published catalogs considered here, but we include this step in all catalog matching because these systematic offsets can be quite large and will result in mismatches (and serious confusion in the literature) if not eliminated before matching.  

Table~\ref{tbl:match_results} lists each MOXC source for the targets in Table~\ref{tbl:other_analyses}, the sequence number and name of the published counterpart we identified (if any), and the distance between the MOXC and published positions (after applying the offsets shown in Table~\ref{tbl:other_analyses} to the published positions).

\begin{deluxetable}{llrr}
\centering 
\tabletypesize{\tiny} \tablewidth{0pt}

\tablecaption{ Previously Published \Chandra Catalogs of MOXC Targets
 \label{tbl:other_analyses}}

\tablehead{
\colhead{Target} & 
\colhead{Reference} & 
\multicolumn{2}{c}{Astrometric Offset\tnm{a}} \\
\cline{3-4}

\colhead{} & 
\colhead{} & 
\colhead{$\alpha_{\rm J2000}$} & 
\colhead{$\delta_{\rm J2000}$} \\
\colhead{} & 
\colhead{} & 
\colhead{\arcsec} & 
\colhead{\arcsec} \\

\numberthecolumn &\numberthecolumn & \numberthecolumn & \numberthecolumn  
\setcounter{column_number}{1}
}

\startdata
NGC 6334           & \citet{Feigelson09} &  0.01 &  0.02 \\ 
NGC 6357           & \citet{Wang07}      &  0.06 &  0.01 \\ 
Eagle Nebula (M16) & \citet{Guarcello12} & -0.06 &  0.01 \\ 
M17                & \citet{Broos07}     &  0.01 &  0.03 \\ 
30 Doradus         & \citet{Townsley06b} & -0.09 &  0.13 \\ 
\enddata
\tablenotetext{a}{The adjustment to the published coordinates required for best alignment with the MOXC coordinates.}
\end{deluxetable}

\begin{deluxetable}{lrlr}
\centering 
\tabletypesize{\tiny} \tablewidth{0pt}

\tablecaption{ MOXC Sources in Published \Chandra Catalogs\tnm{a} 
 \label{tbl:match_results}}

\tablehead{
\multicolumn{1}{c}{MOXC}       & \multicolumn{3}{c}{Published \Chandra Catalog} \\
\multicolumn{1}{c}{\hrulefill} & \multicolumn{3}{c}{\hrulefill} \\

\colhead{Name} & 
\colhead{Seq.} & 
\colhead{Name} & 
\colhead{Offset} \\ 
\colhead{} & 
\colhead{} & 
\colhead{} & 
\colhead{\arcsec} \\

\numberthecolumn & \numberthecolumn & \numberthecolumn & \numberthecolumn  
\setcounter{column_number}{1}
}

\startdata
 171915.47-360226.3 &    1 & 171915.59-360225.5 &  1.6 \\
 171915.55-355559.8 &  -99 &                ... &  NaN \\
 171916.19-355409.2 &    2 & 171916.09-355409.2 &  1.2 \\
 171916.35-355624.5 &  -99 &                ... &  NaN \\
 171916.41-360346.5 &    3 & 171916.41-360346.8 &  0.3 \\
 171917.21-355236.2 &    4 & 171917.12-355235.0 &  1.6 \\
 171917.97-355004.9 &    5 & 171917.93-355005.0 &  0.4 \\
    \enddata
\tablenotetext{a}{This table is available in its entirety in a machine-readable form in the online journal.  A portion is shown here for guidance regarding its form and content.}
\end{deluxetable}


\clearpage

\section{X-RAY CHARACTERIZATIONS OF MOXC TARGETS \label{sec:targets}}

The important results of MOXC take the form of electronic tables, presented in Section~\ref{sec:ptsrccat} above.  Here we wish to add context to these formulaic data products with a few simple comparative figures for the MOXC MSFRs (all shown with celestial J2000 coordinates).  We start by presenting the ACIS exposure maps for each MSFR (Panel (a) in all figures in this section) on a uniform greyscale such that the deepest exposure (490~ks for NGC~3603) is darkest and the shallowest exposure (27~ks for G29.96-0.02) appears lightest.  The exposure time (or a typical exposure for each pointing in the case of mosaicked fields) and total number of ACIS sources is noted on the figure.

Brighter ($\geq$5 net counts) ACIS point sources are overlaid for each field, color-coded by median X-ray energy according to the legend provided.  Faint sources are not shown because their detection is a strong function of off-axis angle due to changing telescope sensitivity \citep{Broos11a}, although they are included in the total number of ACIS sources noted on the figure.  Median energy encodes both the hardness of the intrinsic X-ray spectrum and the obscuration to the source, so it is a poor substitute for an X-ray spectral fit, but such fits are only reliable for brighter X-ray sources and go beyond the scope of MOXC.  In very crowded cluster cores, source symbols overlap so much that not every source can been seen in these plots; symbol layering began with soft sources (red) and ended with the hardest sources (blue), so the number of soft sources in crowded regions may be under-represented by this graphic.

We also provide a multiwavelength perspective on MOXC targets by presenting (in Panel (b) in all figures in this section) an adaptively-smoothed \citep{Townsley03,Broos10} 0.5--7~keV X-ray image of the unresolved emission remaining after all ACIS point sources have been removed (shown in blue; the energy range is truncated at 7~keV to avoid increasing instrumental background at higher energies, which is important for wide-field diffuse images).  This residual X-ray emission certainly includes unresolved X-ray point sources, both lower-mass pre-MS stars in the MSFR and foreground/background components from field stars, the Galactic Ridge Emission, and unresolved AGN.  In many cases, though, it is likely dominated by hot plasma emanating from the MSFRs---the combined effects of massive star winds and, for older complexes, cavity supernova remnants (SNRs) ---thus we will refer to this unresolved emission with the general term ``diffuse.''  We have studied this diffuse X-ray emission in several MYStIX/MOXC targets:  \citet{Townsley03} gave early results for the Rosette Nebula and M17; \citet{Townsley11b} studied the diffuse X-ray components of the Carina Nebula in detail; \citet{Townsley11c} gave basic results for the global diffuse spectra of M17, NGC~3576, NGC~3603, and 30~Doradus and compared them to the global diffuse X-ray spectrum of the Carina Nebula.  

Detailed spectral analysis of diffuse X-ray emission in the MOXC datasets will be reserved for future papers. Here, however, we can briefly illustrate that all MOXC targets contain diffuse X-ray emission by displaying the point-source-excised smoothed X-ray images in the context of \Spitzer\  or {\em WISE} mid-IR images.  Thus Panel (b) in most figures in this section, in addition to showing ACIS diffuse emission (full-band, defined as 0.5--7~keV for diffuse images) in blue, shows \Spitzer/IRAC 8~$\mu$m emission tracing photodissociation regions \citep{Benjamin03} in green, and \Spitzer/MIPS 24~$\mu$m emission tracing ionization fronts \citep{Carey09} in red ({\em WISE} images are used when \Spitzer\ images lack sufficient field coverage).  These images are scaled independently for each target to emphasize diffuse structures; unlike the exposure maps in Panel (a) of each figure, they are {\em not} scaled globally for comparison between MSFRs.  For diffuse X-ray emission, such global comparisons between targets are only relevant in the context of careful X-ray spectral fitting, where the wide range of MSFR absorbing columns can be considered, unresolved point source populations can be modeled, and intrinsic diffuse emission surface brightnesses can be calculated \citep{Townsley11c}.  Although the MOXC paper is primarily a catalog of ACIS point sources, we include this presentation of the morphology of diffuse X-ray emission in MSFRs because it is the winds and explosions of massive stars in these MSFRs that generate much of this diffuse X-ray emission, and seeing it superposed on mid-IR bubbles and alongside the spatial distribution of detected X-ray point sources will help in its interpretation.

In addition to these two standard figure panels shown below for all targets, we show ACIS binned event images for some targets, zoomed in on cluster centers and interesting sources.  These are two-color X-ray images coded by event energy:  soft events (0.5--2~keV) are shown in red; hard events (2--8~keV) are shown in green.  {\em ACIS Extract} source extraction apertures are shown as blue polygons; these typically denote the 90\% enclosed energy contour of the 1.5~keV PSF but may be reduced to smaller enclosed energy fractions for crowded sources, to minimize confusion in the extracted spectra \citep{Broos10}.  

In these short, qualitative vignettes of the 12 MOXC MSFRs, we do not attempt an exhaustive review of the literature or a detailed description of these famous targets.  For readers unfamiliar with these MSFRs, we note that Appendix~A in the MYStIX introductory paper \citep{Feigelson13} briefly describes all MYStIX targets; Table~1 of that paper also notes the chapters in the {\em Handbook of Star Forming Regions} \citep{Reipurth08} that give reviews for most MYStIX targets.  Brief target introductions, in an X-ray context, were given in \citet{Townsley11c} for M17, NGC~3576, NGC~3603, and 30~Doradus.  To better appreciate the physical scales of the zoomed panels, recall that Table~\ref{targets.tbl} gives the conversion from arcminutes to parsecs for the assumed MSFR distance.

Several CCCP papers \citep[e.g.,][]{Townsley11a,Broos11a,Feigelson11} show the ACIS exposure map for the 22-pointing ACIS-I mosaic of the Carina Nebula and describe the spatial distribution of its X-ray point sources; \citet{Townsley11b} study the diffuse X-ray emission in the Carina Nebula in some detail, so that MSFR complex is omitted here.  As noted above, the diffuse X-ray emission seen by ACIS in M17, NGC~3576, NGC~3603, and 30~Doradus was characterized by \citet{Townsley11c}; except for NGC~3576, more extensive ACIS datasets (wider fields and/or deeper exposures) for these MSFRs are shown here.  Diffuse X-ray emission in the Orion Nebula Cluster was studied by \citet{Guedel08} and the ACIS source list was presented by \citet{Getman05}, so that target is also omitted from this section.  

For completeness and for comparison with MOXC targets, Appendix~\ref{othertargets.sec} provides the two standard figure panels described above for several other MYStIX clusters.  \Chandra point source catalogs for those targets were presented by \citet{Kuhn13a} and \citet{Kuhn10} (for W40).  Target descriptions can be found in those papers and in the MYStIX introductory paper \citep{Feigelson13}.


\subsection{NGC~6334 (The Cat's Paw Nebula) \label{sec:n6334}}

NGC~6334 is one of the closest and youngest examples of a giant molecular cloud (GMC) complex engaged in rapid, extensive, nearly coeval multiple MYSC formation \citep{Russeil10,Russeil12,Russeil13}.  NGC~6334 is a giant \hii region (GH{\scriptsize II}R) fueled by a number of MYSCs oriented in a line parallel to the Galactic Plane \citep{Persi10}.  This hierarchical configuration is known as a ``cluster of clusters'' \citep[e.g.,][]{Bastian07,Elmegreen08}.  This mode of star formation may come from the collapse of a cylindrical (rather than spherical) GMC \citep{Jackson10}, as evidenced by the long, dusty filament that threads through the MYSCs in NGC~6334 \citep{Matthews08,Zernickel13}.  As we will show below, we find many cluster-of-clusters examples among the MOXC targets.  

\begin{figure}[htb]
\centering
\includegraphics[width=0.48\textwidth]{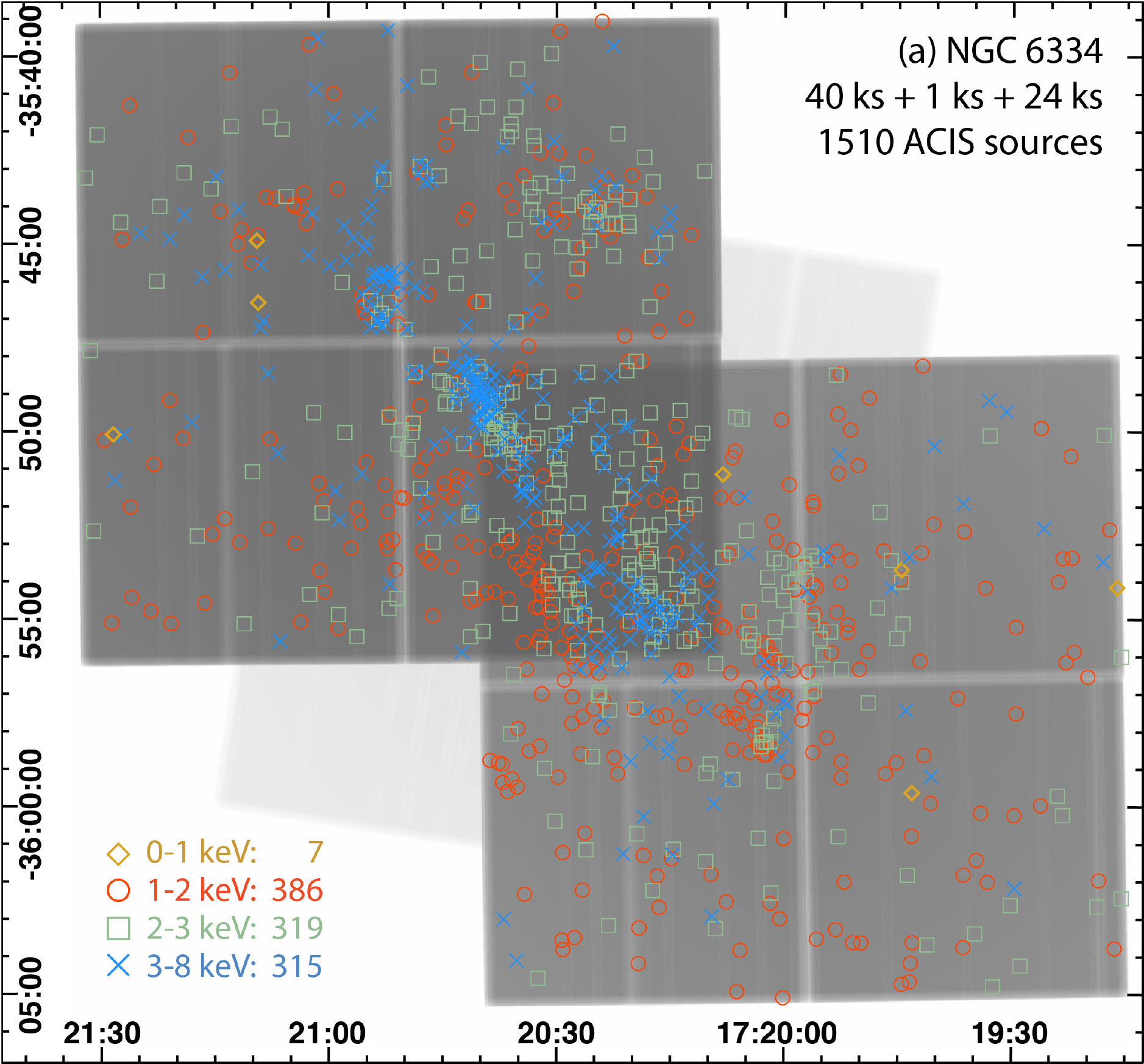}
\includegraphics[width=0.48\textwidth]{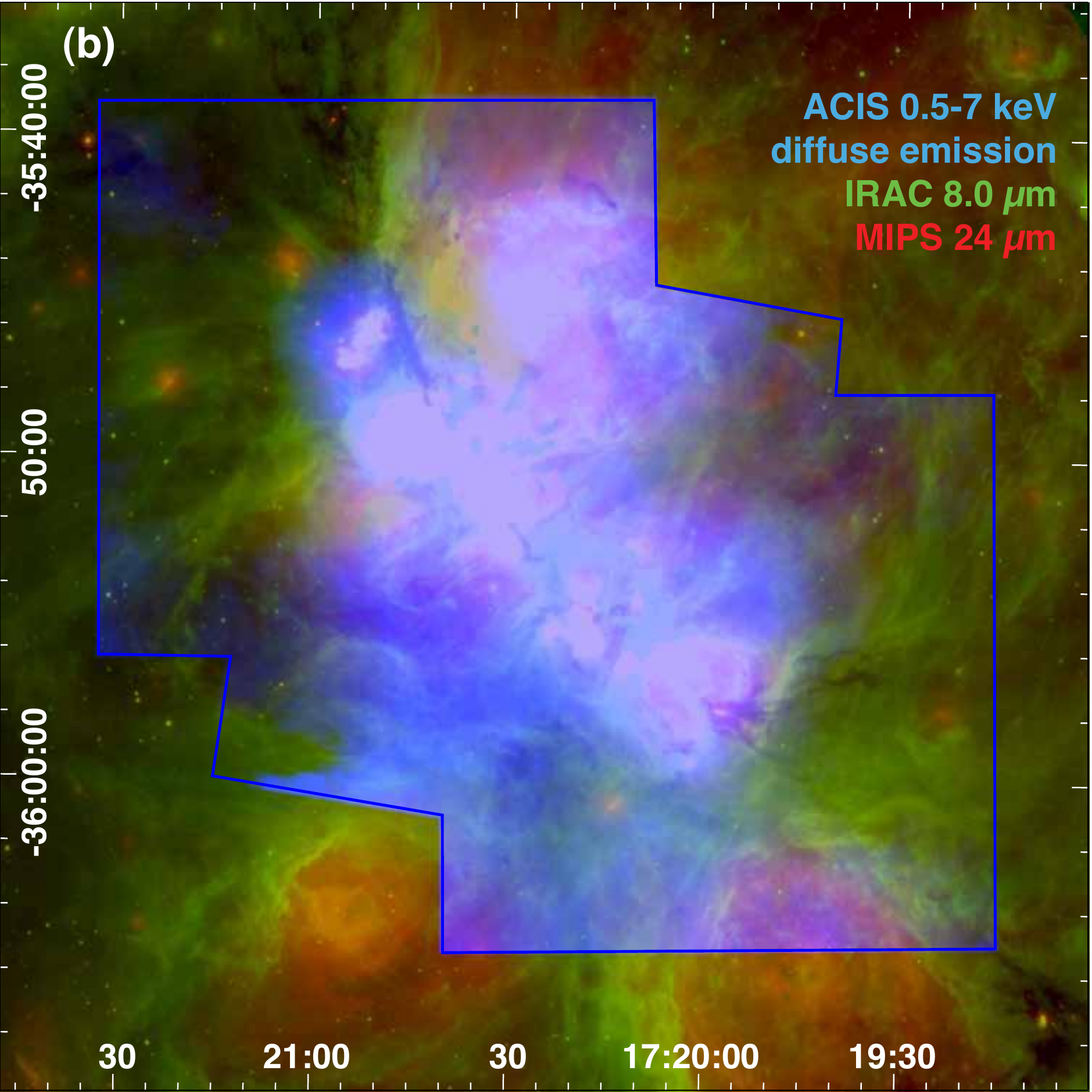}
\caption{NGC~6334 (The Cat's Paw Nebula); all target images are shown in celestial J2000 coordinates.
(a) ACIS exposure map with brighter ($\geq$5 net counts) ACIS point sources overlaid, with symbols and colors denoting median energy for each source.
(b) ACIS diffuse emission (full-band, 0.5--7~keV) in blue, shown in the context of \Spitzer/IRAC 8~$\mu$m emission tracing photodissociation regions (green) and \Spitzer/MIPS 24~$\mu$m emission tracing ionization fronts (red).   
\label{n6334.fig}}
\end{figure}

The original {\em Chandra}/ACIS mosaic of NGC~6334 consisted of two 40-ks ACIS-I pointings \citep{Ezoe06}, but in our analysis the southwest pointing (ObsID 2573) lost 40\% of its exposure due to high background from unsettled space weather.\footnote
{The time variability of the ACIS background is discussed in \S6.16.3 of the {\Chandra Proposers' Observatory Guide} (\url{http://asc.harvard.edu/proposer/POG/}) and in the ACIS Background Memos (\url{http://asc.harvard.edu/cal/Acis/Cal_prods/bkgrnd/current/}).
}
Figure~\ref{n6334.fig}a shows the ACIS exposure map for the mosaic, with the brighter point sources superposed, coded by median energy as described above; the legend shows the number of sources in each median energy bin.  The number of unplotted faint ACIS sources can be calculated by subtracting these median energy tallies from the total number of ACIS sources (given in black at the top of the figure).

Figure~\ref{n6334.fig} only shows the wide-field view of the ACIS mosaic because zoomed images of the pointing centers were presented in \citet{Feigelson09}.  That paper included an ACIS source list obtained with earlier versions of the methods used here (see Section~\ref{sec:other_analyses} above); we completely re-analyzed the ACIS observations of NGC~6334 for MYStIX and present that new source list here for consistency with the other targets in MOXC.  We also included the 1-ks ObsID 8975; while this additional short dataset is not particularly important for X-ray point source studies in NGC~6334, it is surprisingly useful for understanding the extent of the diffuse X-ray emission (Figure~\ref{n6334.fig}b). 

Shown in blue in Figure~\ref{n6334.fig}b is that diffuse X-ray emission; as mentioned above, all ACIS point sources have been excised, then the remaining emission smoothed with an adaptive kernal smoothing routine \citep{Townsley03,Broos10}.  \Chandra shows that NGC~6334 is pervaded by hot plasma and appears to exhibit a large ionized bipolar outflow expanding away from its ``backbone'' of MYSCs (Figure~\ref{n6334.fig}b).  The unresolved X-ray emission is likely dominated by truly diffuse hot plasma rather than by unresolved pre-MS stars in the GH{\scriptsize II}R, since it shows striking complementarity to the famous bubble structures revealed by \Spitzer.  We will see that this is a common theme in all MOXC targets; while the diffuse X-ray emission varies substantially in apparent surface brightness among the different MSFRs (both because of intervening absorption and intrinsic variations in the strength of the hot plasma's X-ray emission), it is always discernable in these \Chandra observations, emanating from cluster cores and filling \Spitzer\ bubbles.

The \Spitzer/GLIMPSE observation (IRAC 8~$\mu$m; green) traces mainly PAH emission and shows the edges of the photon-dominated regions \citep{Heitsch07,Deharveng10}.  The \Spitzer/MIPSGAL observation (MIPS 24~$\mu$m; red---note that light pink regions show missing 24~$\mu$m data due to saturation) traces heated dust \citep{Carey09} and often coincides with H$\alpha$ emission; these show the ionization fronts in the \hii region complex.  The footprint of the 3-pointing ACIS-I mosaic is outlined in blue.

Especially unusual is the fact that NGC~6334 lies just $\sim$$2^{\circ}$ ($\sim$59~pc) away from NGC~6357, another cluster-of-clusters with a prominent bipolar outflow that we have studied with ACIS (see below).  Both of these \GHIIR complexes are thought to have formed from the same GMC \citep{Russeil10}.  


\subsection{NGC~6357 (The War and Peace Nebula) \label{sec:n6357}}

\begin{figure}[htb]
\centering
\includegraphics[width=0.5\textwidth]{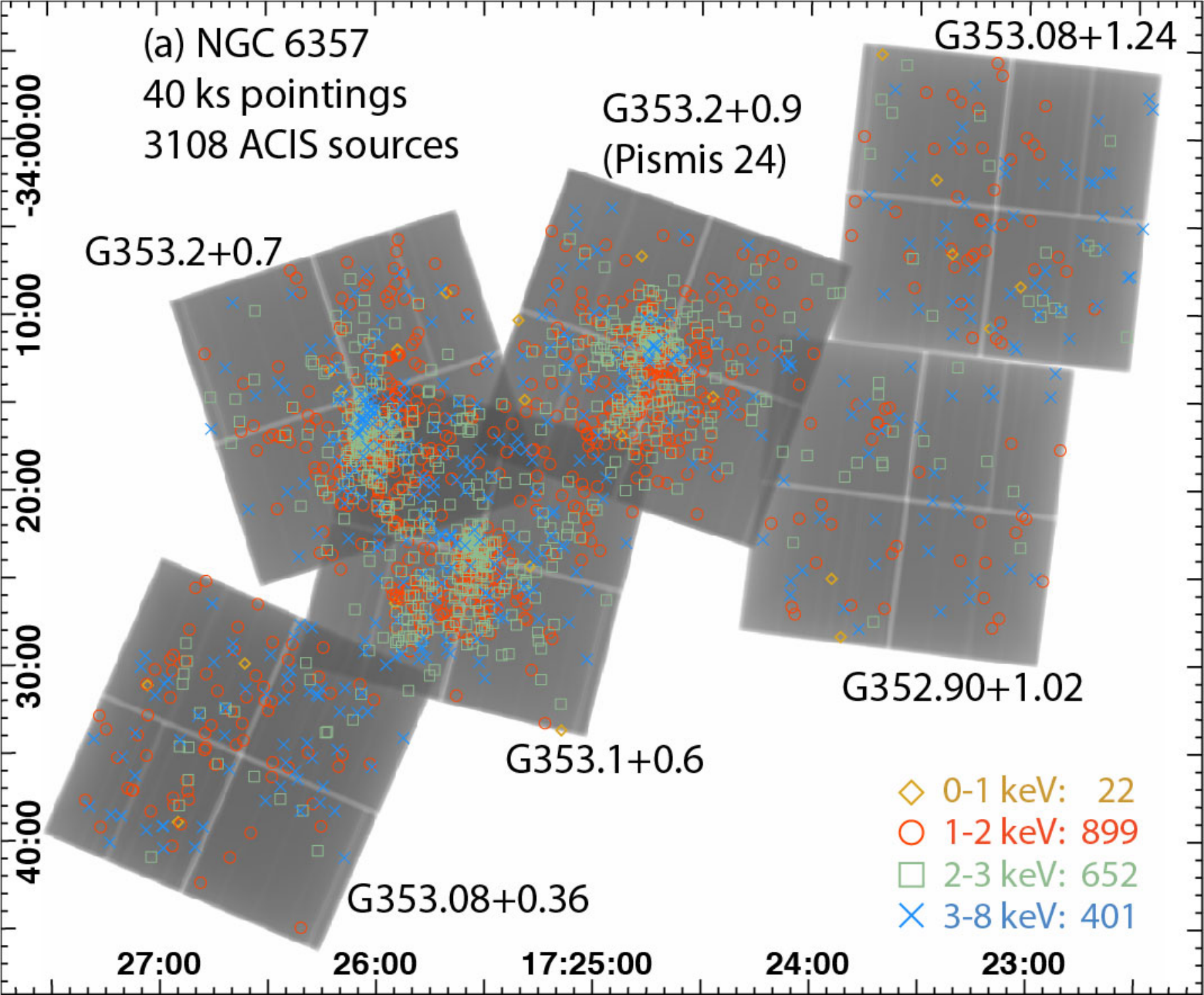}
\includegraphics[width=0.48\textwidth]{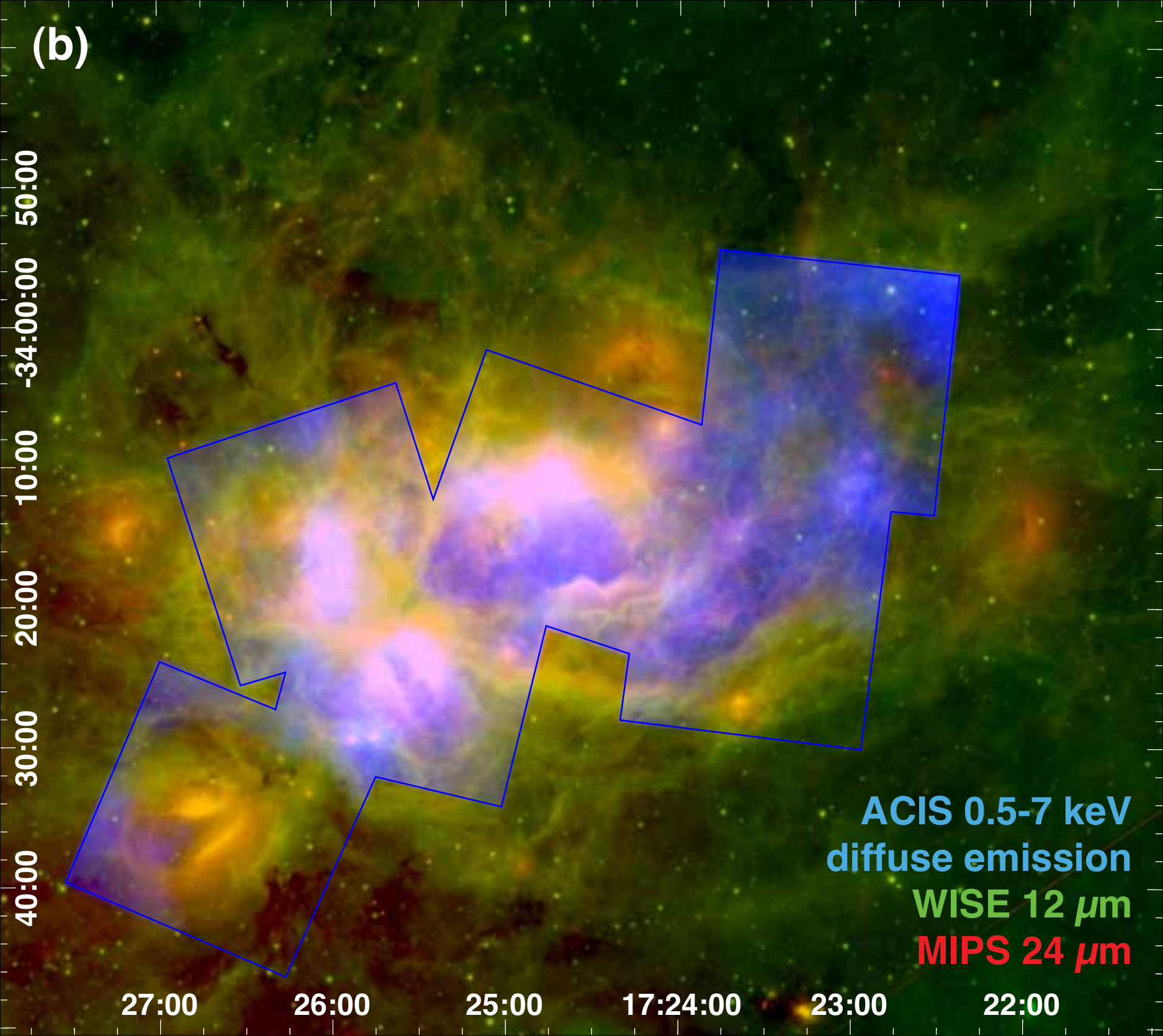}
\includegraphics[width=0.24\textwidth]{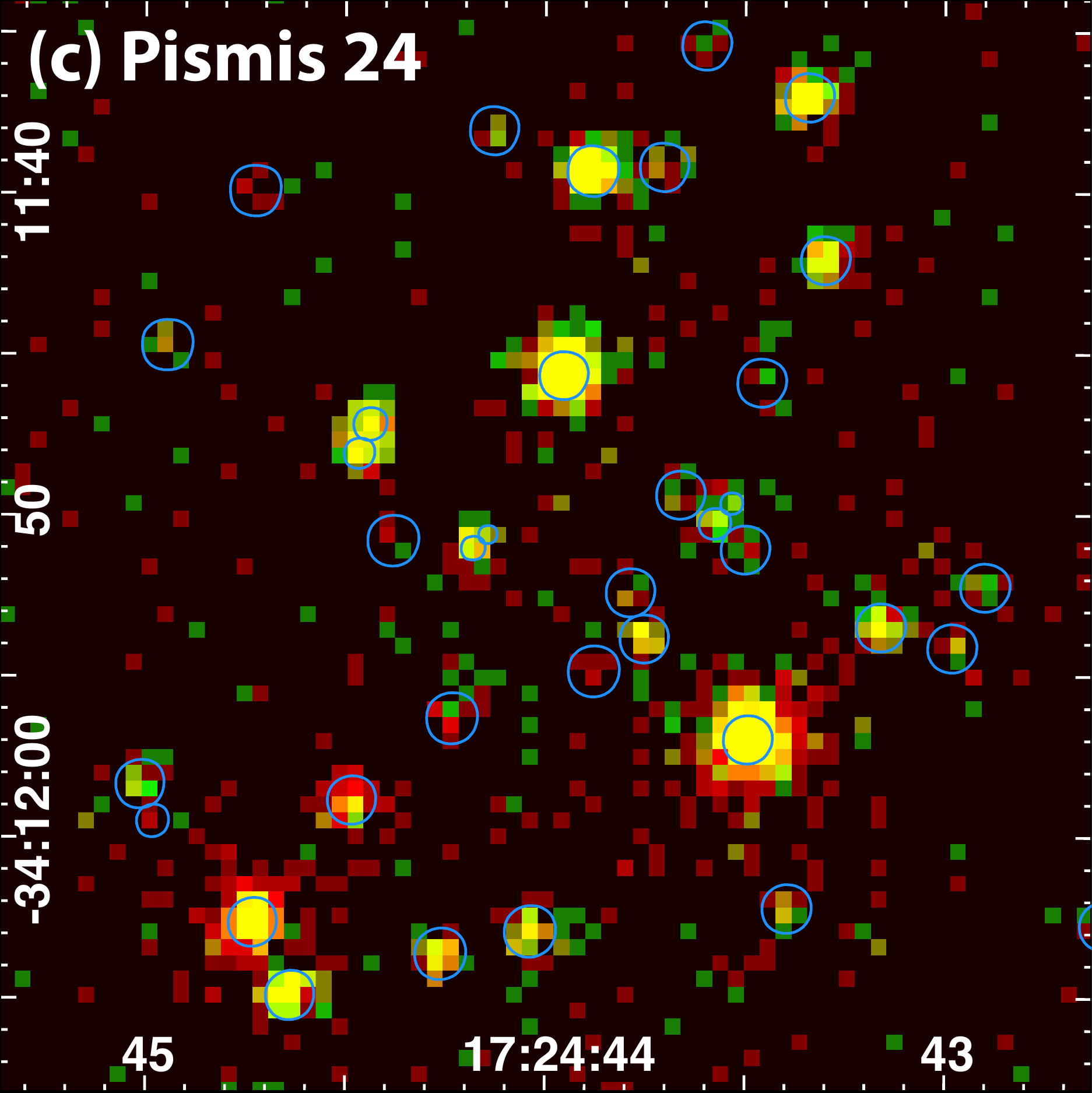}
\includegraphics[width=0.24\textwidth]{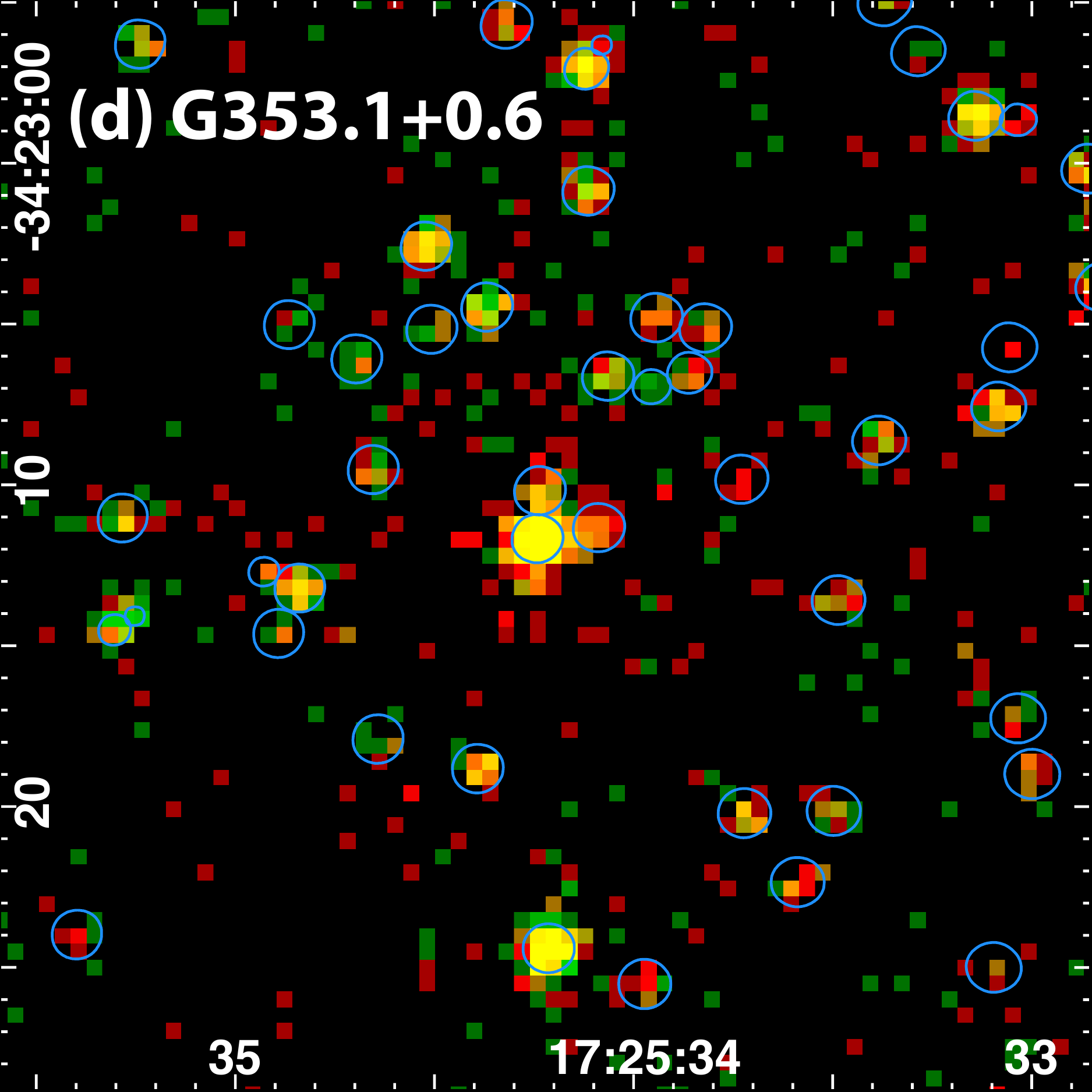}
\includegraphics[width=0.24\textwidth]{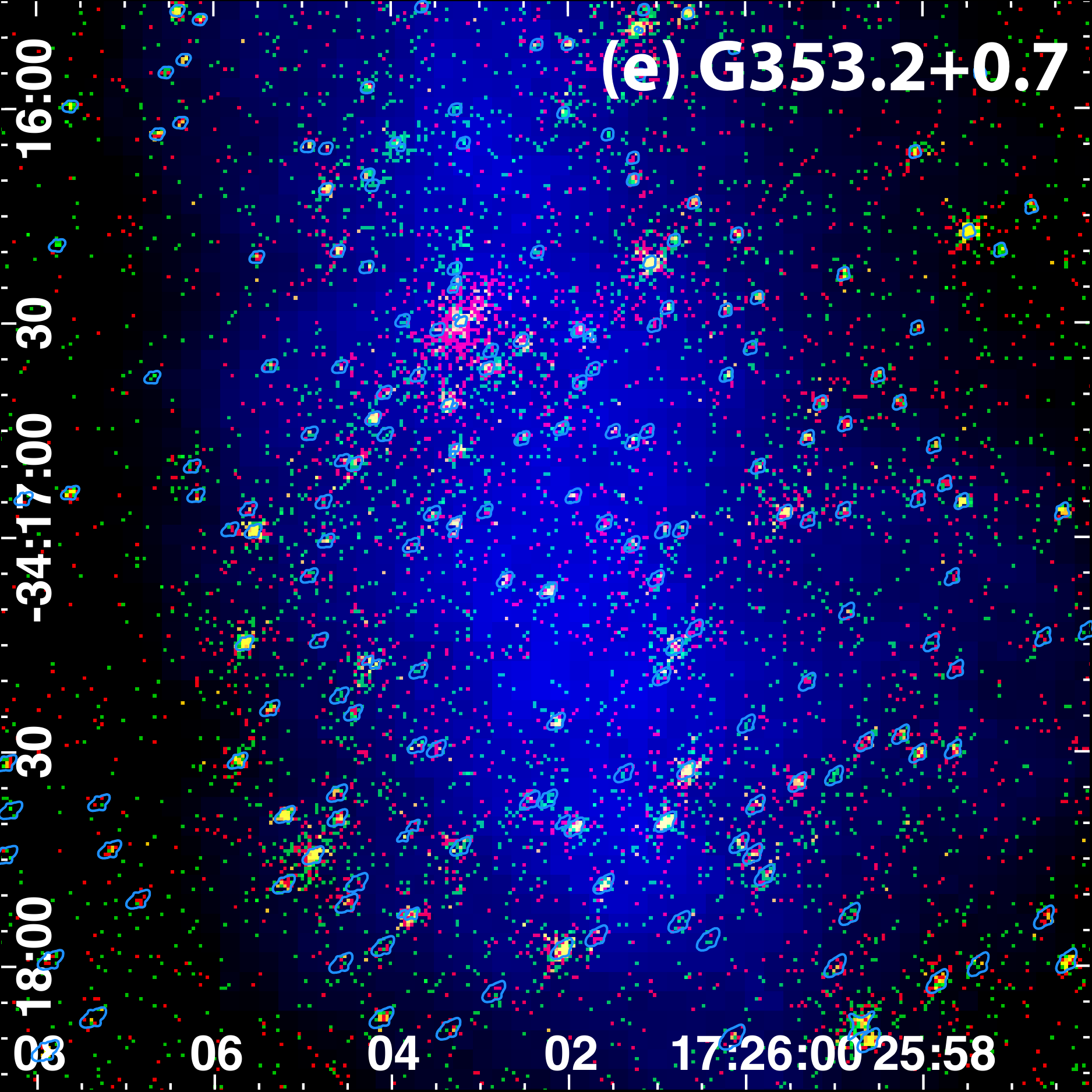}
\includegraphics[width=0.24\textwidth]{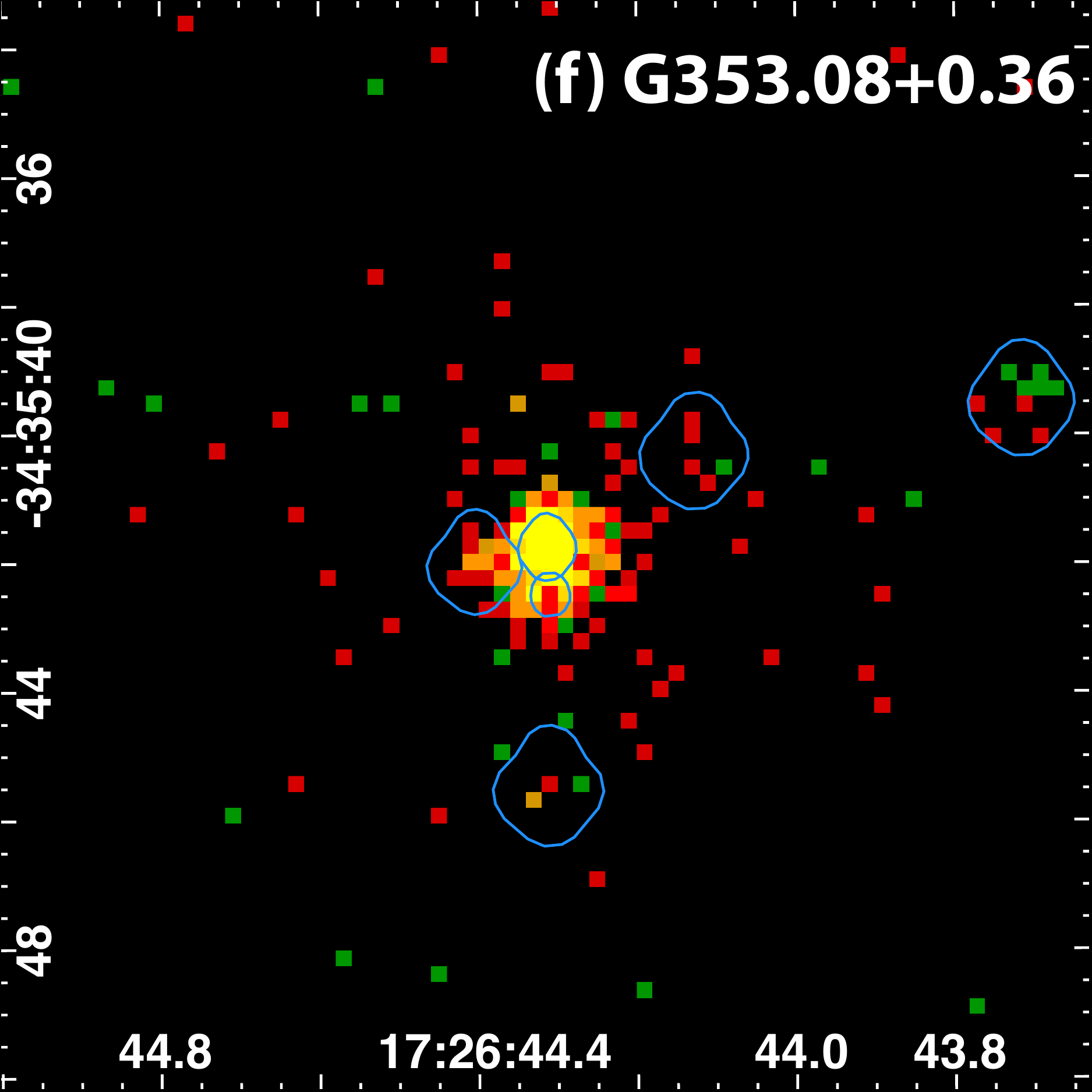}
\caption{NGC~6357 (The War and Peace Nebula).
(a) ACIS exposure map with brighter ($\geq$5 net counts) ACIS point sources overlaid, with symbols and colors denoting median energy for each source.
(b) ACIS diffuse emission (full-band, 0.5--7~keV) in blue, {\em WISE} Band 3 (12~$\mu$m) emission in green (\Spitzer/IRAC data do not cover the full ACIS field), and \Spitzer/MIPS 24~$\mu$m emission in red.  
(c--f) ACIS binned event images centered on the eastern four pointings, with soft events (0.5--2~keV) shown in red and hard events (2--8~keV) in green.  Point source extraction regions are overlaid in blue.
\label{n6357.fig}}
\end{figure}

The \GHIIR NGC~6357 has produced at least three MYSCs that have blown parsec-scale bubbles.  Overall it appears to be slightly older than its neighbor \GHIIR NGC~6334, with a prominent 60$\arcmin$-diameter shell seen in H$\alpha$ opening away from the Galactic Plane \citep{Lortet84,Cappa11}, which lies southeast of the complex (Figure~\ref{n6357.fig}a).  The brightest \hii region in NGC~6357 is G353.2+0.9, ionized by the MYSC Pismis~24 \citep{Bohigas04,Maiz07}.  Our original 40-ks ACIS GTO observation of this MYSC \citep{Wang07} revealed $\sim$800 X-ray point sources (Figure~\ref{n6357.fig}a,c).  As noted above, recent millimeter continuum work \citep{Munoz07,Russeil10} has shown that NGC~6357 and NGC~6334 formed at opposite ends of the same GMC; at a distance of just 1.7~kpc, these two clusters-of-clusters (one oriented parallel to the Galactic Plane, the other perpendicular to it) merit close attention.  NGC~6357's H$\alpha$ shell may outline a proto-superbubble blown by an older MYSC that has now dissolved into a ``distributed population'' of young stars.  The presence of a Wolf-Rayet star inside the shell and the wide distribution of X-ray-emitting pre-MS stars around Pismis~24 \citep{Wang07} are indirect evidence for such an older MYSC; its cavity supernovae might have helped to expand its wind-blown bubble into today's large H$\alpha$ shell.    

East and southeast of G353.2+0.9 appear two smaller bubbles, G352.2+0.7 and G353.1+0.6 (Figure~\ref{n6357.fig}a, b).  G353.1+0.6 was known to contain a poorly-studied MYSC with O3 stars, known as AH03~J1725-34.4 \citep{Dias02,Russeil12} or as [BDSB2003]~101 \citep{Damke06,Borissova05,Bica03}; our 40-ks ACIS-I observation reveals another $\sim$800 X-ray sources there (Figure~\ref{n6357.fig}a,d).  The region around G352.2+0.7 was serendipitously captured 20$^{\prime}$ off-axis (on the ``S2'' detector) in the original Pismis~24 ACIS-I dataset, where we noticed a diffuse patch of hard X-rays coincident with a bright {\em IRAS} ``confused region.''  This prompted us to propose an ACIS-I pointing on G352.2+0.7; this new 40-ks observation revealed yet a third MYSC in NGC~6357, also with $\sim$800 X-ray sources (Figure~\ref{n6357.fig}a,e).  The wide field of Figure~\ref{n6357.fig}e shows the populous X-ray cluster plus full-band (0.5--7 keV) ACIS diffuse emission in blue.  X-ray spectral analysis is necessary to determine if this is dominated by unresolved point sources or diffuse shock emission from massive star winds.

H$\alpha$, mid-IR, and millimeter continuum \citep{Russeil10} features suggest that NGC~6357 probably contains other young clusters in its lower superbubble lobe, which is expanding more slowly into the dense GMC towards the Galactic Plane.  One of these regions contains hot cores and other signposts of massive star formation, and has V-shaped ionized bars filled with H$\alpha$ emission and bright at 24~$\mu$m.  We observed this region, which we call G353.08+0.36, with ACIS-I for 60~ks of GTO time (easternmost pointing in Figure~\ref{n6357.fig}a,b) and were thoroughly surprised to find that no rich cluster resides here; rather we find a single bright X-ray source at the center of the field (ACIS source name J172644.31-343541.7), with a few faint companions (Figure~\ref{n6357.fig}f).  A simple X-ray spectral fit shows that this bright source is obscured by $A_V \sim 15$~mag of extinction and is dominated by a soft thermal plasma ($kT \sim 0.4$~keV); its absorption-corrected 0.5--8~keV X-ray luminosity is $L_X \sim 5 \times 10^{32}$~erg~s$^{-1}$.  Details of this fit and more sophisticated X-ray spectral fitting will be featured in a future paper, but here we can assert that this soft plasma, high X-ray luminosity, and the source's $K$-band magnitude of 6.9 (from 2MASS) all indicate that this is a massive star.

V-shaped 24~$\mu$m features similar to what we see in G353.08+0.36 are common in the \Spitzer/MIPSGAL survey \citep{Carey09}; perhaps they are signposts of massive star formation, but apparently they do not necessarily signal the presence of a MYSC.  In this case, we are likely seeing the alternate scenario---a massive star that has formed in (near) isolation.  It is unlikely to be a runaway from one of the nearby MYSCs, given its cohort of fainter X-ray (thus likely lower-mass pre-MS) companions.

Once the $>$3100 ACIS point sources are removed from the ACIS mosaic, hot plasma filling the bubbles blown by NGC~6357's MYSCs can easily be seen (Figure~\ref{n6357.fig}b).  The morphology of the diffuse X-ray emission lends a more three-dimensional sense to the mid-IR view of the complex; areas that lack diffuse X-ray emission either have no hot plasma (displacement) or they have enough intervening cold material that the soft X-rays generated by this plasma are absorbed (shadowing).  Two new ACIS-I pointings obtained last year expand the mosaic to the west and northwest, enhancing our exploration of the diffuse X-ray emission in NGC~6357.  They sample hot plasma threading through channels and fissures in the degree-sized shell; this bowl-shaped structure is seen clearly in the \Spitzer\ and {\em WISE} data in Figure~\ref{n6357.fig}b, opening to the northwest.


\subsection{M16 (The Eagle Nebula) \label{sec:m16}}

\begin{figure}[htb]
\centering
\includegraphics[width=0.48\textwidth]{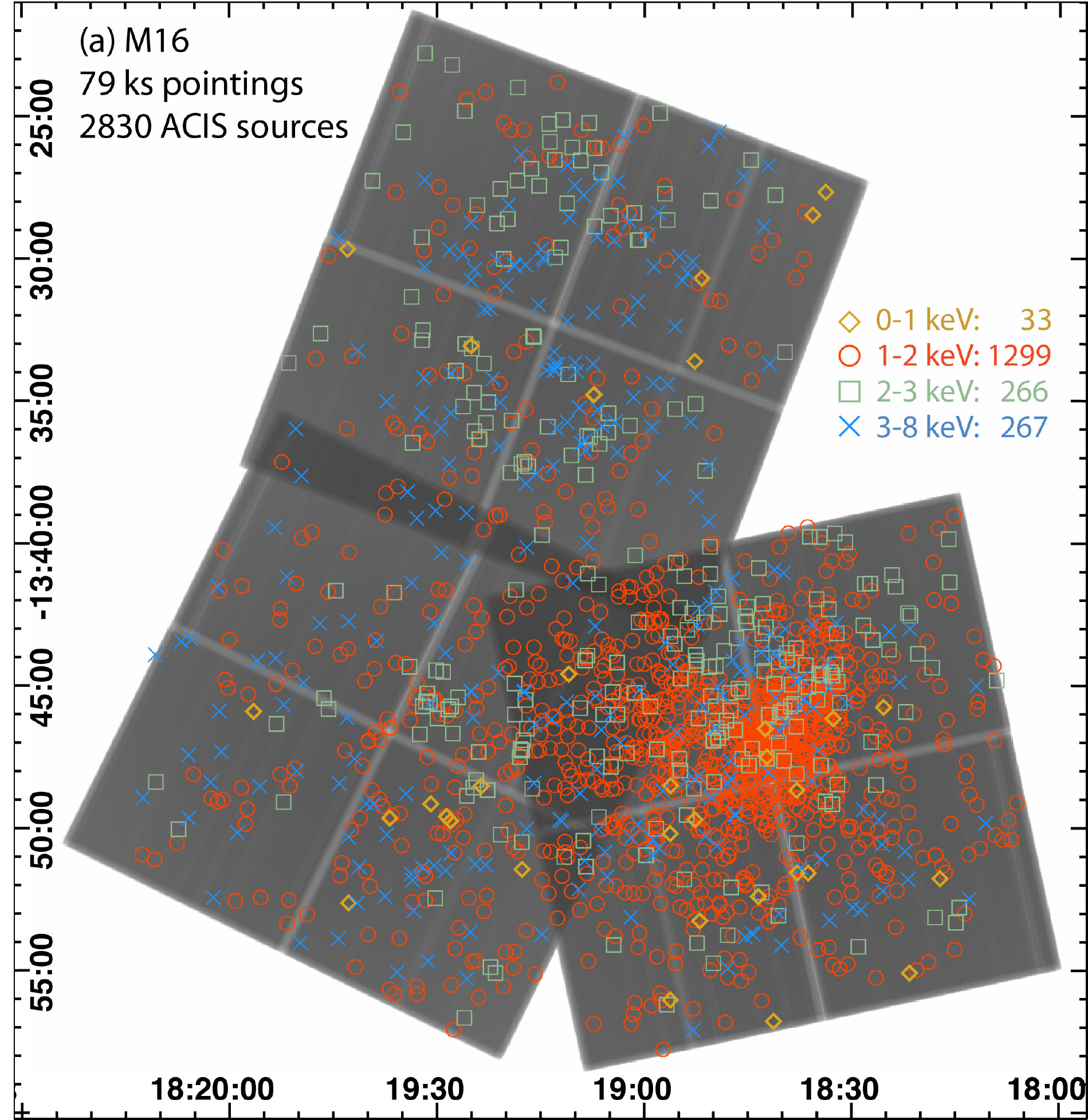}
\includegraphics[width=0.48\textwidth]{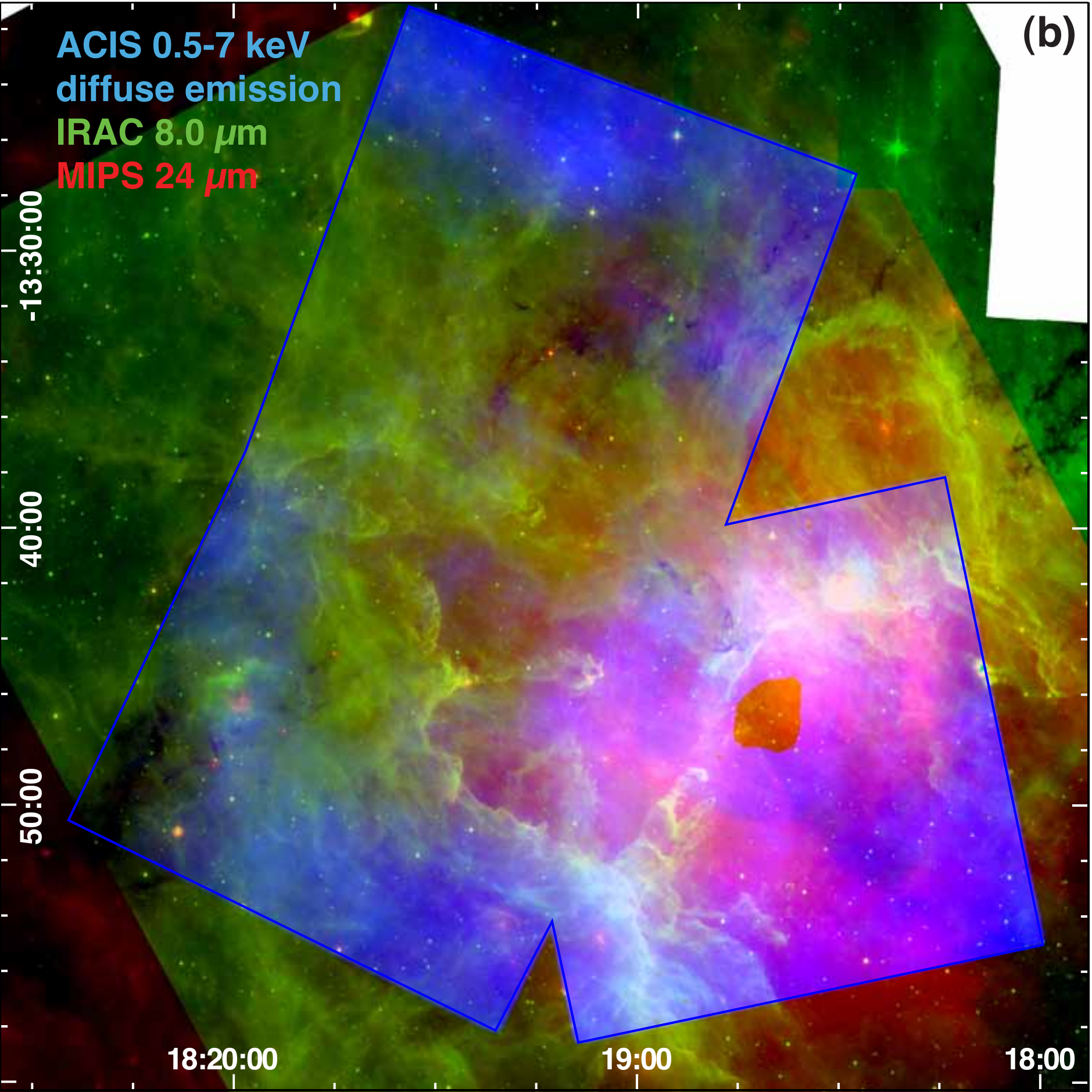}
\includegraphics[width=0.48\textwidth]{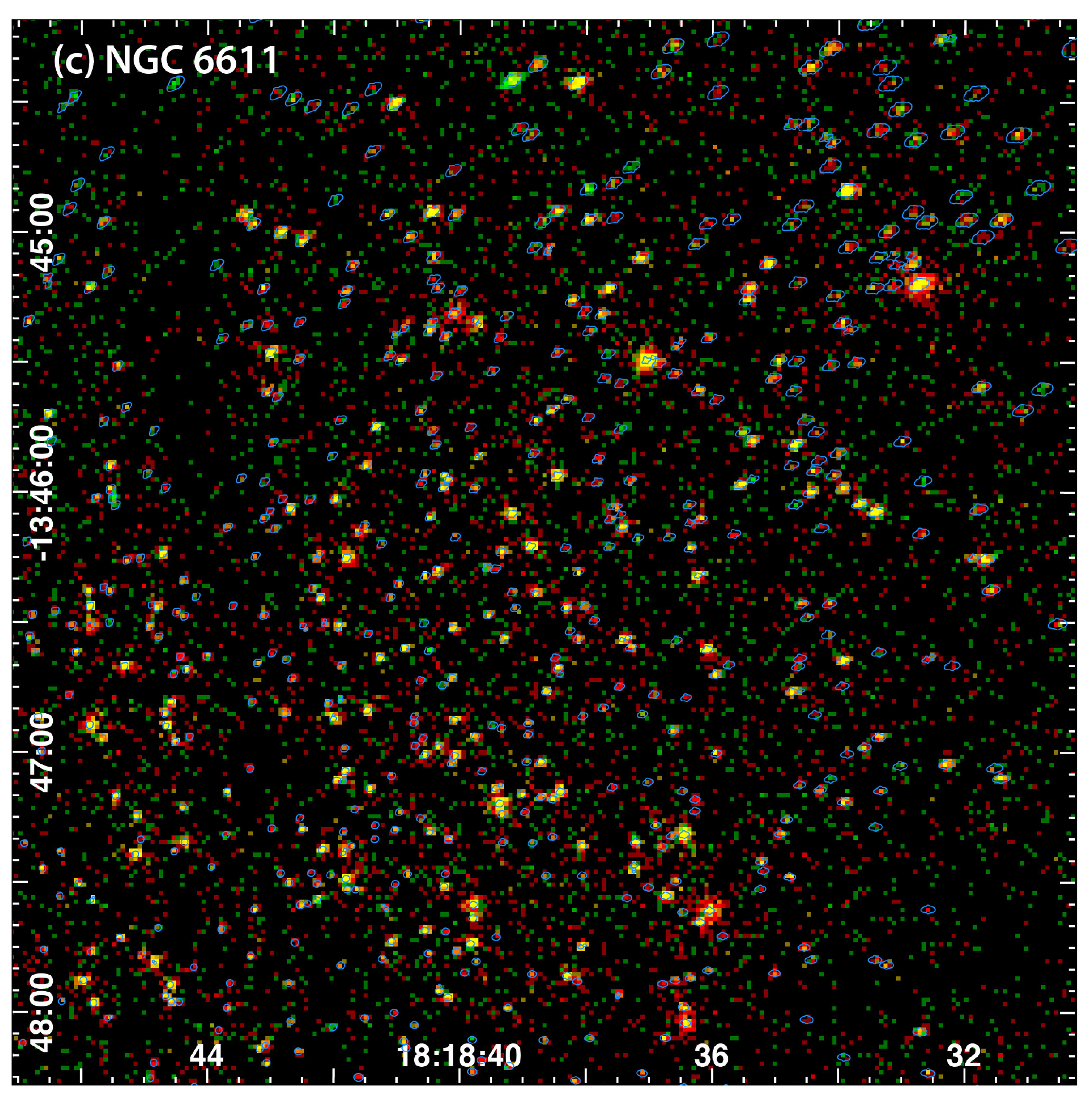}
\includegraphics[width=0.48\textwidth]{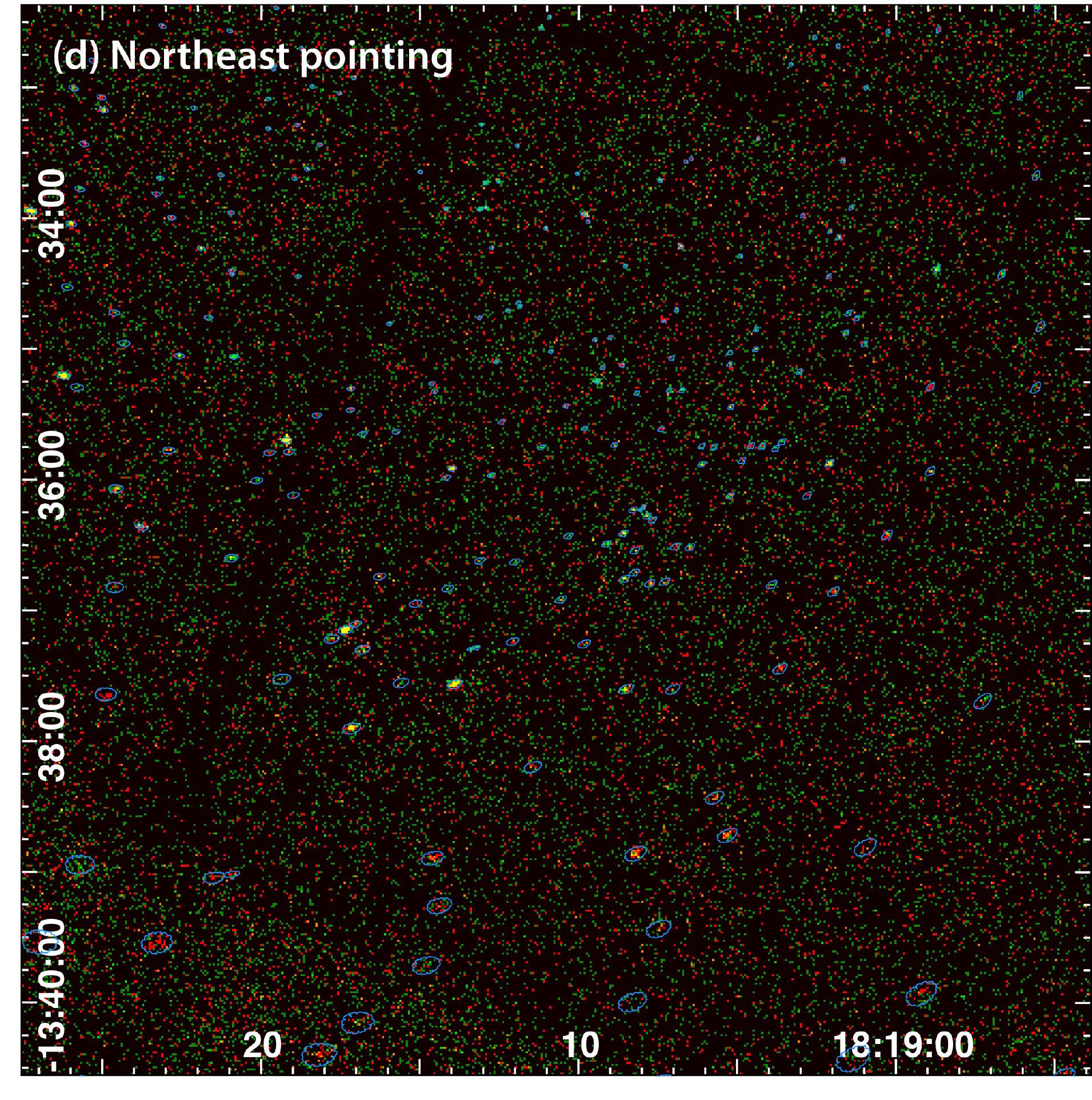}
\caption{M16 (The Eagle Nebula, NGC~6611).
(a) ACIS exposure map with brighter ($\geq$5 net counts) ACIS point sources overlaid, with symbols and colors denoting median energy for each source.
(b) ACIS diffuse emission (full-band, 0.5--7~keV) in blue, \Spitzer/IRAC 8~$\mu$m emission in green, and \Spitzer/MIPS 24~$\mu$m emission in red.  The core of NGC~6611 has been masked in the ACIS image, since it is likely to be dominated by unresolved point sources.   
(c,d) ACIS binned event images, with soft events (0.5--2~keV) shown in red and hard events (2--8~keV) in green.  Point source extraction regions are overlaid in blue.  
\label{m16.fig}}
\end{figure}

In contrast to the ``clusters-of-clusters'' NGC~6334 and NGC~6357, the famous Eagle Nebula (M16) is dominated by the single, monolithic MYSC NGC~6611 (Figure~\ref{m16.fig}).  The ACIS-I mosaic of this region consists of the original pointing on NGC~6611 \citep{Linsky07} plus two eastern pointings obtained to examine disk fractions in pre-MS stars far from the ionizing massive stars \citep{Guarcello10}.  \citet{Guarcello12} find 1755 X-ray sources; our methods push deeper, yielding 60\% more sources from the same observations.  Figure~\ref{m16.fig}c shows part of NGC~6611, illustrating the large spatial extent of this cluster.  Figure~\ref{m16.fig}d illustrates part of the ACIS field centered on the embedded northeast cluster described in \citet{Indebetouw07} and noted by \citet{Guarcello12}.


The \Spitzer\ data on M16 were analyzed by \citet{Flagey11}, who find a prominent heated dust shell that could be caused by the winds and radiation from NGC~6611's massive stars, or from an old supernova event from a very massive star in the cluster.  They analyze the \Chandra data on NGC~6611 and find evidence for faint diffuse X-ray emission; our analysis clearly confirms this finding (Figure~\ref{m16.fig}b).  Recent work \citep{DeMarchi13} gives evidence for a $\sim$16-Myr-old stellar population in the Eagle Nebula; if this older population is confirmed, it would provide a natural explanation for supernova activity in the region.  In our future detailed spectral analysis of M16's diffuse X-ray emission, we will search for abundance variations and for evidence of non-equilibrium ionization in the plasma components; these may be signs of supernova activity.  Based on the diffuse X-ray emission seen in regions likely to have experienced supernovae \citep[e.g., Carina,][]{Townsley11b} compared to those too young to have seen such activity \citep[e.g., M17,][]{Townsley03}, we predict that winds from massive stars can generate sufficient shocks to explain the luminosity of the diffuse X-ray emission in M16 and that supernovae in such \hii region cavities thermalize and fade quickly and are thus easily hidden in the wind-generated plasma emission.  In all likelihood, then, both mechanisms described by \citet{Flagey11} are at work in M16.

\clearpage

\subsection{M17 (The Omega Nebula) \label{sec:m17}}

\begin{figure}[htb]
\centering
\includegraphics[width=0.48\textwidth]{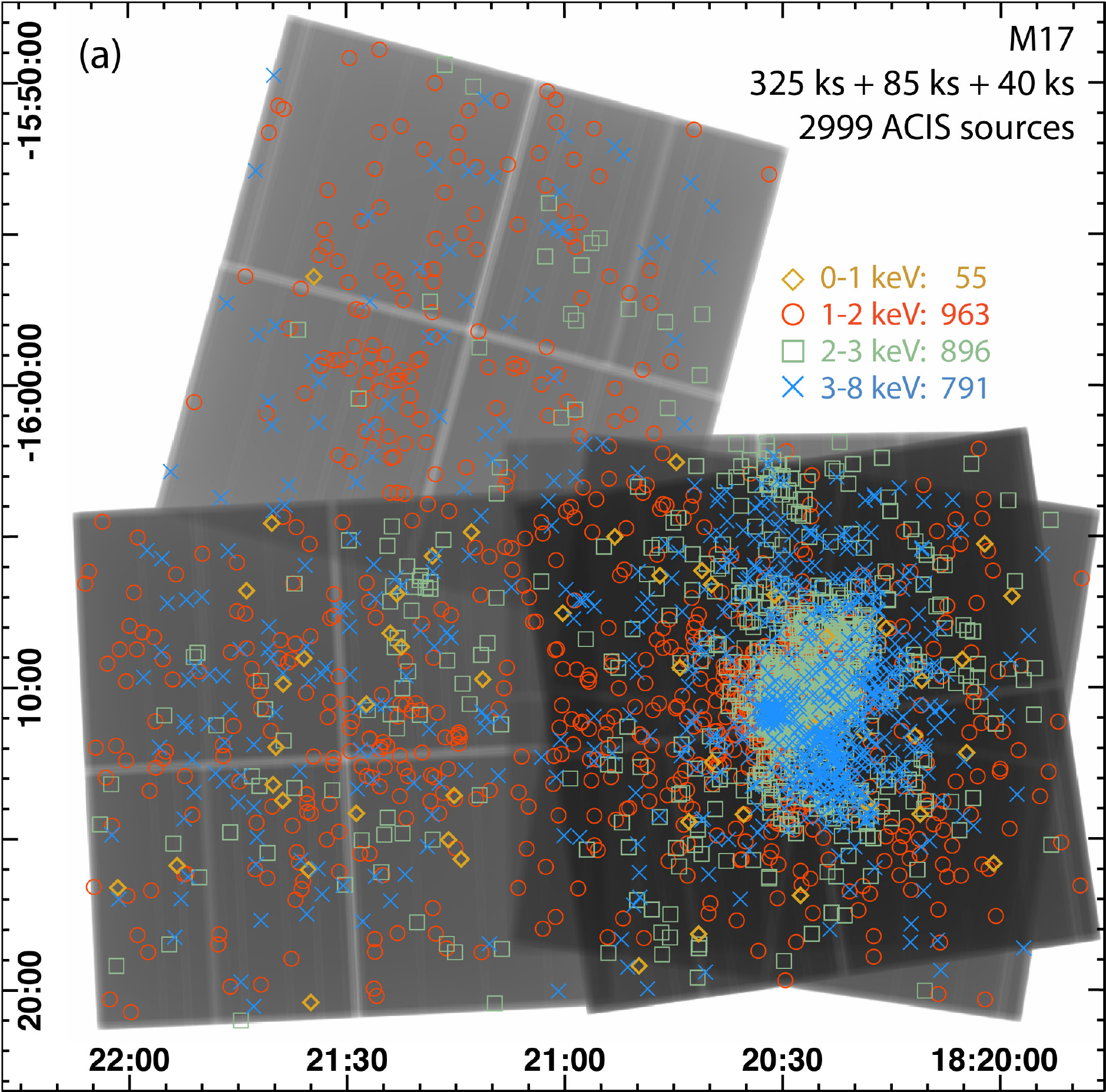}
\includegraphics[width=0.48\textwidth]{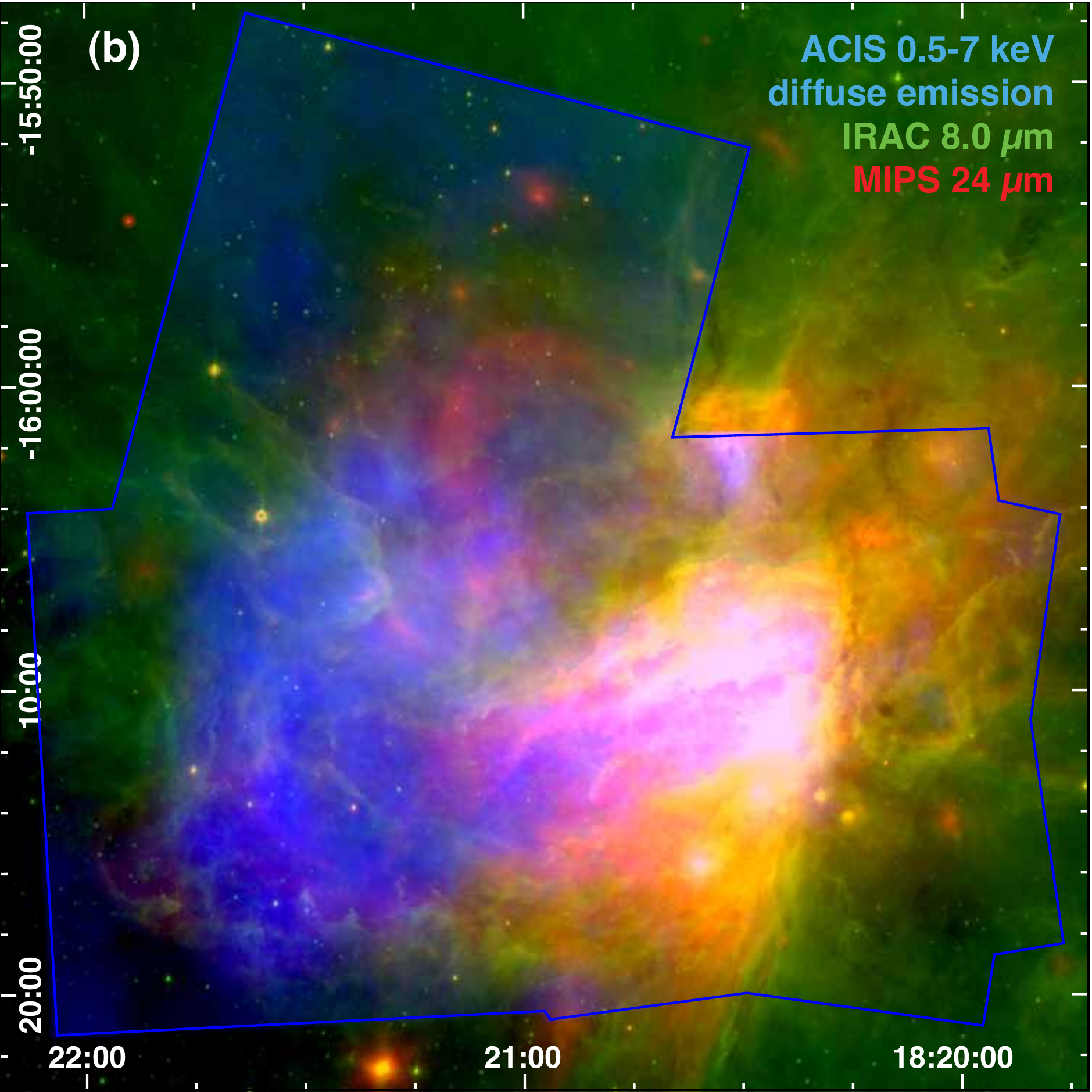}
\includegraphics[width=0.32\textwidth]{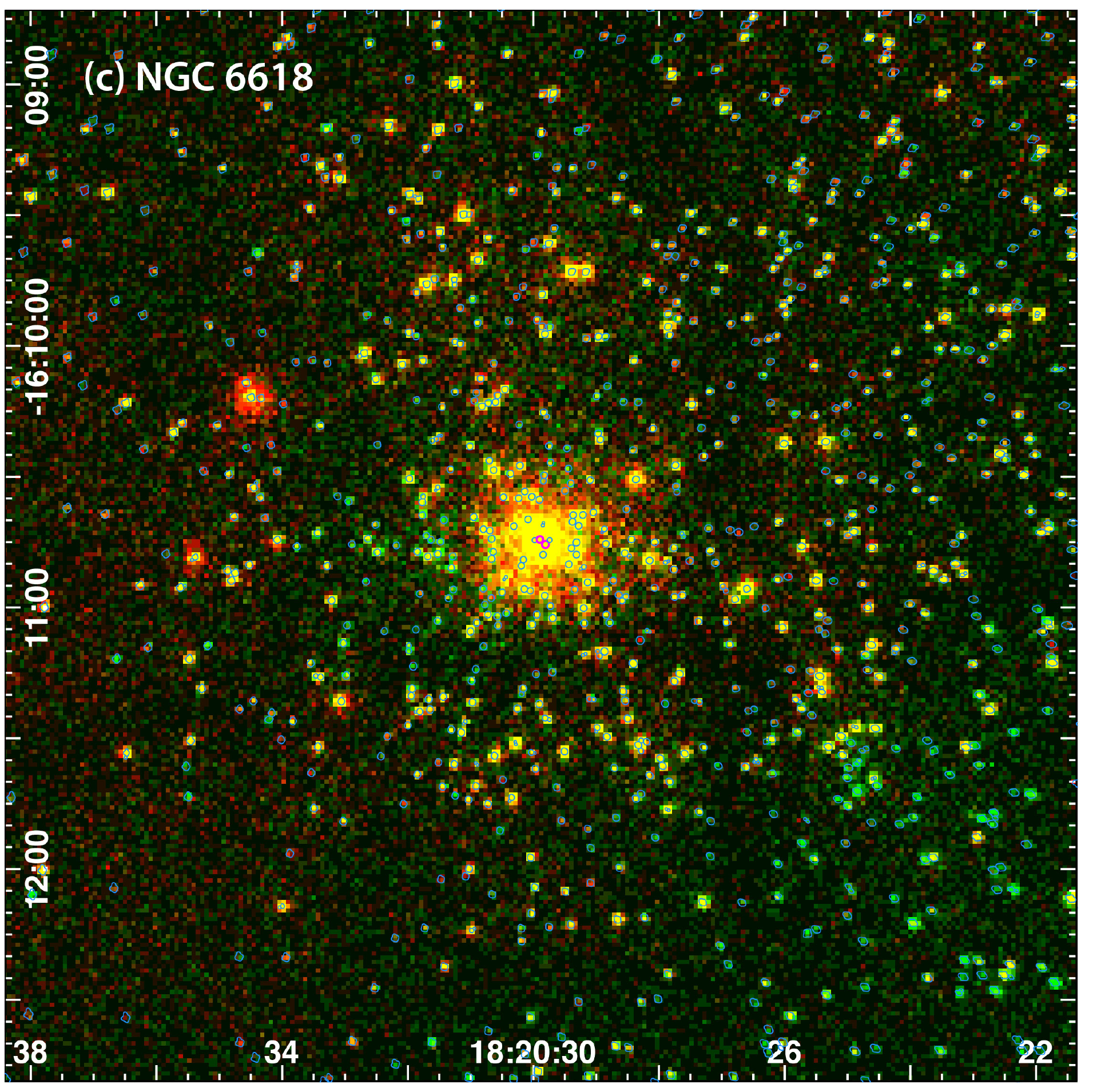}
\includegraphics[width=0.325\textwidth]{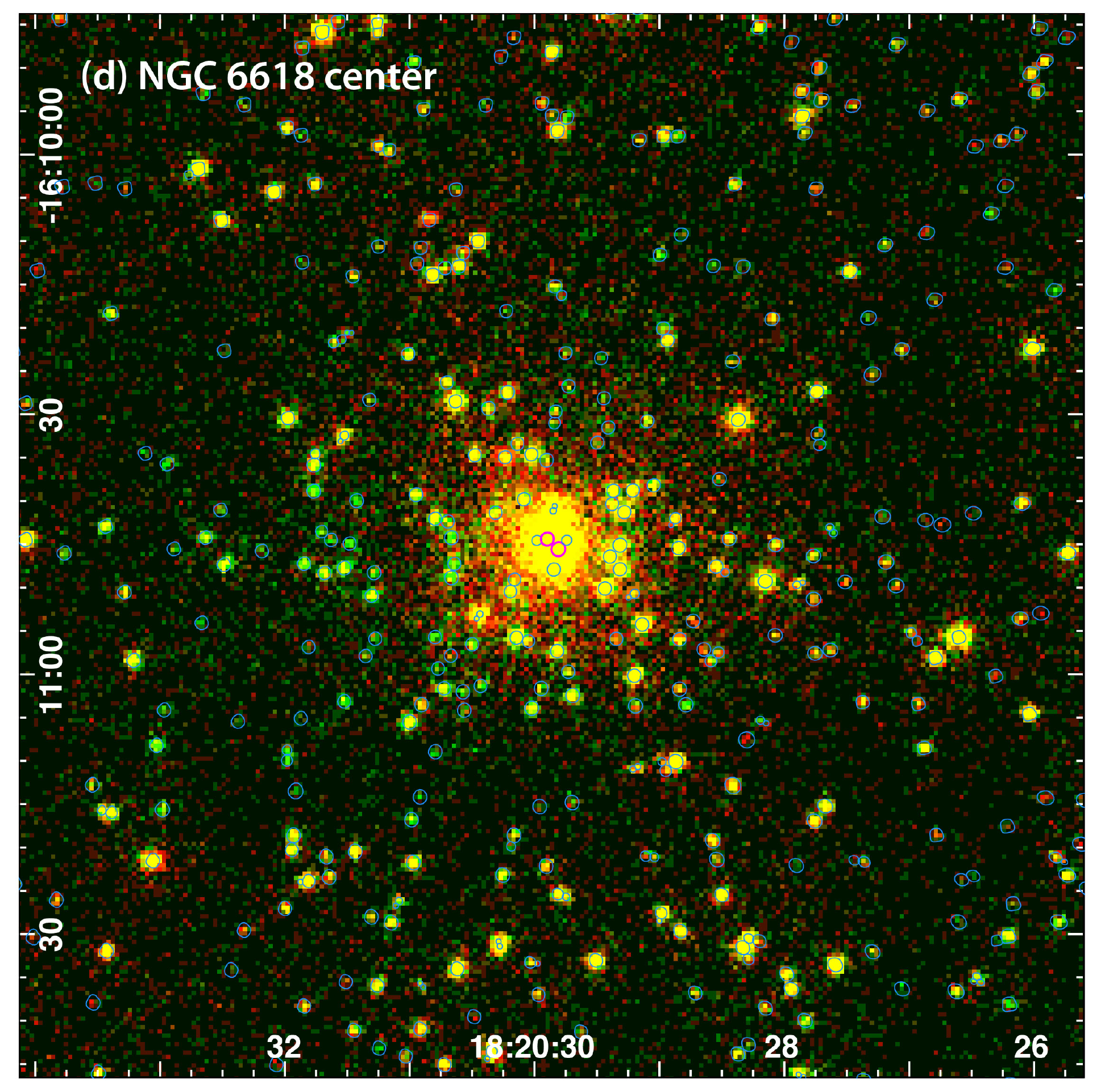}
\includegraphics[width=0.34\textwidth]{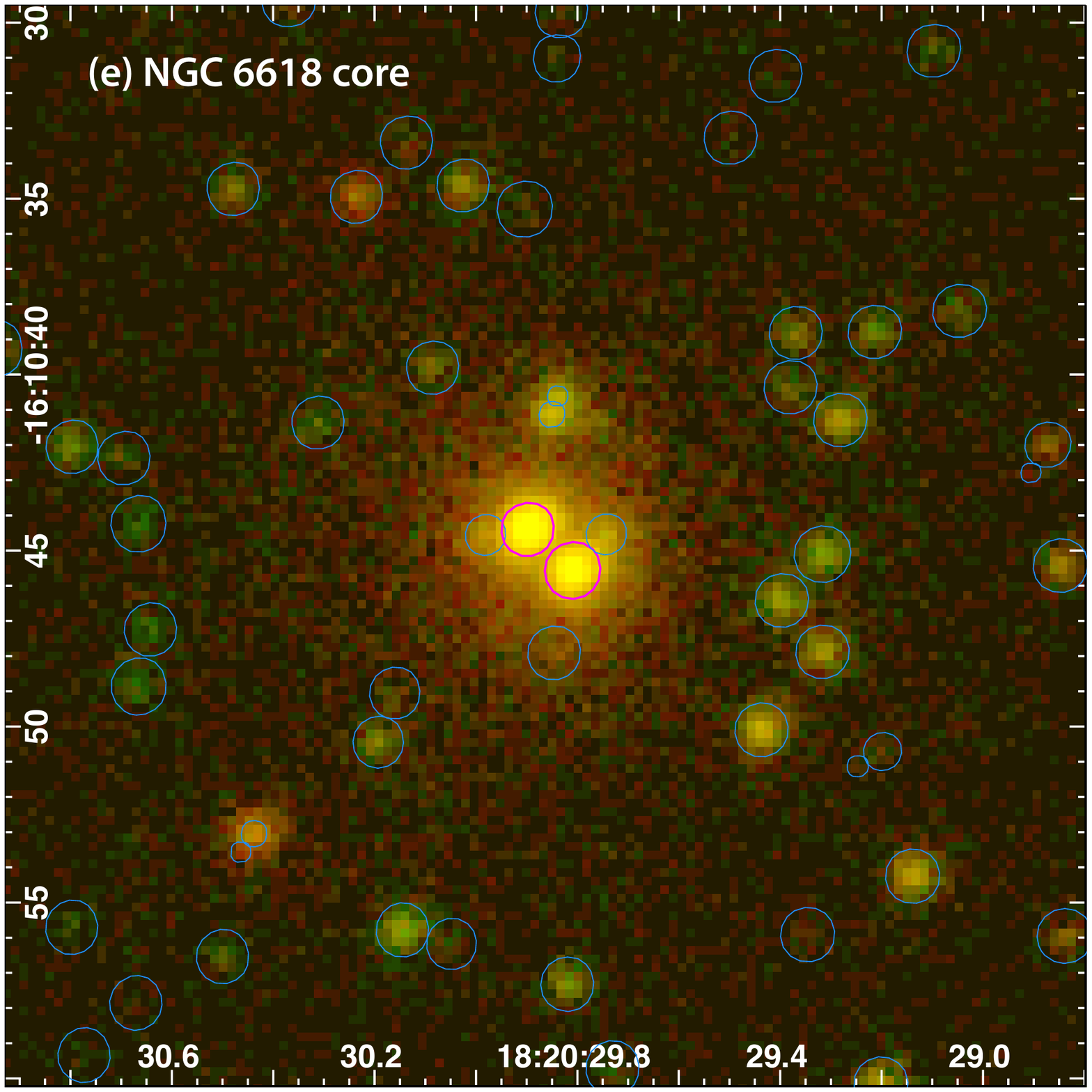}
\caption{M17 (The Omega Nebula, NGC~6618).
(a) ACIS exposure map with brighter ($\geq$5 net counts) ACIS point sources overlaid, with symbols and colors denoting median energy for each source.
(b) ACIS diffuse emission (full-band, 0.5--7~keV) in blue, \Spitzer/IRAC 8~$\mu$m emission in green, and \Spitzer/MIPS 24~$\mu$m emission in red.  
(c--e) ACIS binned event images, with soft events (0.5--2~keV) shown in red and hard events (2--8~keV) in green.  Point source extraction regions are overlaid in blue.  The CEN1 O4 stars are noted by magenta extraction regions (both X-ray sources are piled up at least part of the time).
\label{m17.fig}}
\end{figure}

Like M16, the M17 \GHIIR is powered by a monolithic massive cluster, NGC~6618 (Figure~\ref{m17.fig}a).  This cluster generates a dramatic outflow of hot plasma \citep{Dunne03,Townsley03} seen as bright diffuse X-ray emission (Figure~\ref{m17.fig}b) either shadowed or confined by the IR-bright \citep{Povich07} V-shaped ionized bars that surround the cluster.  Our original 40-ks ACIS GTO observation of M17 yielded studies of this X-ray outflow \citep{Townsley03} and $>$900 X-ray point sources in NGC~6618 \citep{Broos07}.  We also compared M17's diffuse X-ray emission to that seen in the Carina Nebula \citep{Townsley11c}.  A recent analysis of the {\em XMM-Newton} data on M17 \citep{Mernier13} nicely illustrates the X-ray variability of pre-MS stars; obtained 20 months later than our original ACIS-I observation, the {\em XMM-Newton} data have very few X-ray sources in common with the ACIS observation \citep{Broos07}.

The secondary pointings in the ACIS mosaic were obtained to sample the prominent diffuse X-ray emission (eastern pointing) and the sources powering a large, older \Spitzer\ bubble (northern pointing) that may be triggering further star formation \citep{Povich09}.  The addition of these pointings to our M17 ACIS-I mosaic completes the coverage of the hot plasma outflow.  Additional faint diffuse X-ray emission is present in the northern bubble.  Careful spatial decomposition and spectral fitting is warranted for M17's diffuse X-ray emission; spectral fit parameter maps can then be compared to the more evolved, cluster-of-clusters Carina complex \citep{Townsley11b}.

Our \Chandra Large Project data on NGC~6618 provides much deeper coverage of its pre-MS population.  This observation reveals the richness of NGC~6618; high median energies of the X-ray sources illustrate that this MYSC is still largely obscured by its natal cloud (Figure~\ref{m17.fig}a).  Figure~\ref{m17.fig}c shows a wide view of NGC~6618; note the variety of source hardnesses, especially towards the southwest as obscuration in the southwest bar absorbs soft X-rays.  Figure~\ref{m17.fig}d zooms in on the central part of NGC~6618, showing a concentrated cluster.  Finally, Figure~\ref{m17.fig}e shows the central massive stars, the O4-O4 pair CEN1 \citep{Chini80}; each component is now thought to be an O4-O4 binary itself \citep{Hoffmeister08}, thus four O4 stars reside at the center of NGC~6618.  Each of these O4-O4 binaries exhibits X-ray emission from a hard thermal plasma, likely indicating that they are colliding-wind binaries \citep[e.g.,][]{Pittard10} or have substantial magnetic fields that are generating magnetically-channeled wind shocks \citep{Babel97,Gagne05}.  These bright X-ray sources show a complicated mix of variability and photon pile-up; time-resolved X-ray spectral analysis and spatio-spectral pile-up reconstruction is necessary to fully exploit the ACIS observations of these two important O4-O4 binaries.

\clearpage

\subsection{W3 \label{sec:w3}}

The large ($3^{\circ} \times 1.5^{\circ}$) W4/W3/HB3 complex contains one of the most massive GMCs in the outer Galaxy \citep{Heyer98}, massive embedded protostars \citep{Megeath96}, every known type of \hii region (hypercompact to diffuse), and HB3, one of the biggest SNRs in the Galaxy \citep{Routledge91}.  The W3 GMC is highly clumped and is being compressed by the W4 superbubble to its east \citep{Moore07}.  It is strongly turbulent and filamentary and has probably experienced both spontaneous and triggered star formation distributed throughout the cloud \citep{Rivera11}.  A recent {\em Herschel} study suggests that this active environment, fueled by massive star feedback, might ensure a continuous supply of flowing cloud material to facilitate MYSC formation \citep{Rivera13}.

\begin{figure}[htb]
\centering
\includegraphics[width=0.49\textwidth]{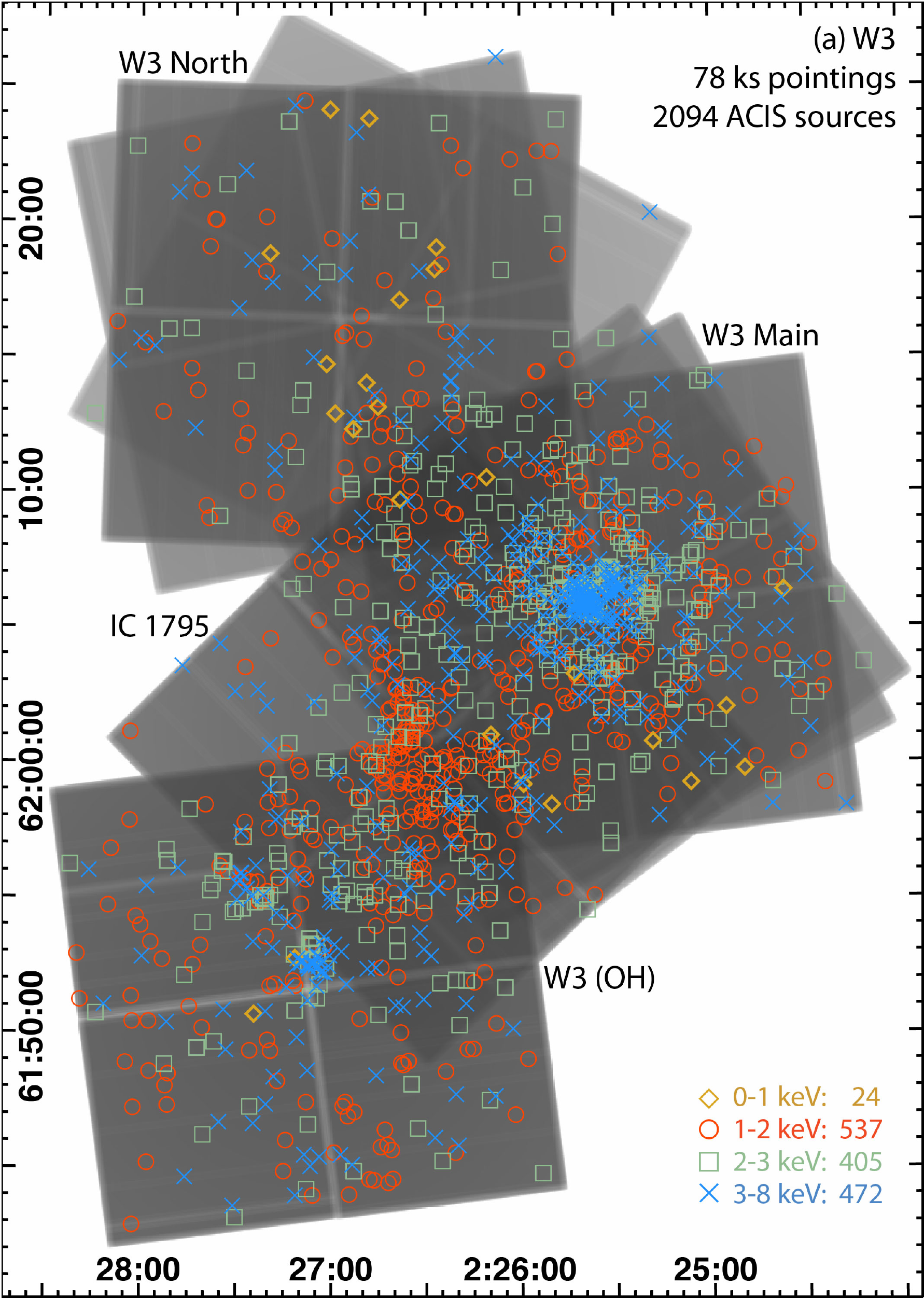}
\includegraphics[width=0.48\textwidth]{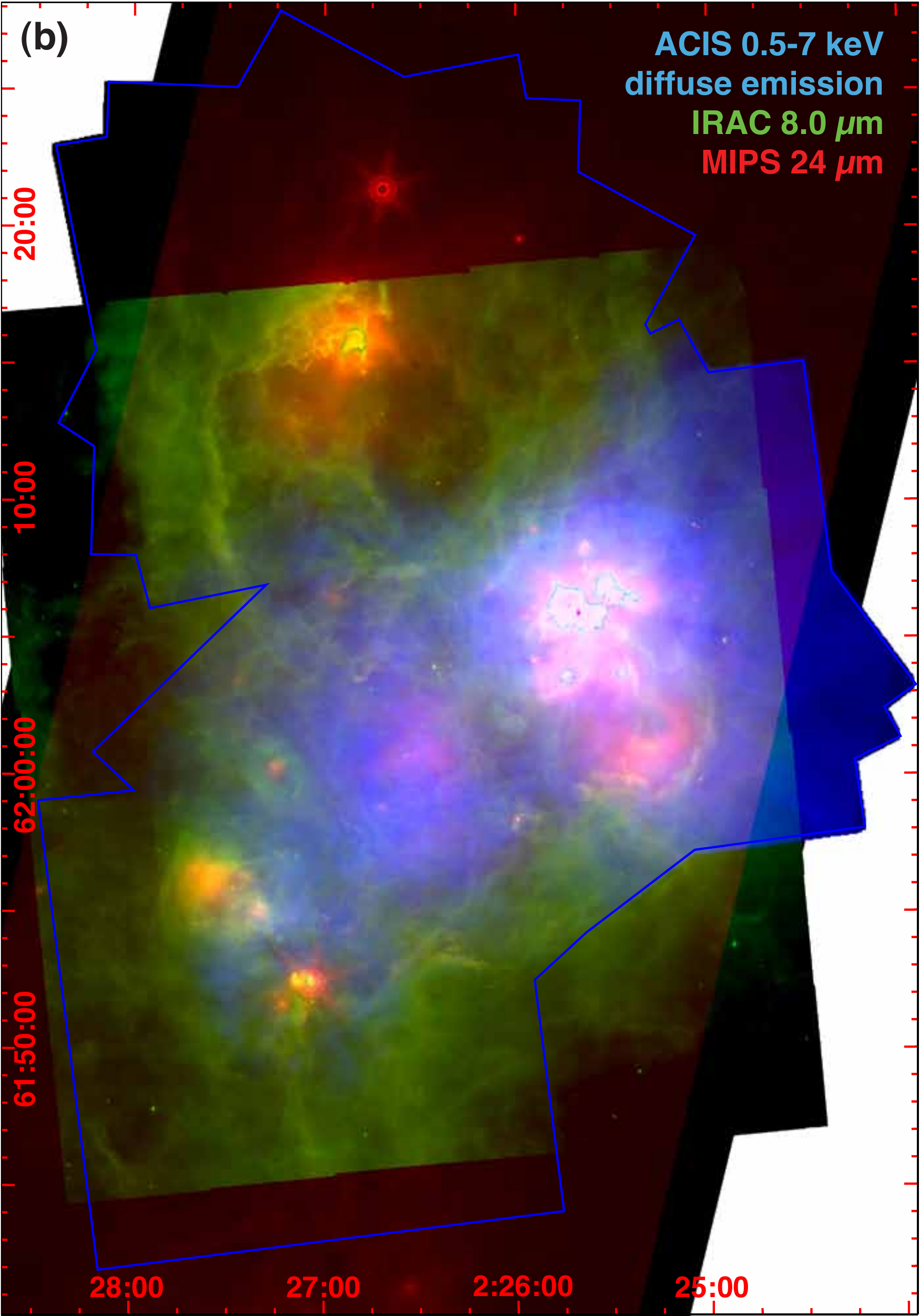}
\caption{W3 (IC~1795).
(a) ACIS exposure map with brighter ($\geq$5 net counts) ACIS point sources overlaid, with symbols and colors denoting median energy for each source.
(b) ACIS diffuse emission (full-band, 0.5--7~keV) in blue, \Spitzer/IRAC 8~$\mu$m emission in green, and \Spitzer/MIPS 24~$\mu$m emission in red.  Sharp green and cyan rims around IR-bright regions (in this and subsequent multiwavelength images) are artifacts associated with \Spitzer\ image saturation.
\label{w3.fig}}
\end{figure}

The W3 MSFR has a cluster-of-clusters morphology, with clear age differences between clusters \citep{Tieftrunk97} and what appears to be a substantial population of ``distributed'' young stars (Figure~\ref{w3.fig}).  Detailed images of the three original \Chandra pointings were given in \citet{Feigelson08} and are not repeated here, although we again note that the prominent \hii region in W3~North is ionized by a single massive star that lacks a surrounding rich cluster (perhaps similar to the massive star seen in the southeastern ACIS-I pointing on NGC~6357) and that we detect the ionizing sources for the famous ultracompact \hii region (UCH{\scriptsize II}R) W3(OH) and the hypercompact \hii region W3~Main~IRS5.  These young massive stars are seen through very large absorbing columns because they produce hard X-ray emission, as we saw in the M17 O4-O4 binaries above.  Now a fourth, 49-ks ACIS-I pointing on the older, more revealed cluster IC~1795 \citep{Roccatagliata11} is included in our ACIS mosaic and in the X-ray source list that we present in MOXC.  

Despite comparatively deep ACIS observations, the diffuse X-ray component in W3 is quite faint in apparent surface brightness compared to other MOXC MSFRs; while discernable around W3~Main and IC~1795, it is barely detectable in the northern pointing or south of W3(OH).  This will limit our future efforts to map its physical parameters through X-ray spectral fitting, as large areas will have to be averaged together to generate a spectrum of sufficient quality to constrain those parameters.  Perhaps the most important result will be the absence of diffuse emission around the ionizing O star in W3~North (note that the ACIS mosaic extends far above the \Spitzer\ data in Figure~\ref{w3.fig}b).  Given the ubiquity of such diffuse X-ray emission in other MSFRs (as demonstrated by this paper) and the ability of \Chandra to detect that emission even with short observations of distant, obscured regions, its absence in W3~North is quite surprising and leaves us with a mystery to inspire future work. 

\clearpage

\subsection{W4 \label{sec:w4}}

It is thought that past episodes of OB star formation in W4 \citep{Oey05} produced its well-known chimney, the first directly imaged in our Galaxy \citep{Normandeau96}, and its huge superbubble detected in H$\alpha$ and extending $>$1000~pc above the Galactic plane \citep{Reynolds01}. The most recent star formation episode has left the \GHIIR IC~1805 and its MYSC (OCl~352) at the center of W4 \citep[e.g.,][]{Wolff11}.  IC~1805 is fueling the W4 chimney and ionizing the superbubble cavity \citep{Lagrois09a}.

W4 also exhibits a 1$^\circ$ loop of emission to the south of IC~1805 \citep{Lagrois09b}, where the bubble created by IC~1805's predecessors is encountering denser material in the Galactic Plane.  IR and radio work \citep{Terebey03} has shown that the W4 shell is very inhomogeneous, allowing 40\% of the MYSC's ionizing photons to leak out of the superbubble and ionize the surrounding ISM.  Detailed modeling of the W4 superbubble \citep{Basu99} predicted that both cavities should be filled with T=5$\times 10^6$~K (0.43~keV) gas.

\begin{figure}[htb]
\centering
\includegraphics[width=0.48\textwidth]{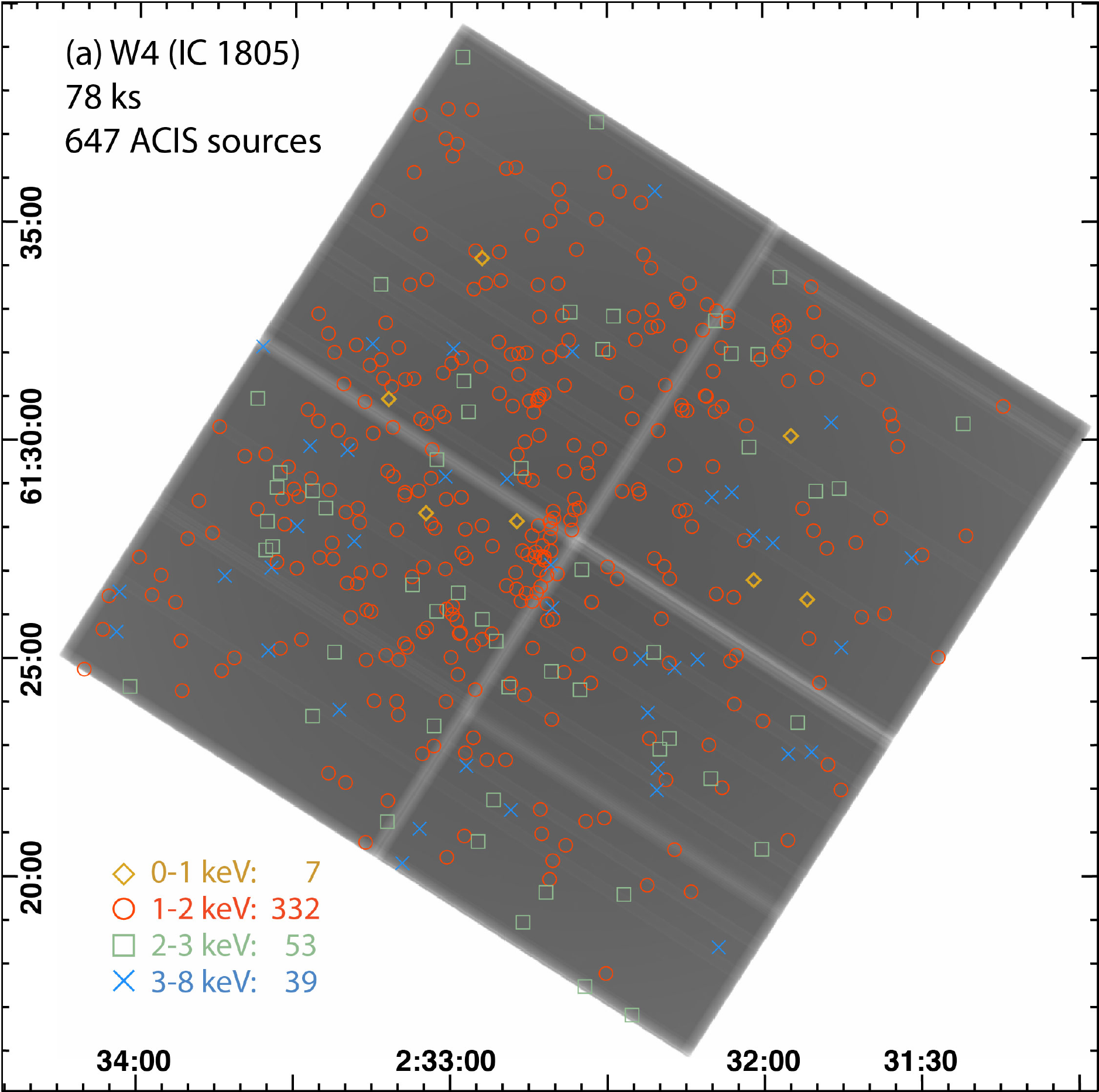}
\includegraphics[width=0.495\textwidth]{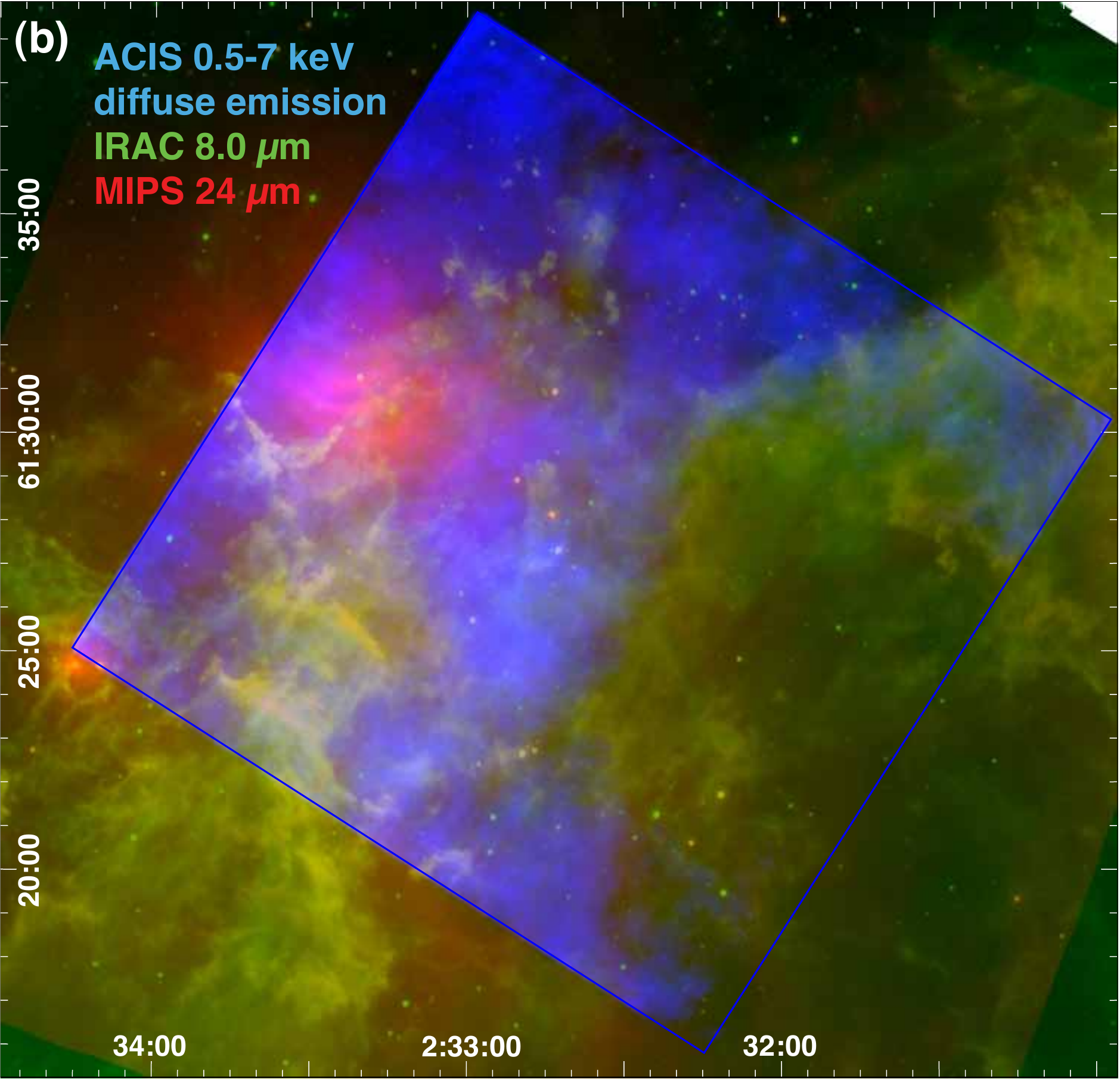}
\caption{W4 (IC~1805).
(a) ACIS exposure map with brighter ($\geq$5 net counts) ACIS point sources overlaid, with symbols and colors denoting median energy for each source.
(b) ACIS diffuse emission (full-band, 0.5--7~keV) in blue, \Spitzer/IRAC 8~$\mu$m emission in green, and \Spitzer/MIPS 24~$\mu$m emission in red.   
\label{w4.fig}}
\end{figure}

There are at least 60 OB stars in the IC~1805 cluster \citep{Shi99}.  Our 80-ksec ACIS-I observation was centered on the binary massive star HD~15558 \citep{DeBecker06}.  The distribution of X-ray sources (Figure~\ref{w4.fig}a) shows a fairly relaxed, unobscured cluster resembling its neighbor IC~1795 in W3.

IC~1805 provides a striking example of hot plasma interfacing with cold clouds (Figure~\ref{w4.fig}b).  The diffuse X-ray emission threads through this field, strongly shadowed or displaced by the cold material west of the cluster center.  The high-mass X-ray binary LSI+61$^{\circ}$303 lies just east of IC~1805 and is thought to be associated with that MYSC or with an earlier generation of massive star formation in W4 \citep{Mirabel04}.  This is strong evidence that W4 has seen supernova activity.  As for Carina \citep{Townsley11b}, a search for signs of supernova activity will be done via careful spectral analysis of the diffuse X-ray emission in IC~1805.  In all likelihood though, based on the diffuse X-ray emission we see in other MYSCs too young to have hosted supernovae, the hot plasma found by \Chandra in IC~1805 can be explained primarily by the winds from its massive stars.


\subsection{NGC~3576 \label{sec:n3576}}

\begin{figure}[htb]
\centering
\includegraphics[width=0.48\textwidth]{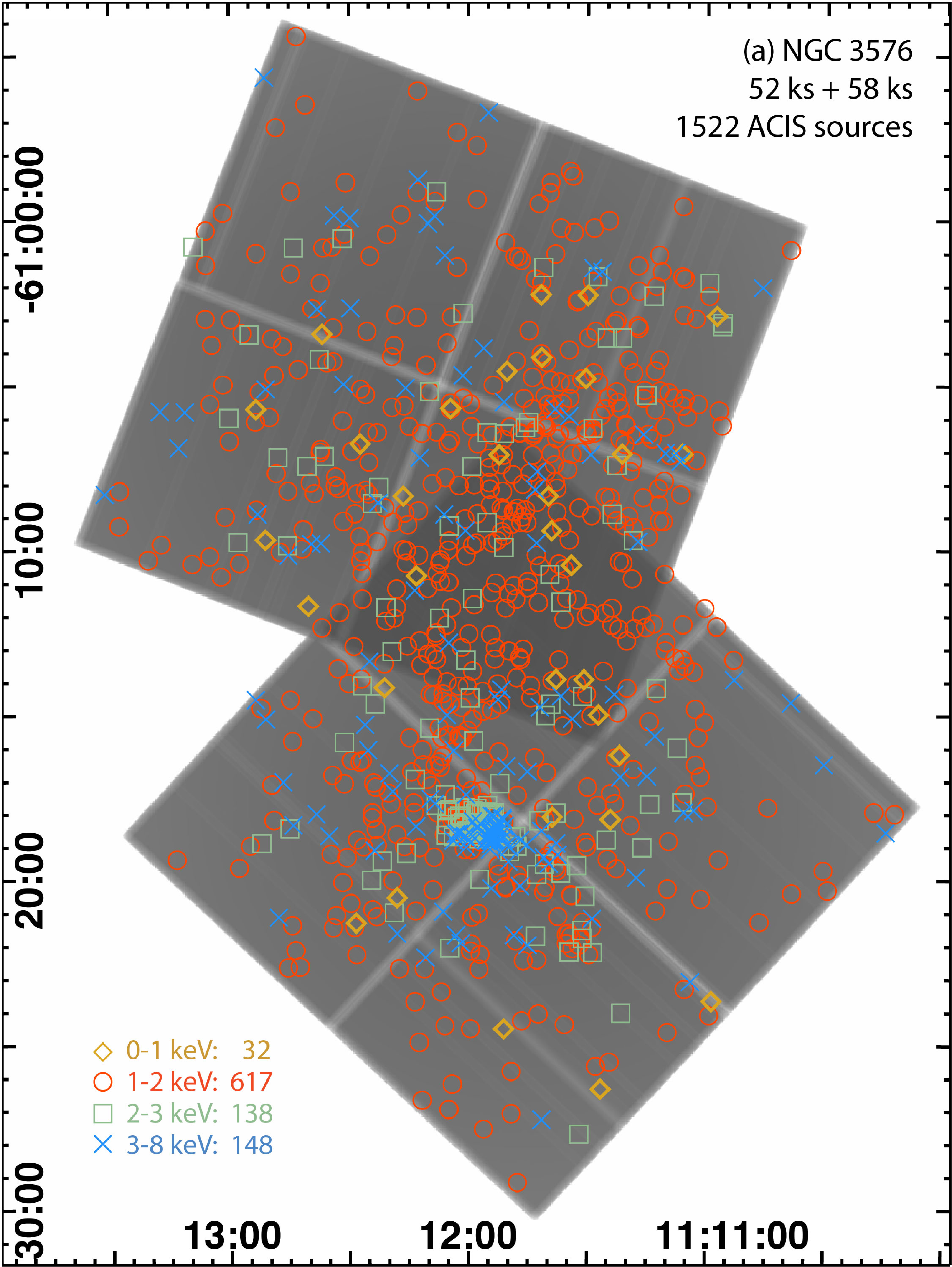}
\includegraphics[width=0.48\textwidth]{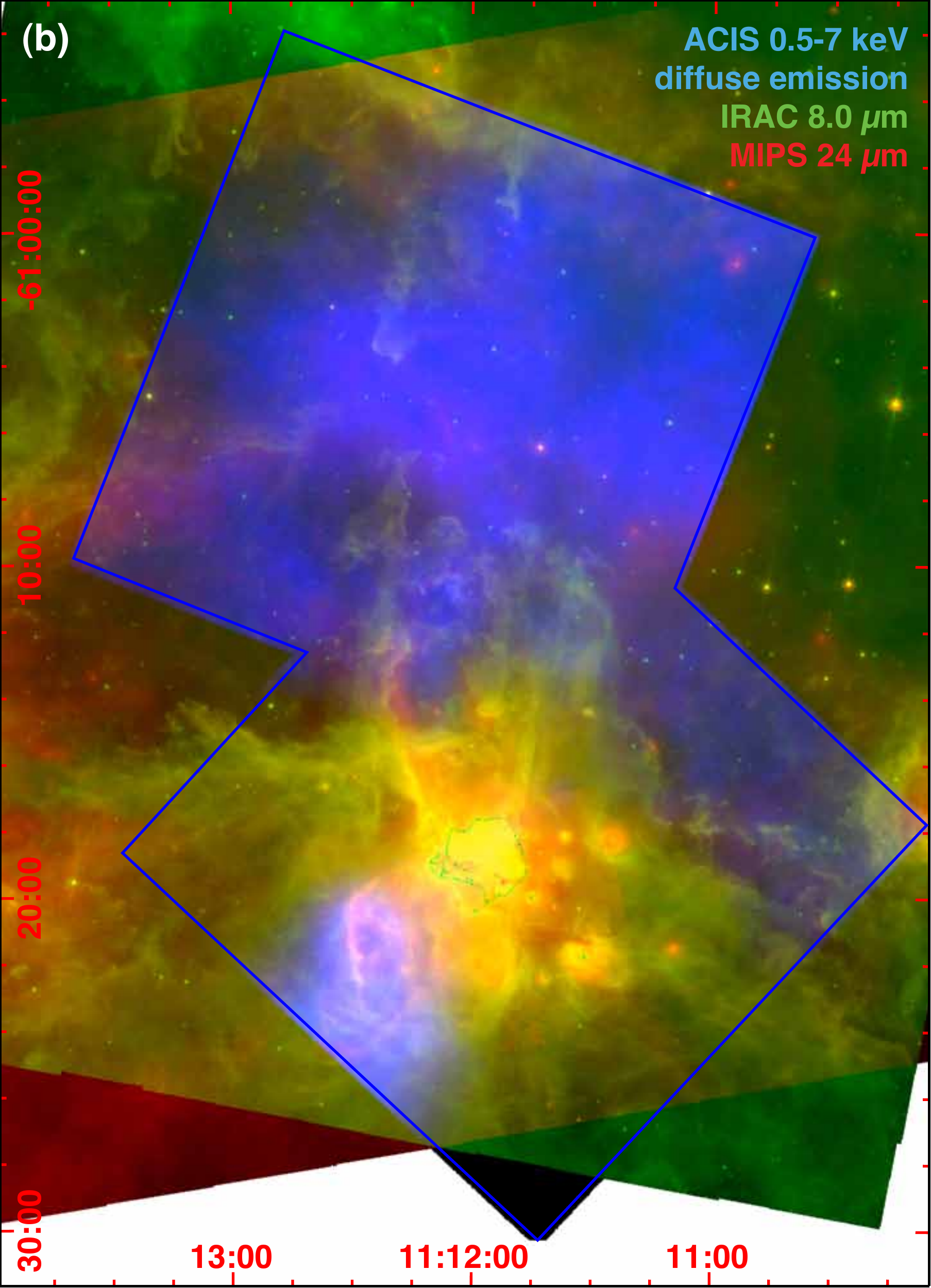}
\includegraphics[width=0.48\textwidth]{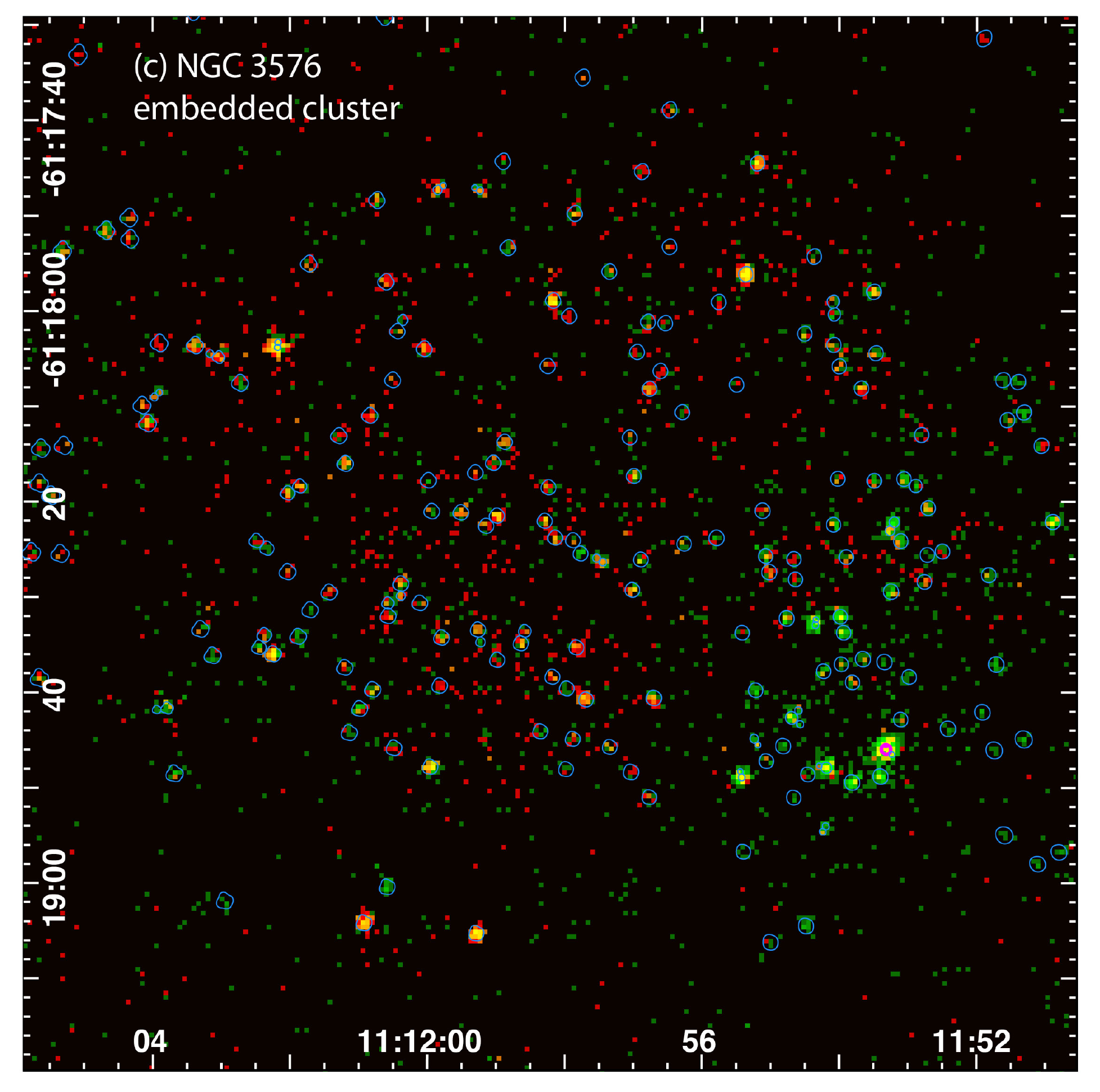}
\includegraphics[width=0.48\textwidth]{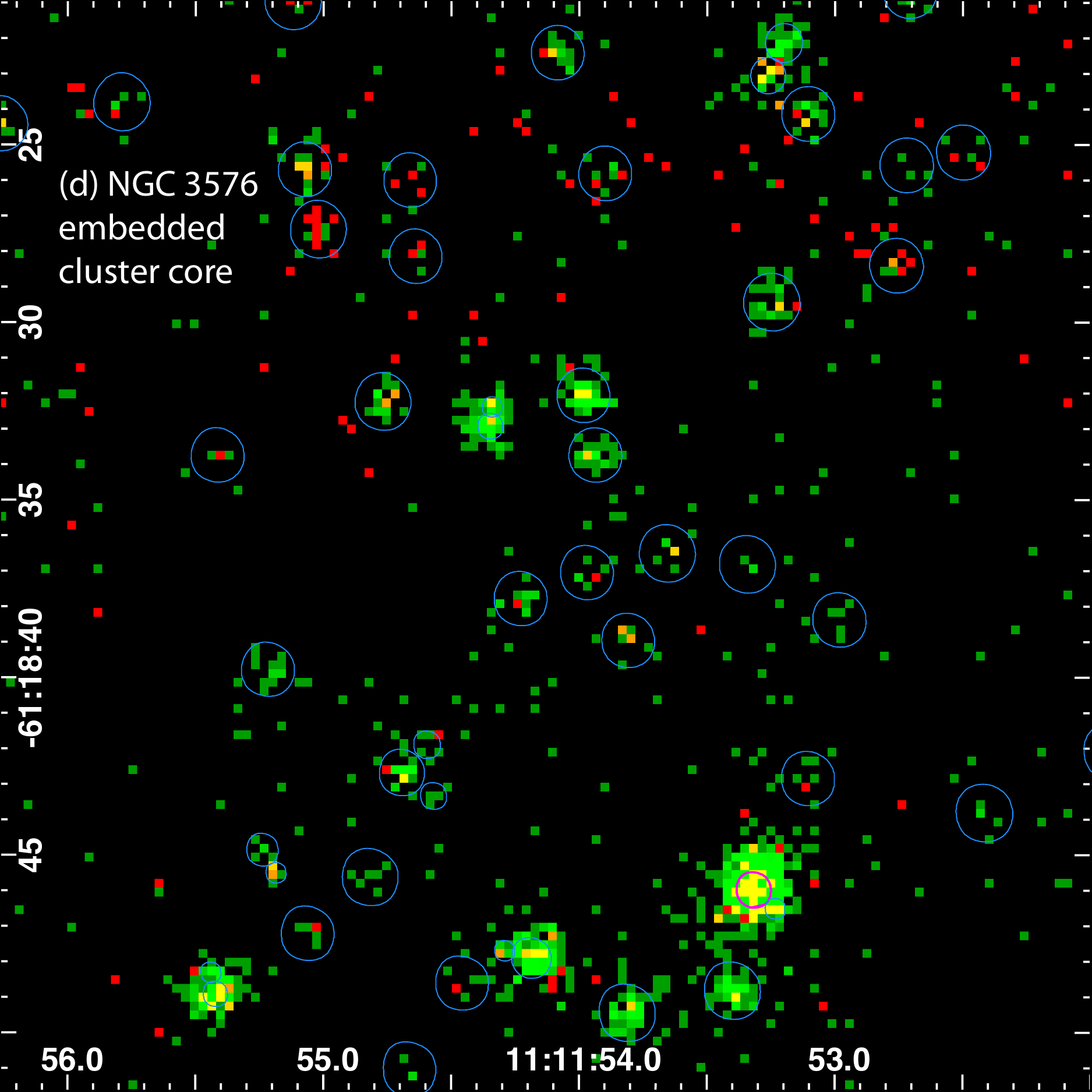}
\caption{NGC~3576.
(a) ACIS exposure map with brighter ($\geq$5 net counts) ACIS point sources overlaid, with symbols and colors denoting median energy for each source.
(b) ACIS diffuse emission (full-band, 0.5--7~keV) in blue, \Spitzer/IRAC 8~$\mu$m emission in green, and \Spitzer/MIPS 24~$\mu$m emission in red. 
(c,d) ACIS binned event images, with soft events (0.5--2~keV) shown in red and hard events (2--8~keV) in green.  Point source extraction regions are overlaid in blue; the magenta region indicates a piled-up source.  
\label{n3576.fig}}
\end{figure}

In \citet{Townsley11c}, we described the \Chandra observations of the NGC~3576 \GHIIR and the NGC~3576 OB Association to its north (Figure~\ref{n3576.fig}).  ACIS showed that this complex consists of an older, revealed, relaxed cluster whose most massive members have probably already been lost as supernovae, next to an embedded, concentrated MYSC ionizing a GH{\scriptsize II}R, perhaps indicating sequential massive star formation \citep[e.g.,][]{Garcia94}.  The cavity to the north of the GMC hosting the embedded cluster is filled with diffuse X-ray emission.  It includes a substantial hard thermal (or non-thermal) component likely related to the pulsar found in this region; our ACIS data also revealed a pulsar wind nebula around this source.  Additionally, soft plasma appears to be flowing out of a crevice or fissure in the GMC towards us \citep{Rogers13}; this is clearly seen to the southeast of the embedded MYSC.

Here, we report the $>$1500 X-ray point sources found in these data.  This study provides the first detailed look at the older northern cluster and significantly improves the census of members in the embedded MYSC, including many highly-obscured but luminous X-ray sources (one even exhibiting photon pile-up) that are likely this cluster's massive stars ionizing its GH{\scriptsize II}R.  As we have seen in other very young, embedded MYSCs, these massive stars are detected in X-rays behind very large absorbing columns because they are hard X-ray emitters; apparently hard X-ray emission from magnetically-channeled wind shocks \citep{Babel97,Gagne05} and/or colliding-wind binaries \citep{Pittard10} is the norm rather than the exception in these regions.  None of the seven massive protostars found by \citet{Andre08} are detected in our \Chandra observations.  This implies that the mechanism generating hard X-ray emission in the massive stars powering the \GHIIR has not turned on yet in these protostellar sources.  

Figure~\ref{n3576.fig}c reveals the embedded cluster powering NGC~3576.  Note the gradient in event energy from northeast to southwest; this is due to a known gradient in the extinction in front of the cluster \citep{Persi94}.  Figure~\ref{n3576.fig}d zooms into the highly-obscured southwest corner of the cluster.  The heavy magenta extraction region indicates the piled-up ACIS source J111153.31-611845.9 (p1\_605); see Table~\ref{pile-up_risk.tbl}.  

\clearpage

\subsection{G333.6-0.2 \label{sec:g333.6-0.2}}

\begin{figure}[htb]
\centering
\includegraphics[width=0.49\textwidth]{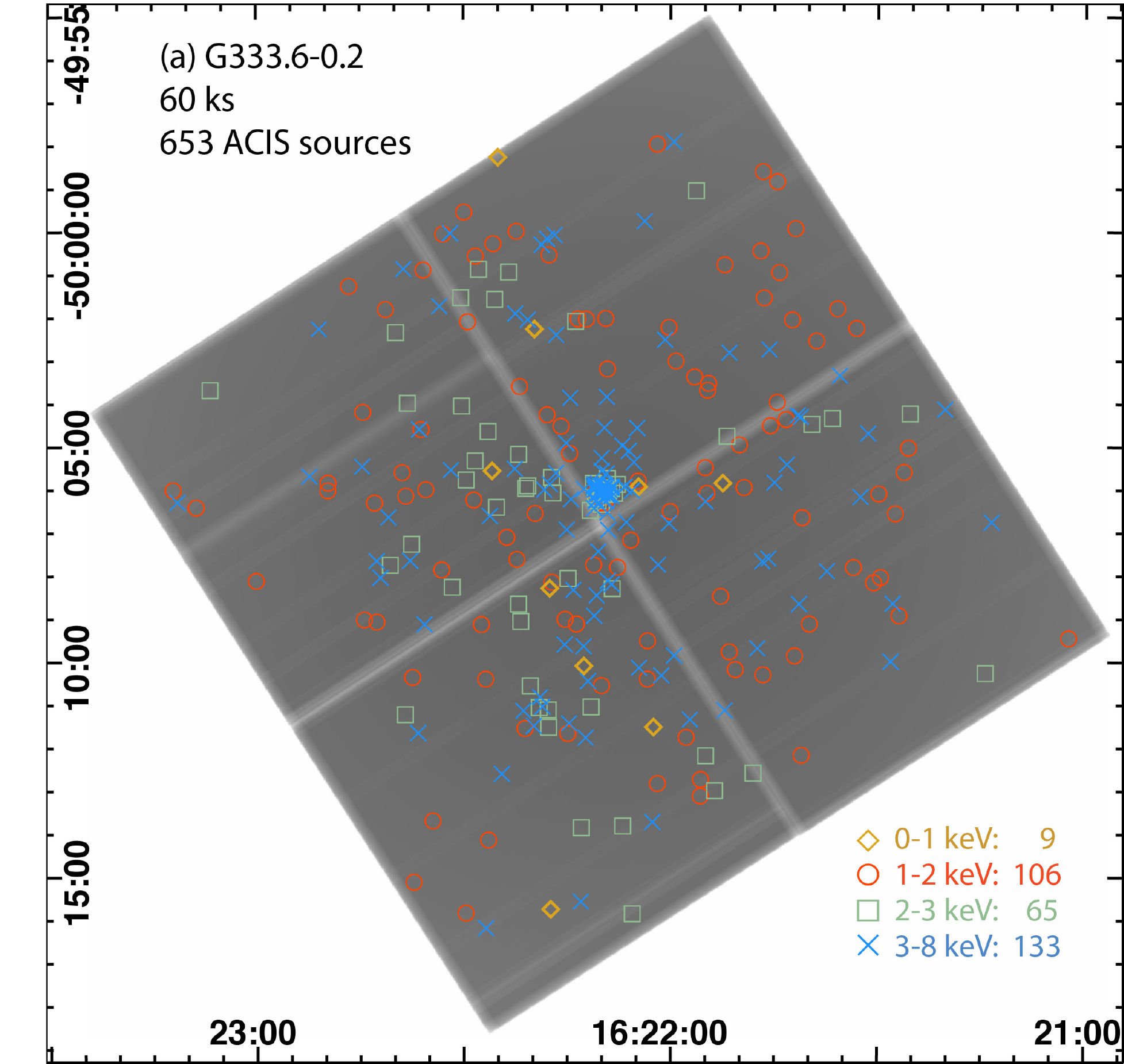}
\includegraphics[width=0.48\textwidth]{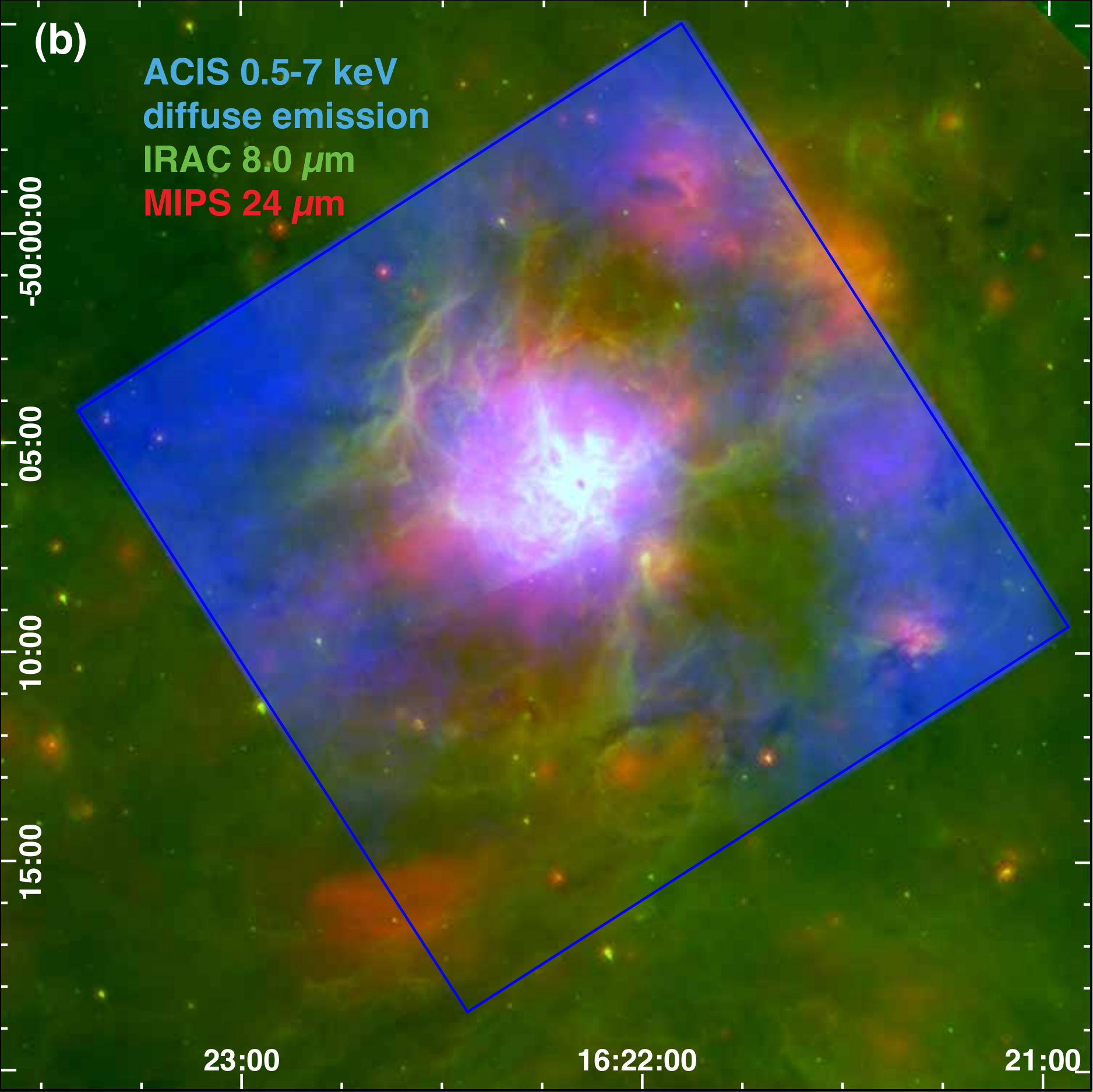}
\includegraphics[width=0.48\textwidth]{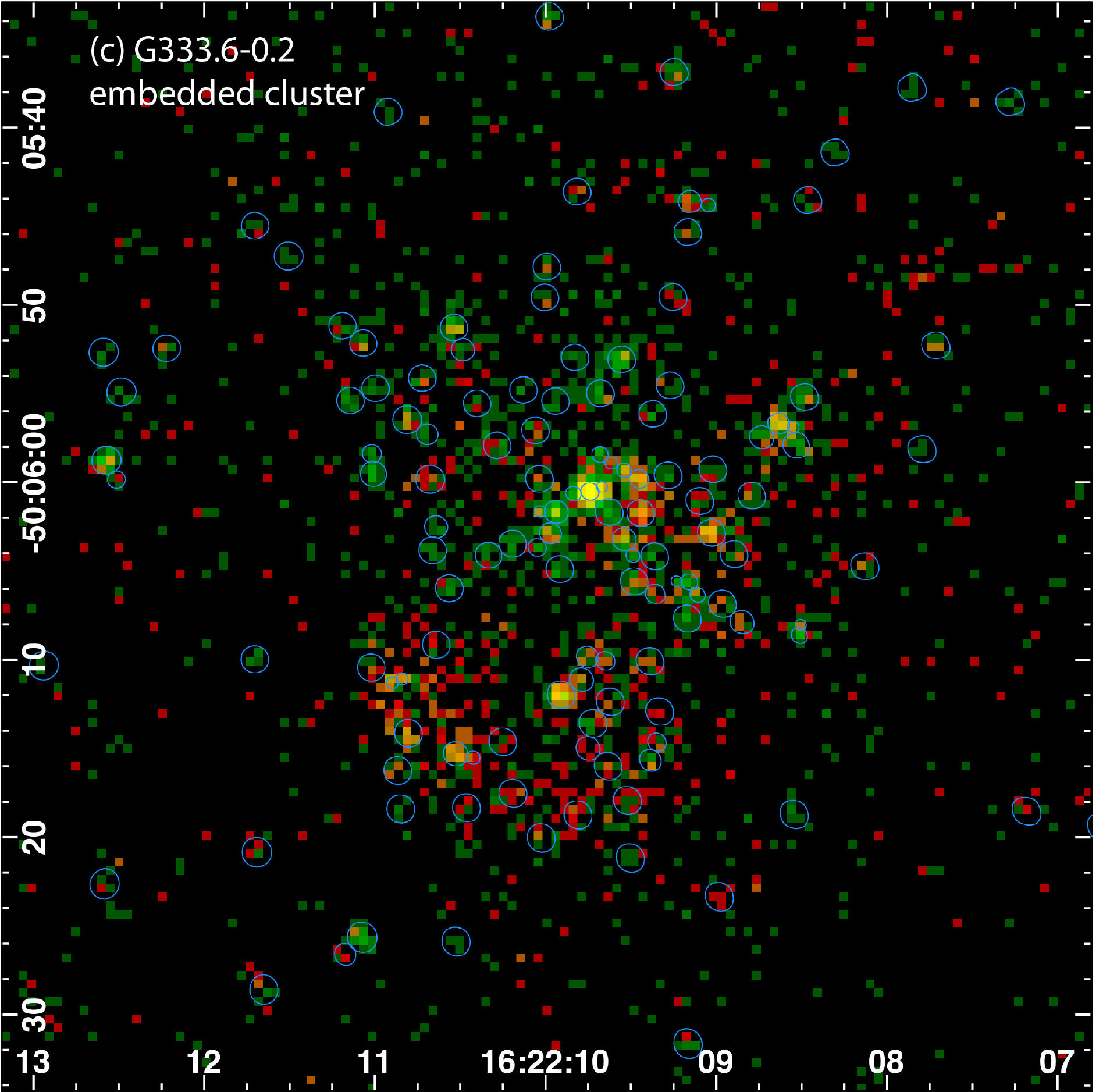}
\includegraphics[width=0.5\textwidth]{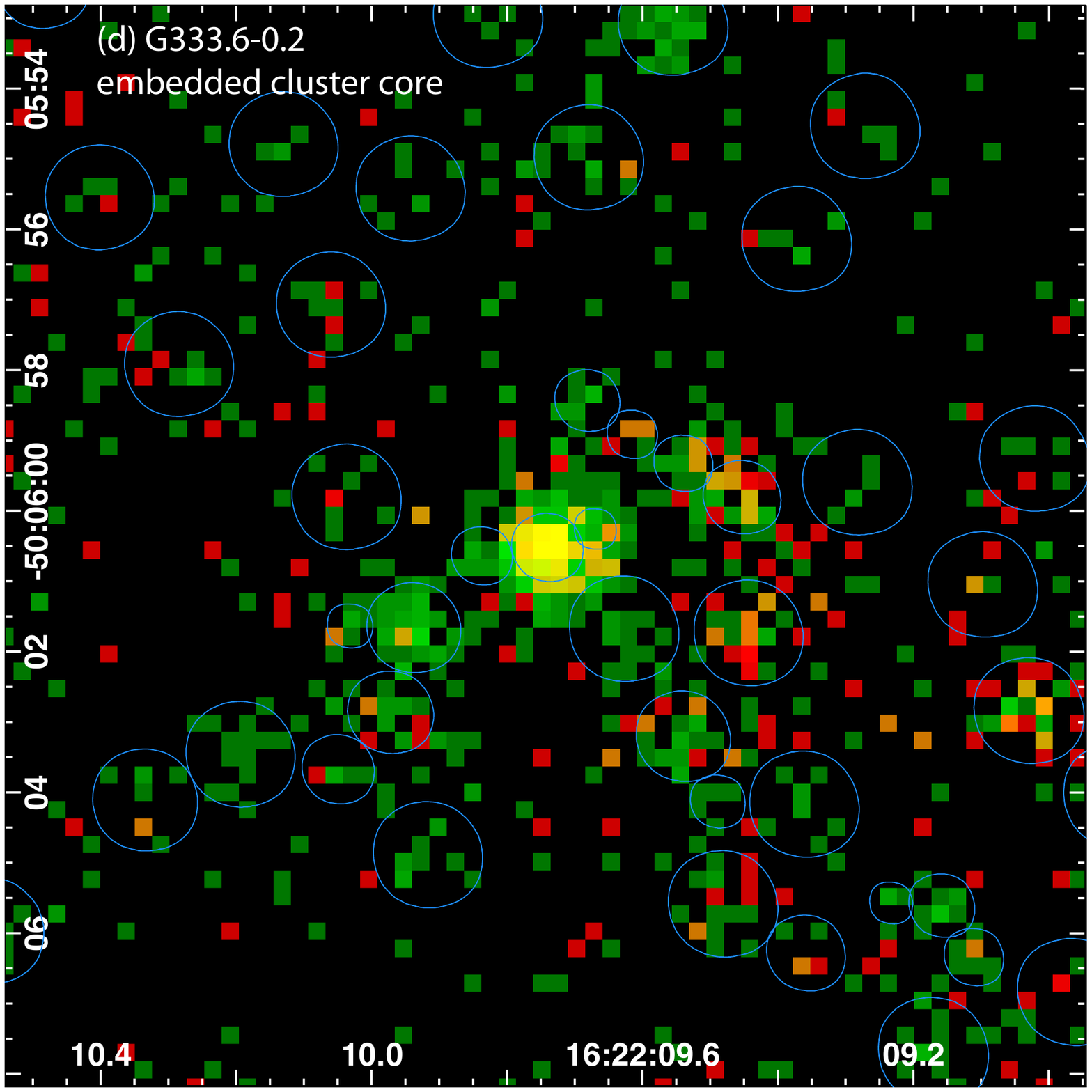}
\caption{G333.6-0.2.
(a) ACIS exposure map with brighter ($\geq$5 net counts) ACIS point sources overlaid, with symbols and colors denoting median energy for each source.
(b) ACIS diffuse emission (full-band, 0.5--7~keV) in blue, \Spitzer/IRAC 8~$\mu$m emission in green, and \Spitzer/MIPS 24~$\mu$m emission in red.   
(c,d) ACIS binned event images, with soft events (0.5--2~keV) shown in red and hard events (2--8~keV) in green.  Point source extraction regions are overlaid in blue.  C:  central cluster, shown with 0.5$\arcsec$ pixels.  D:  cluster core with 0.25$\arcsec$ pixels, showing the crowded central sources contributing to the high {\em IRAS} flux in this region.
\label{g333.6-0.2.fig}}
\end{figure}

The GMC known as G333 hosts at least four highly-obscured MYSCs \citep{Figueredo05, Bains06} plus the slightly more revealed, multi-clustered star-forming complex RCW~106 \citep{Russeil05}.  G333 has a conspicuously elongated morphology, extending over 80~pc along the Galactic Plane but with a width of only $\sim$15~pc.  The \GHIIR G333.6-0.2 (Figure~\ref{g333.6-0.2.fig}) is the best-studied MSFR in G333; an OB population of $\sim$100 stars has been predicted for it, based on its molecular gas mass \citep{Fujiyoshi05}.  The central embedded MYSC is young enough ($10^{5-6}$~yr) for its massive stars to retain circumstellar disks \citep{Sollins04a}.  Complex velocity structure that may indicate champagne flows and a strong extinction gradient ($A_{V}$$\sim$12--36~mag) across the MYSC were found in a radio recombination line study \citep{Fujiyoshi06}.  On larger scales, {\em Spitzer} reveals an 11-pc bipolar bubble centered on the MYSC (Figure~\ref{g333.6-0.2.fig}b) and likely blown by its massive stars.  High-opacity CO clouds circumscribe these bubbles \citep{Wong08} and coincide with methanol maser sources and small IRDCs \citep{Breen07}; these likely indicate the next generation of star formation being triggered by G333.6-0.2.  

{\em Chandra} reveals the massive stellar engine powering G333.6-0.2:  in 60~ks we detected $>$100 X-ray point sources in the MYSC (Figure~\ref{g333.6-0.2.fig}c).  This embedded young cluster is remarkably similar to the one powering NGC~3576 (Figure~\ref{n3576.fig}c above), even showing a similar absorption gradient.  In the IR-saturated cluster center (Figure~\ref{g333.6-0.2.fig}d) the ACIS data become critical, revealing a single bright source surrounded by 2 close, fainter companions within a $1\arcsec$ radius and 19 in the central $10\arcsec$, all contributing to the very large {\em IRAS} flux.  A single-temperature {\em apec} fit to the bright central source yields $N_{H}$=6$\times 10^{22}$~cm$^{-2}$ ($A_{V}$=30~mag), kT=2.4~keV, and intrinsic $L_{X}$=3$\times 10^{32}$~erg/s, confirming that it is indeed an embedded massive star.  The source spectrum likely includes an additional soft X-ray component that is intrinsically brighter than the hard component but totally obscured, thus X-ray spectral fitting cannot constrain it.  Several other hard, bright X-ray sources are seen in the MYSC; once again these massive stars are seen only because they emit hard X-rays that penetrate the deep obscuration.   

Diffuse X-ray emission pervades the MYSC and extends into the outer regions of the GMC (Figure~\ref{g333.6-0.2.fig}b), suggesting that hot plasma is leaking out of the embedded cluster; the surrounding natal cloud is being eroded by hot star winds even though its central stars remain highly obscured.  The diffuse spectrum exhibits two dominant soft plasma components similar to what we see in other MSFRs \citep{Townsley11b, Townsley11c} and a faint emission line at 6.3 keV consistent with fluorescent neutral iron from the cold clouds that surround the MYSC.

Over 500 additional X-ray sources are found across the wider ACIS-I field (Figure~\ref{g333.6-0.2.fig}a).  As in NGC~6357 and Carina, this ``distributed'' population is likely part of an older generation of young stars inhabiting the GMC.  We have new ACIS data in hand that sample the rest of the G333 GMC, including all four MYSCs and RCW~106.  The completed ACIS mosaic will bring new understanding of this prominent cluster-of-clusters complex.

\clearpage

\subsection{W51A \label{sec:w51}}

With W51, we begin MOXC's examination of MSFRs substantially more distant and massive than the MYStIX sample.  W51A is an excellent example of distributed massive star formation, containing $>$20 distinct radio \hii regions of every known type, from hypercompact to diffuse \citep{Mehringer94}.  These \hii regions are spread out linearly along the Galactic Plane and cover several parsecs, reminiscent of the other cluster-of-clusters complexes described above.  W51A sports the very massive young stellar object W51~IRS2E showing infall \citep{Sollins04b} that might be an example of a ``quenched'' \hii region \citep{Keto03}.  Just the main MYSC complex, G49.5-0.4, has $>$30 O stars \citep{Okumura00}.

\begin{figure}[htb]
\centering
\includegraphics[width=0.48\textwidth]{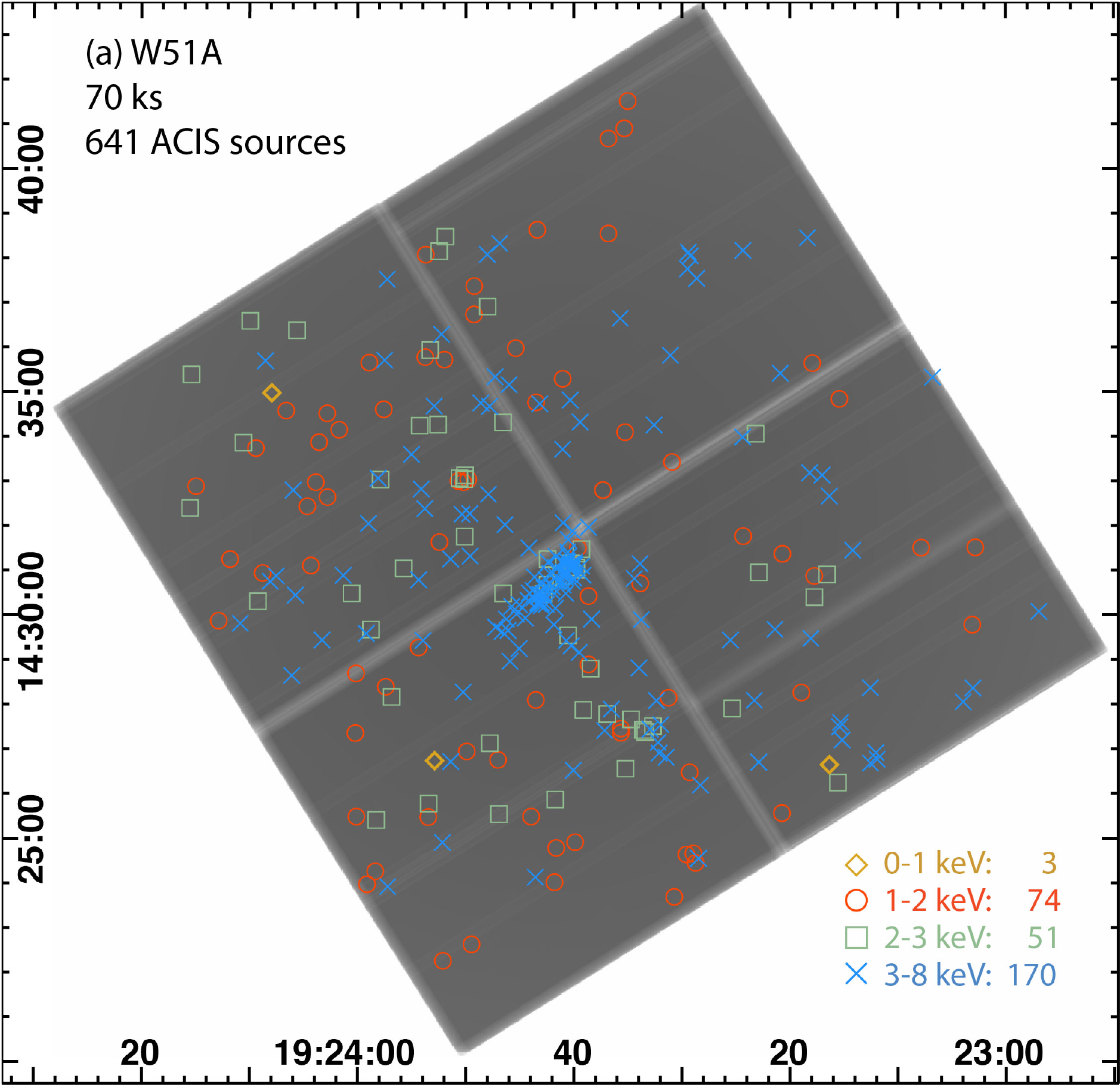}
\includegraphics[width=0.48\textwidth]{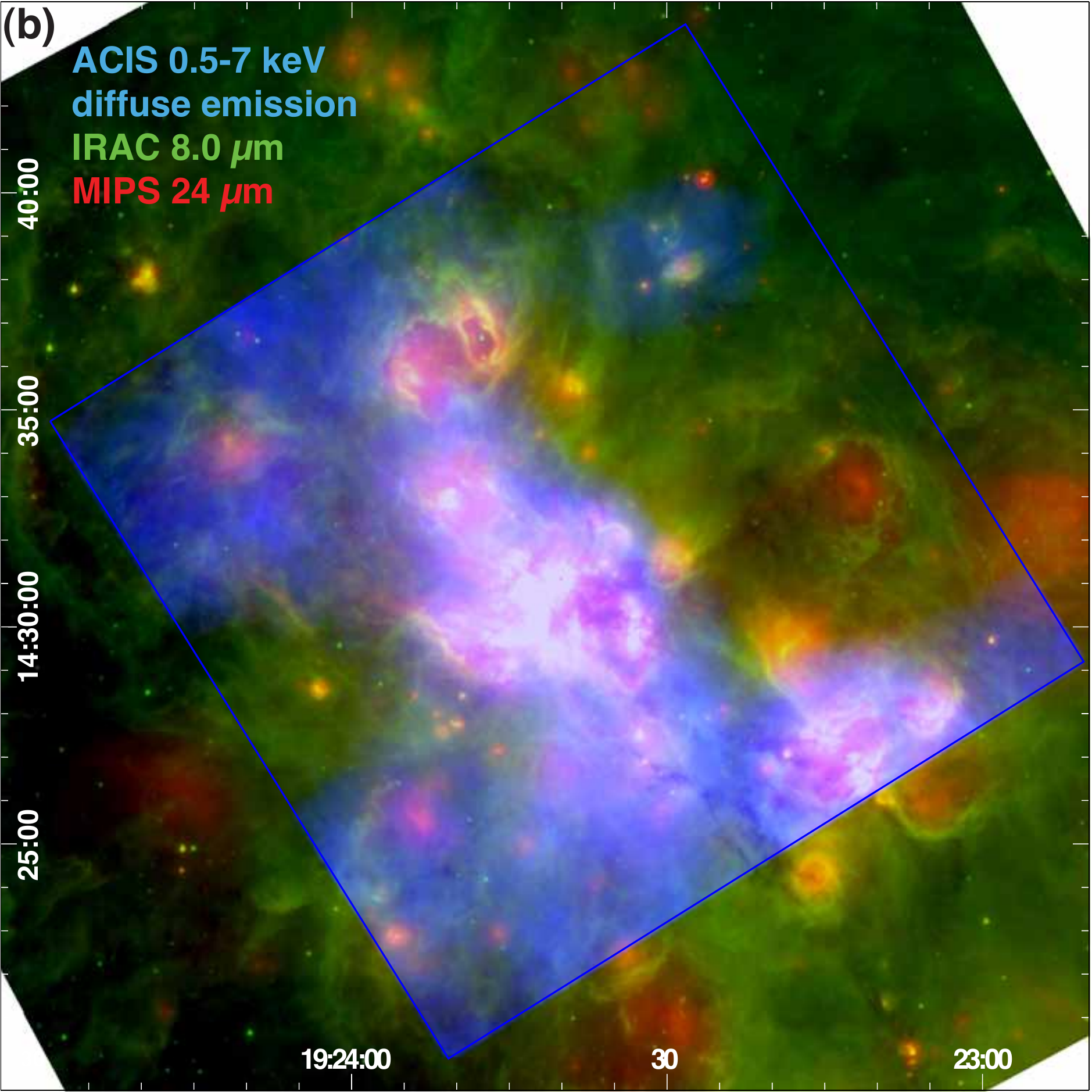}
\includegraphics[width=0.32\textwidth]{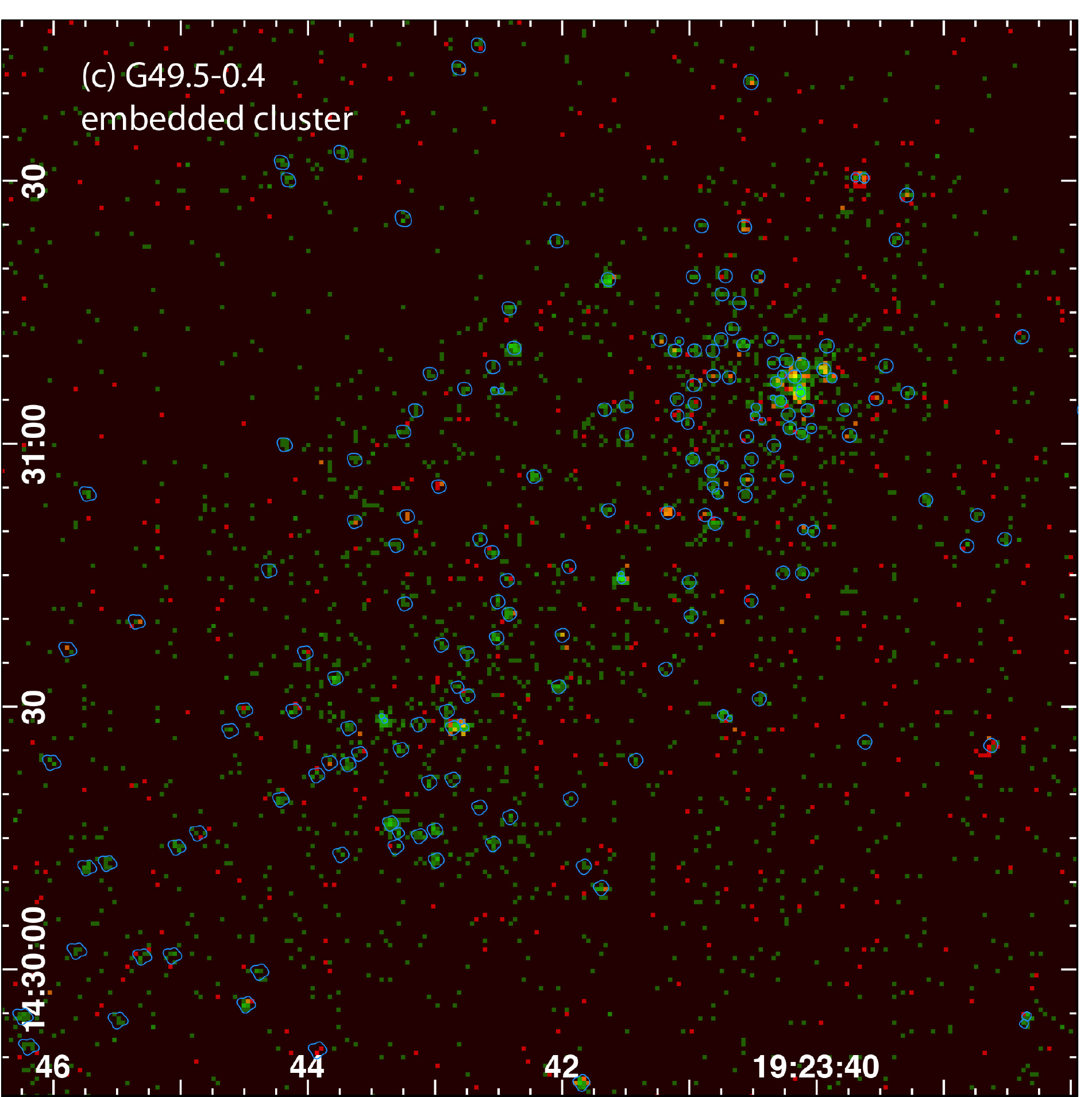}
\includegraphics[width=0.32\textwidth]{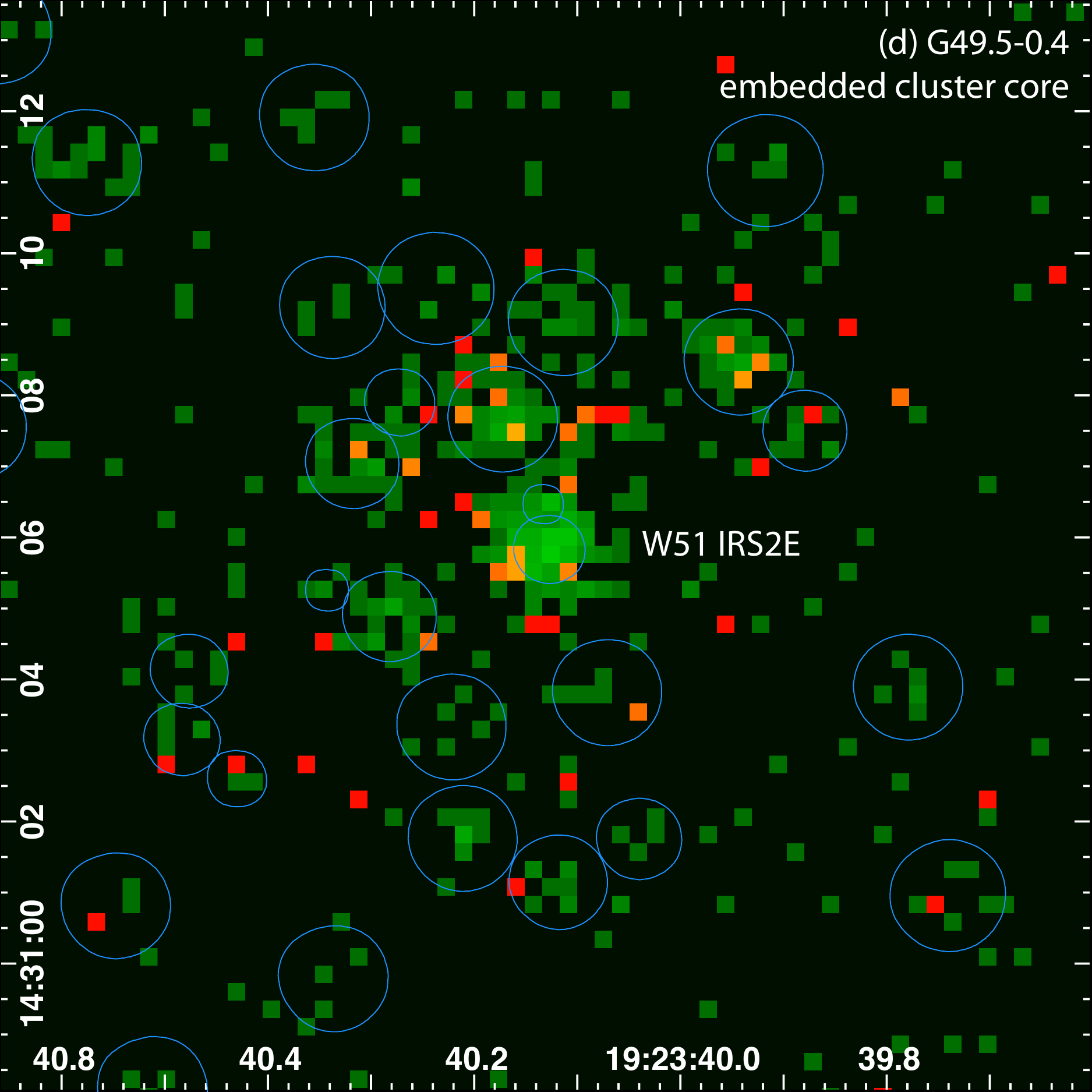}
\includegraphics[width=0.32\textwidth]{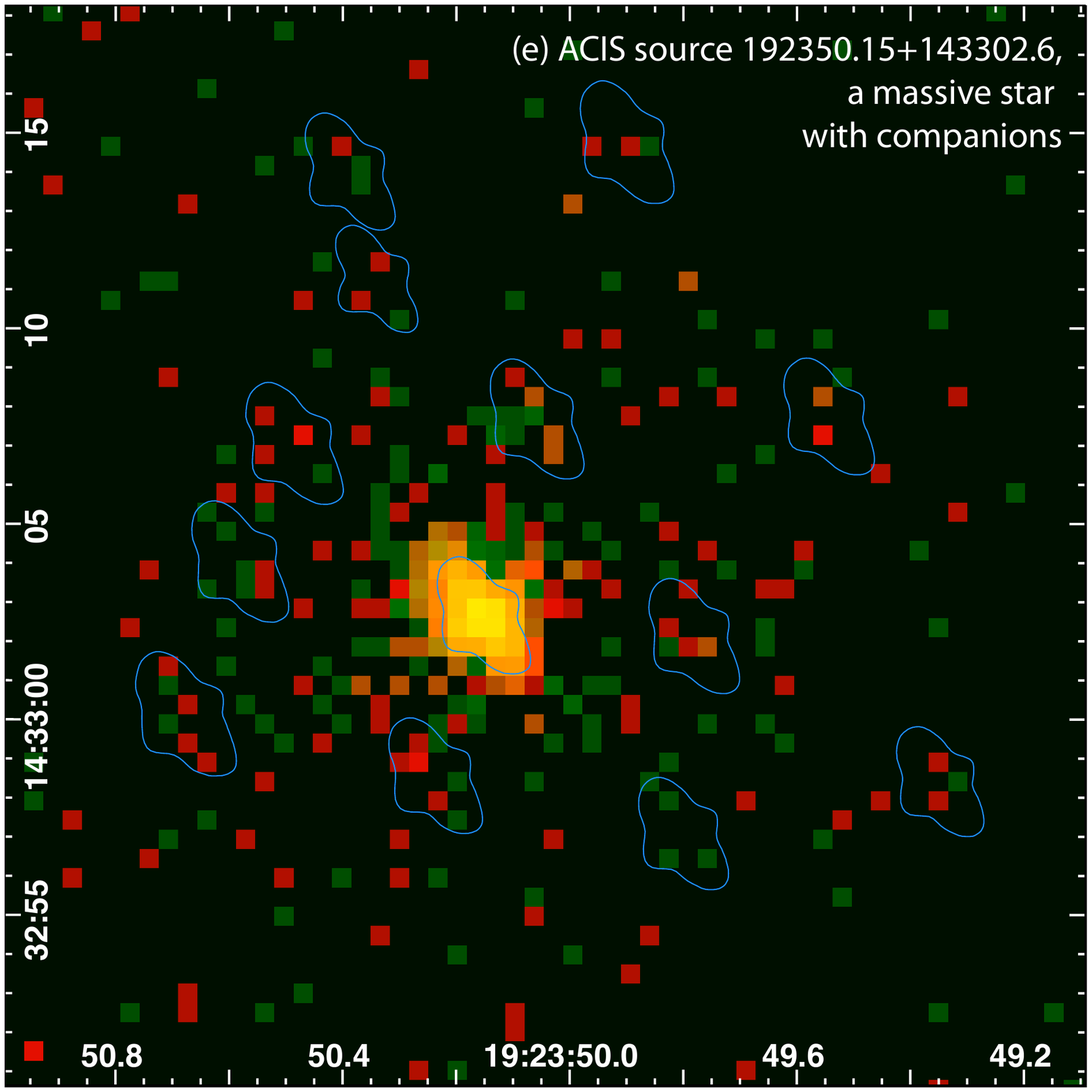}
\caption{W51A.
(a) ACIS exposure map with brighter ($\geq$5 net counts) ACIS point sources overlaid, with symbols and colors denoting median energy for each source.
(b) ACIS diffuse emission (full-band, 0.5--7~keV) in blue, \Spitzer/IRAC 8~$\mu$m emission in green, and \Spitzer/MIPS 24~$\mu$m emission in red.  
(c--e) ACIS binned event images, with soft events (0.5--2~keV) shown in red and hard events (2--8~keV) in green.  Point source extraction regions are overlaid in blue.  
\label{w51.fig}}
\end{figure}

Our 72-ks ACIS-I observation found 641 point sources in multiple clumps scattered around the field (Figure~\ref{w51.fig}a), with a clear concentration of $>$100 sources in the elongated central embedded cluster G49.5-0.4 (Figure~\ref{w51.fig}c).  The soft diffuse X-ray emission (Figure~\ref{w51.fig}b) is strikingly cut off in the northwestern part of the field and in a swath southeast of G49.5-0.4.  Clearly there is ample hot plasma suffusing this young MSFR complex, either being shadowed or displaced by dense bays of molecular material.

W51~IRS2E is a unique ACIS source (Figure~\ref{w51.fig}d) in the MOXC sample, emitting most of its X-ray photons in a broad 6.5-keV line due to fluorescing neutral Fe in its surrounding dense clouds; it is seen through heavy extinction only because of its amazingly hard X-ray emission, which varies by a factor of 2 in less than a day \citep{Townsley05}.  It is part of the crowded G49.5-0.4 core (Figure~\ref{w51.fig}c).  W51~IRS2E is exceptional even compared to the other very young, embedded massive stars with hard X-ray emission that we have described in the MOXC sample; in order for its emission to be concentrated in this reprocessed fluorescent neutral Fe line, the ionizing source itself must be producing most of its high-energy photons above the Fe absorption edge at 7.1~keV.  A similar source is found in the MSFR IRAS~20126+4104 \citep{Anderson11}.

Figure~\ref{w51.fig}e shows a bright X-ray source with a K$<$10 2MASS counterpart, likely the mid-O type massive star ionizing radio \hii region ``g'' \citep{Mehringer94}, called source \#4168 in ``Region 2'' defined by \citet{Okumura00}.  Note the concentration of fainter X-ray sources around it, similar to the southeast pointing we examined in NGC~6357 above; such groupings are not uncommon \citep[e.g.\ QZ~Car in the Carina Nebula,][]{Townsley11a}.  Again, they seem to indicate that massive stars can form in near isolation, with just a handful of lower-mass companion pre-MS stars rather than the full complement of hundreds that would be expected from a standard initial mass function.  


\subsection{G29.96-0.02 \label{sec:g29.96}}

\begin{figure}[htb]
\centering
\includegraphics[width=0.48\textwidth]{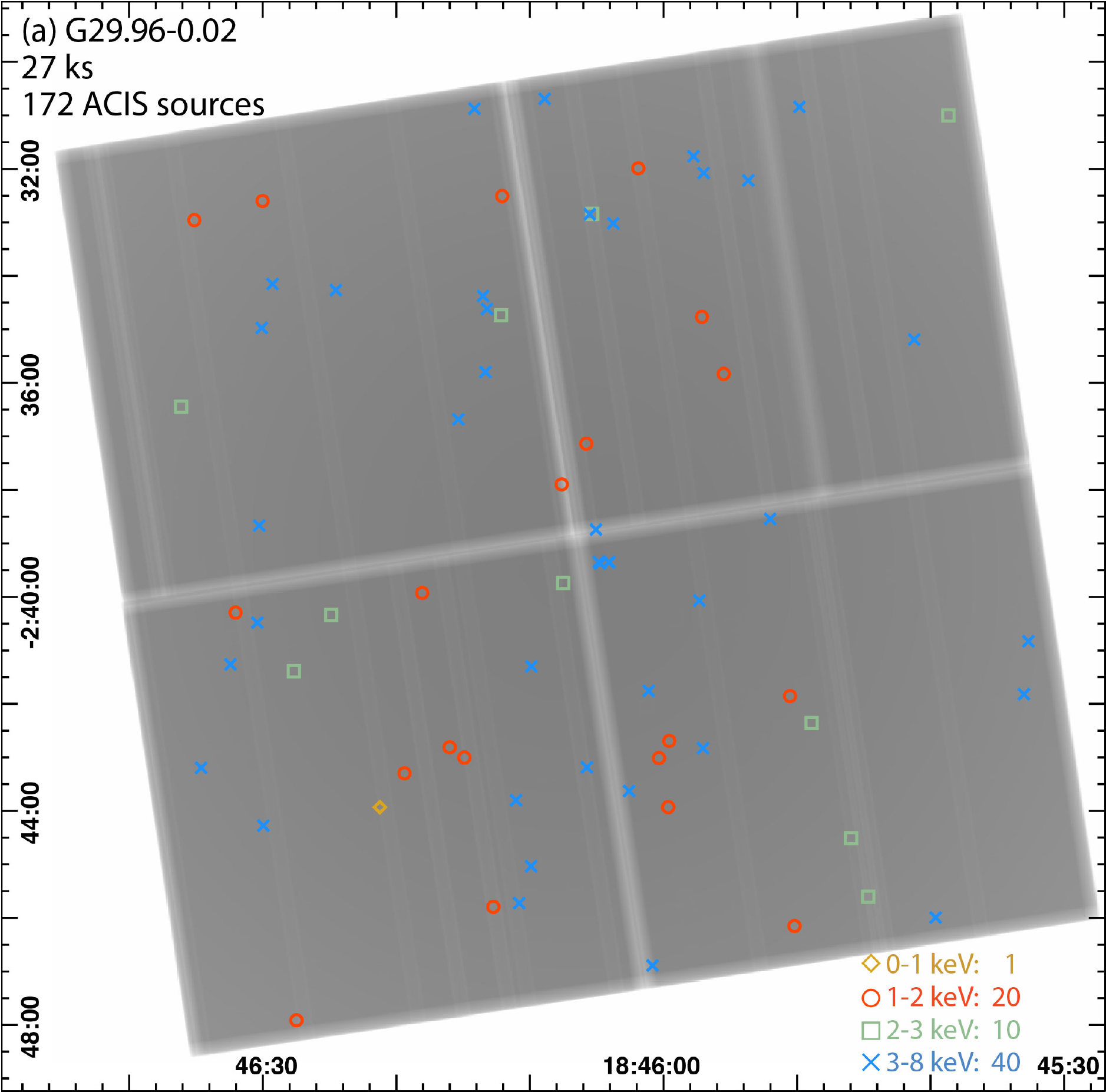}
\includegraphics[width=0.48\textwidth]{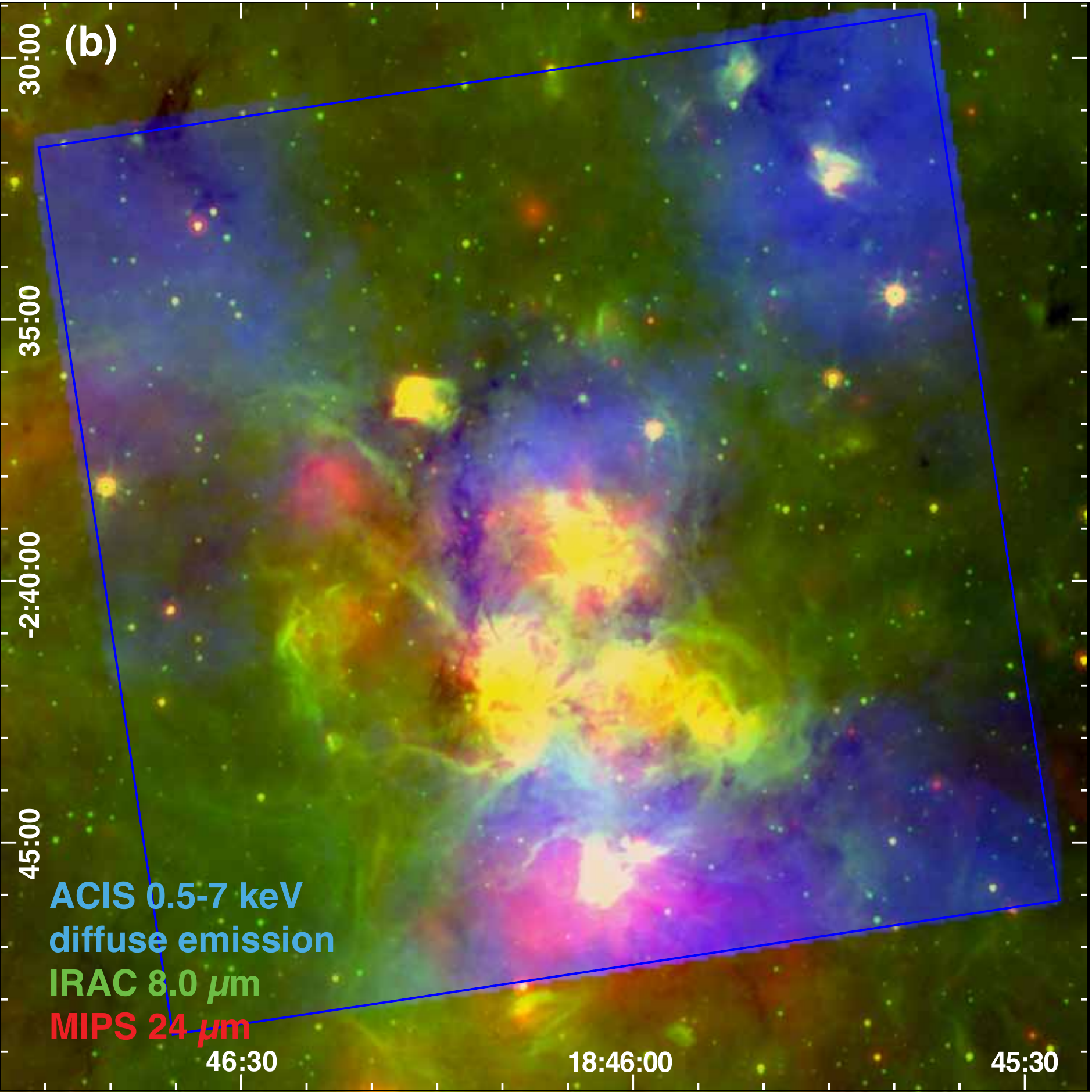}
\caption{G29.96-0.02.
(a) ACIS exposure map with brighter ($\geq$5 net counts) ACIS point sources overlaid, with symbols and colors denoting median energy for each source.
(b) ACIS diffuse emission (full-band, 0.5--7~keV) in blue, \Spitzer/IRAC 8~$\mu$m emission in green, and \Spitzer/MIPS 24~$\mu$m emission in red.   
\label{g29.96.fig}}
\end{figure}

This MSFR hosts a famous cometary \UCHIIR \citep{Wood89}.  Its distance has long been highly uncertain, but \citet{Russeil11} has recently used \Herschel\ data to place it at $\sim$6.2~kpc.  A more recent \Herschel\ study of this region shows that it is very young, with clumps just now forming stars \citep{Beltran13}.  The ACIS exposure of this field (Figure~\ref{g29.96.fig}a) is too shallow to tell us much about the MYSC, although the massive star ionizing the \UCHIIR is clearly detected.  It has a hard spectrum ($kT \sim 2$~keV) and is seen through a large absorbing column ($A_V \sim 75$~mag), similar to other \UCHIIR and \GHIIR ionizing sources detected in X-rays, such as those described above.

The real surprise here is the large-scale diffuse X-ray emission (Figure~\ref{g29.96.fig}b) surrounding the embedded cluster; its anticoincidence with \Spitzer\ bubbles is a clear indication that it is associated with this MSFR (patches of diffuse X-rays in the northeast and northwest corners of the ACIS field are not necessarily associated with the MYSC).  The fact that this diffuse X-ray emission is detectable across such a large distance with such a modest ACIS-I exposure is quite interesting; even at this early evolutionary stage, massive star winds in this embedded MYSC are perforating their natal clouds and contributing to the hot ISM.


\subsection{NGC~3603 \label{sec:n3603}}

NGC~3603 is a spectacular monolithic MYSC, one of the most massive MYSCs known in the Galaxy.  Its dense concentration of massive stars, ionization fronts, and dust pillars constitute one of the Galaxy's most prominent and powerful starburst clusters \citep{Westmoquette13}.  Despite its large distance, it is far less obscured \citep[$A_{V}$$\sim$4.6~mag,][]{Rochau10} than the Galaxy's more massive MYSCs, making NGC~3603 accessible to visual telescopes. \citet{Melena08} presented spectroscopy for 16 bright stars, for a total of 38 typed massive stars:  3 WRs, 2 older supergiants, and 33 O stars; they estimate that $>$60\% of NGC~3603's O stars still lack spectral types.  While some recent work \citep{Kudryavtseva12,Pang13} finds that the MYSC formed in a near-instantaneous burst 1--2~Myr ago, other studies \citep{Beccari10,Correnti12} come to the opposite conclusion, saying that star formation in NGC~3603 occurred nearly continuously over the range of 2--30~Myr ago.

NGC~3603 was observed for 49~ks in the first year of the \Chandra mission (ObsID~0633); in the original analysis \citep{Moffat02}, $>$40 massive stars were seen in X-rays, including 3 WR stars, plus $>$300 other X-ray sources and diffuse X-ray emission.  \citet{Sung04} also analyzed this \Chandra dataset, finding $>$2000 sources and using their spatial distribution to estimate a cluster radius of $\sim$2$\arcmin$; they note the extreme X-ray source confusion in the cluster center.  Our own analysis of ObsID~0633, using methods similar to the ones used for MOXC, yielded 1328 X-ray sources on the ACIS-I array \citep{Townsley11c}.

A decade later, this original \Chandra dataset was augmented with an additional 450-ks ACIS-I observation of the field; MOXC includes the full complement of usable ACIS data, totaling $>$494~ks (see Table~\ref{tbl:obslog}).  Three newly-identified O2 stars, called WR~42e \citep{Roman12}, MTT~58 \citep{Roman13a}, and MTT~68 \citep{Roman13b}, are found to be bright X-ray sources in the ACIS data; many new massive star candidates will be nominated from our studies of the ACIS data.  Despite the source confusion in this dense MYSC, MOXC catalogs almost 4000 \Chandra X-ray point sources.  This \Chandra Large Project study of NGC~3603 provides an important contrast to the CCCP; NGC~3603 contains as many massive stars as the entire Carina complex but concentrated into a single dense MYSC rather than spread over several less massive clusters.

\begin{figure}[htb]
\centering
\includegraphics[width=0.48\textwidth]{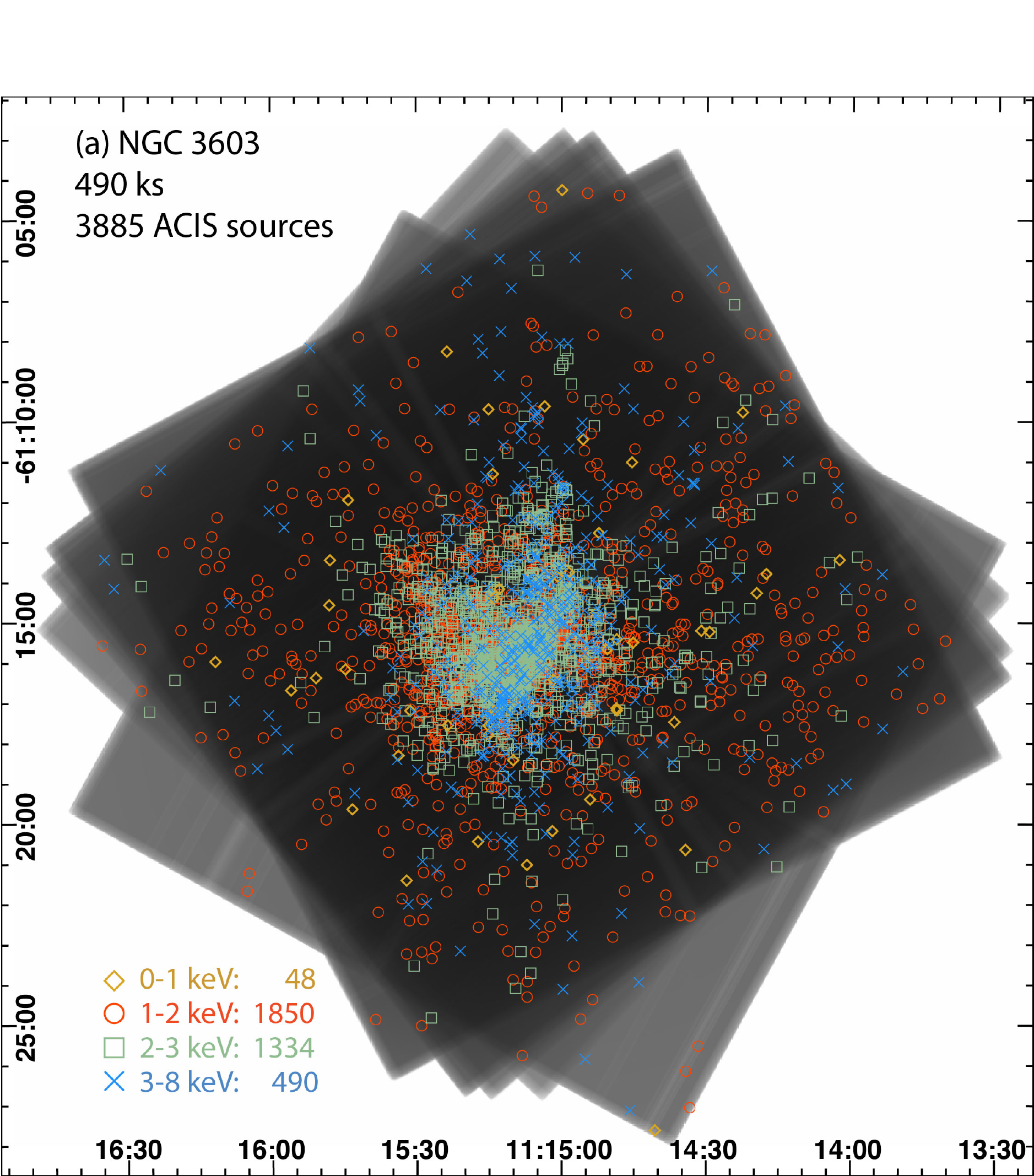}
\includegraphics[width=0.48\textwidth]{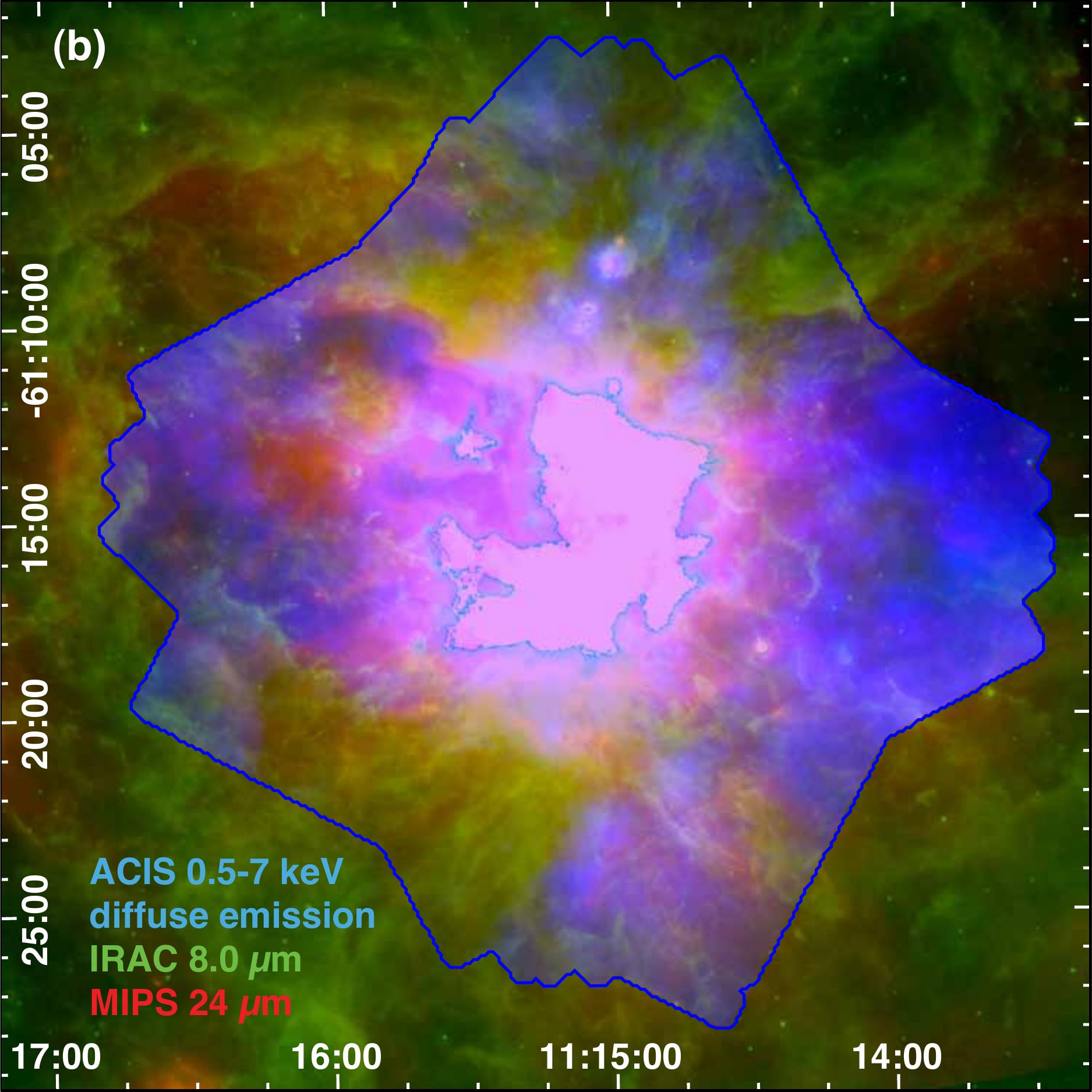}
\includegraphics[width=0.48\textwidth]{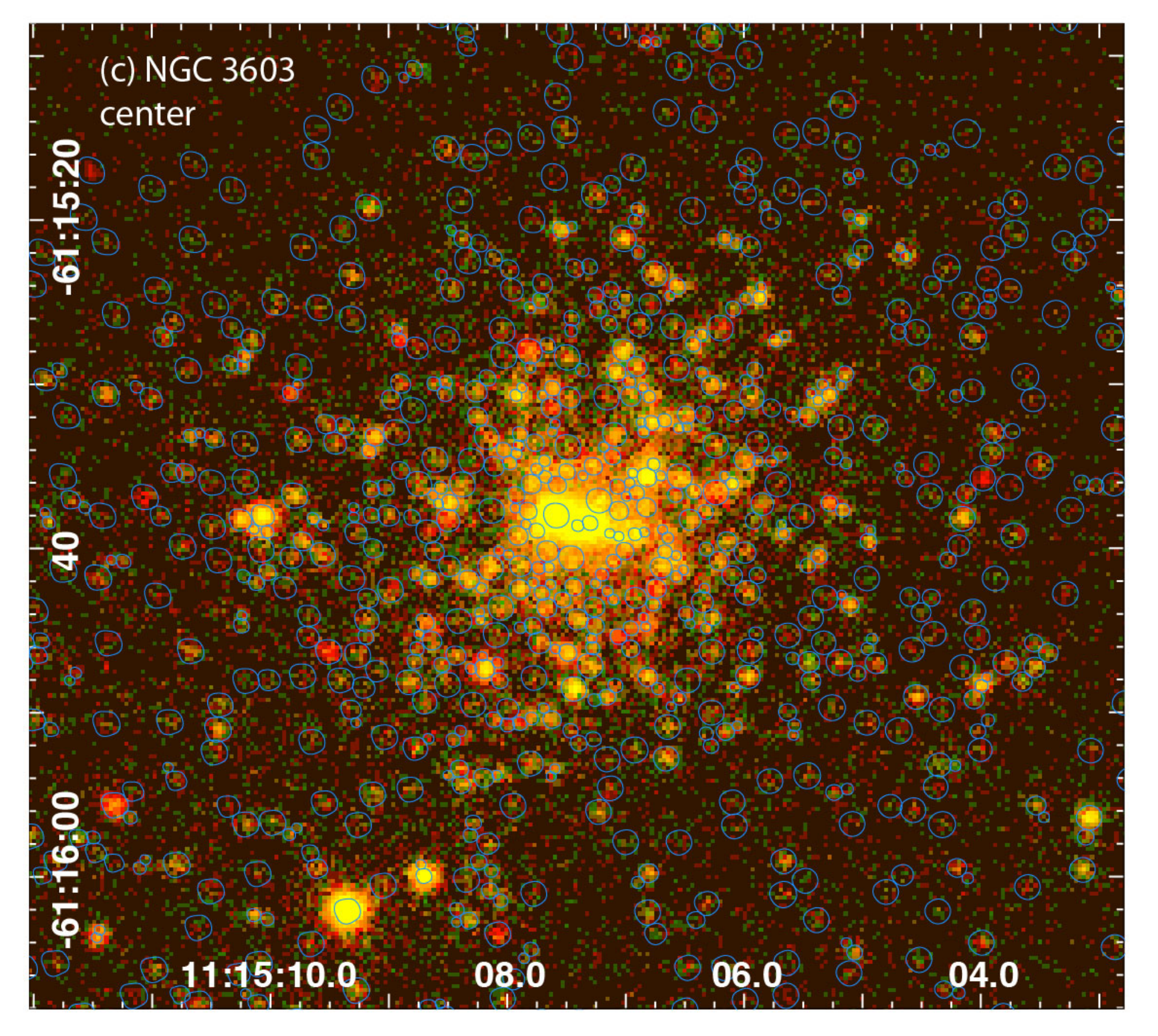}
\includegraphics[width=0.51\textwidth]{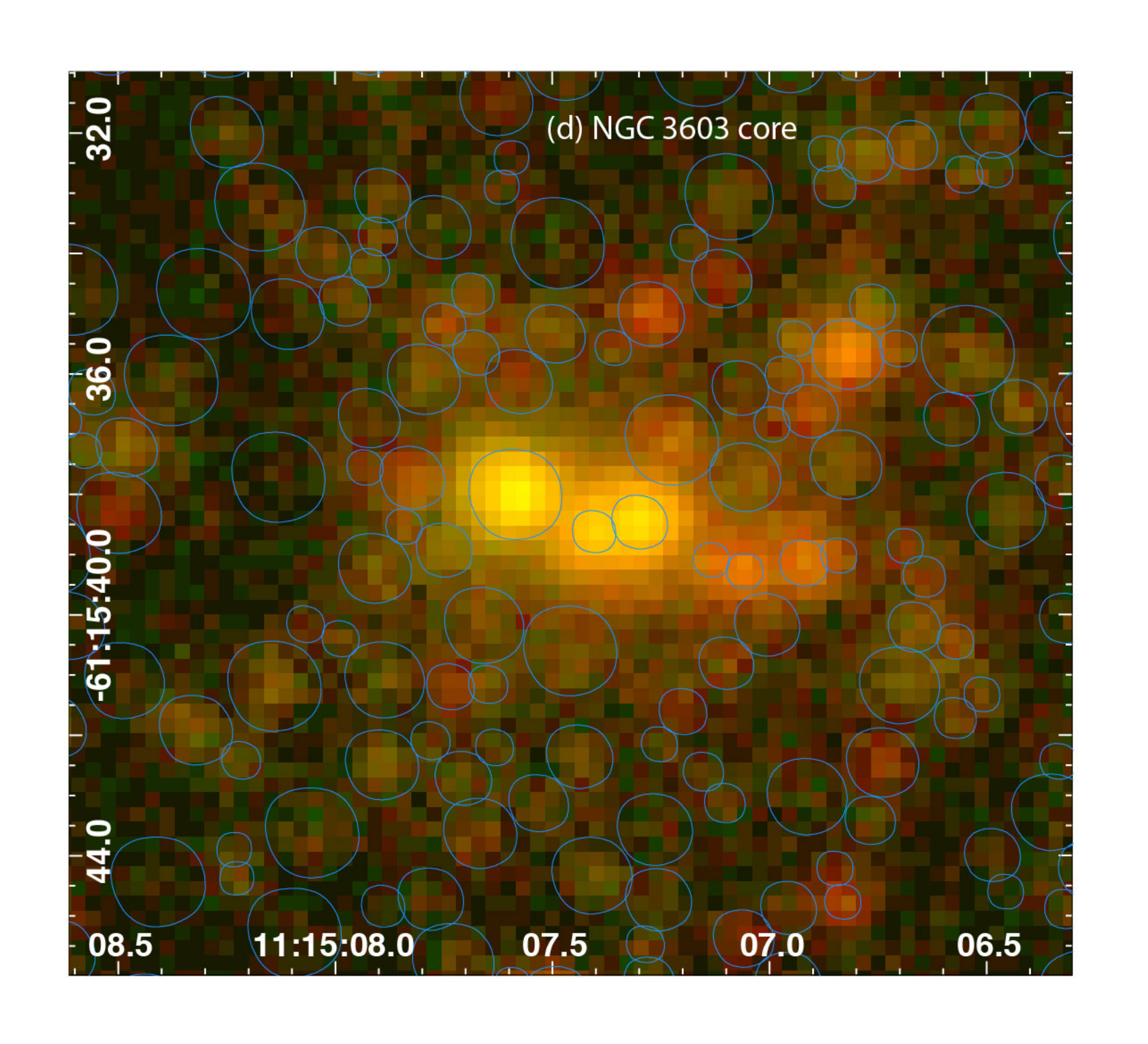}
\caption{NGC~3603.
(a) ACIS exposure map with brighter ($\geq$5 net counts) ACIS point sources overlaid, with symbols and colors denoting median energy for each source.
(b) ACIS diffuse emission (full-band, 0.5--7~keV) in blue, \Spitzer/IRAC 8~$\mu$m emission in green, and \Spitzer/MIPS 24~$\mu$m emission (heavily saturated here) in red.
(c,d) ACIS binned event images, with soft events (0.5--2~keV) shown in red and hard events (2--8~keV) in green.  Point source extraction regions are overlaid in blue.
\label{n3603.fig}}
\end{figure}

Figure~\ref{n3603.fig}c illustrates the extreme crowding in this ACIS observation, where the central arcminute of the NGC~3603 MYSC contains $\sim$700 ACIS sources.  In just the central $\sim$15$\arcsec$ of the field we find $\sim$130 ACIS sources (Figure~\ref{n3603.fig}d), including the 3 WR stars, which show variability and pile-up that will require sophisticated spatio-spectral modeling, as for the O4-O4 binaries in M17 shown above.  Clearly many more point sources are present in the ACIS data but are too crowded to be individually resolved; these contribute to the high X-ray background seen in this image.  More sophisticated methods of image reconstruction and source extraction would be required to do full justice to this rich but complicated ACIS dataset.

CO data show that NGC~3603's MYSC is interacting with its natal cloud \citep{Rollig11}; dynamical studies of the surrounding pillars show that they are being eroded by the MYSC's powerful winds \citep{Westmoquette13}.  Thus it is perhaps not surprising that NGC~3603 exhibits spectacular diffuse X-ray emission, found even in the original short observation \citep{Moffat02,Townsley11c} but really showcased in the \Chandra Large Project data presented here (Figure~\ref{n3603.fig}b).  Our X-ray spectral fitting of diffuse emission in the original ACIS dataset \citep{Townsley11c} showed that there was still a large contribution to the X-ray spectrum from unresolved pre-MS stars.  We expect that the new dataset, 10 times deeper, will give a global spectrum more indicative of the hot plasma X-ray emission in this field, and will provide enough photons for tessellated spectral fits and parameter mapping \citep{Townsley11b}.

\clearpage

\subsection{30 Doradus (The Tarantula Nebula) \label{sec:30dor}}

30~Doradus, in the Large Magellanic Cloud, is the most massive and luminous MSFR in the Local Group, containing several thousand massive stars plus several hundred thousand lower-mass pre-MS stars.  Its central MYSC, R136, has 1000 times the ionizing radiation of Orion \citep{Conti08} and at least 5 times as many early-O stars as Carina \citep{Evans11}; recent {\it HST} observations suggest that it is undergoing a merger with a slightly older cluster \citep{Sabbi12}.  

30~Dor was one of the first \Chandra targets, observed for $\sim$22~ks in 1999 as part of the ACIS Instrument Team's GTO program \citep{Townsley06a,Townsley06b}, and observed again for 92-ks in 2006 (Table~\ref{tbl:obslog}).  Due to significant calibration differences between those observations\footnote
{
The 1999 data were obtained with the ACIS camera at a warmer focal plane temperature (-110C) than that used for most of the mission (-120C).
}, the short 1999 observation is not included in the MOXC analysis.

\begin{figure}[htb]
\centering
\includegraphics[width=0.48\textwidth]{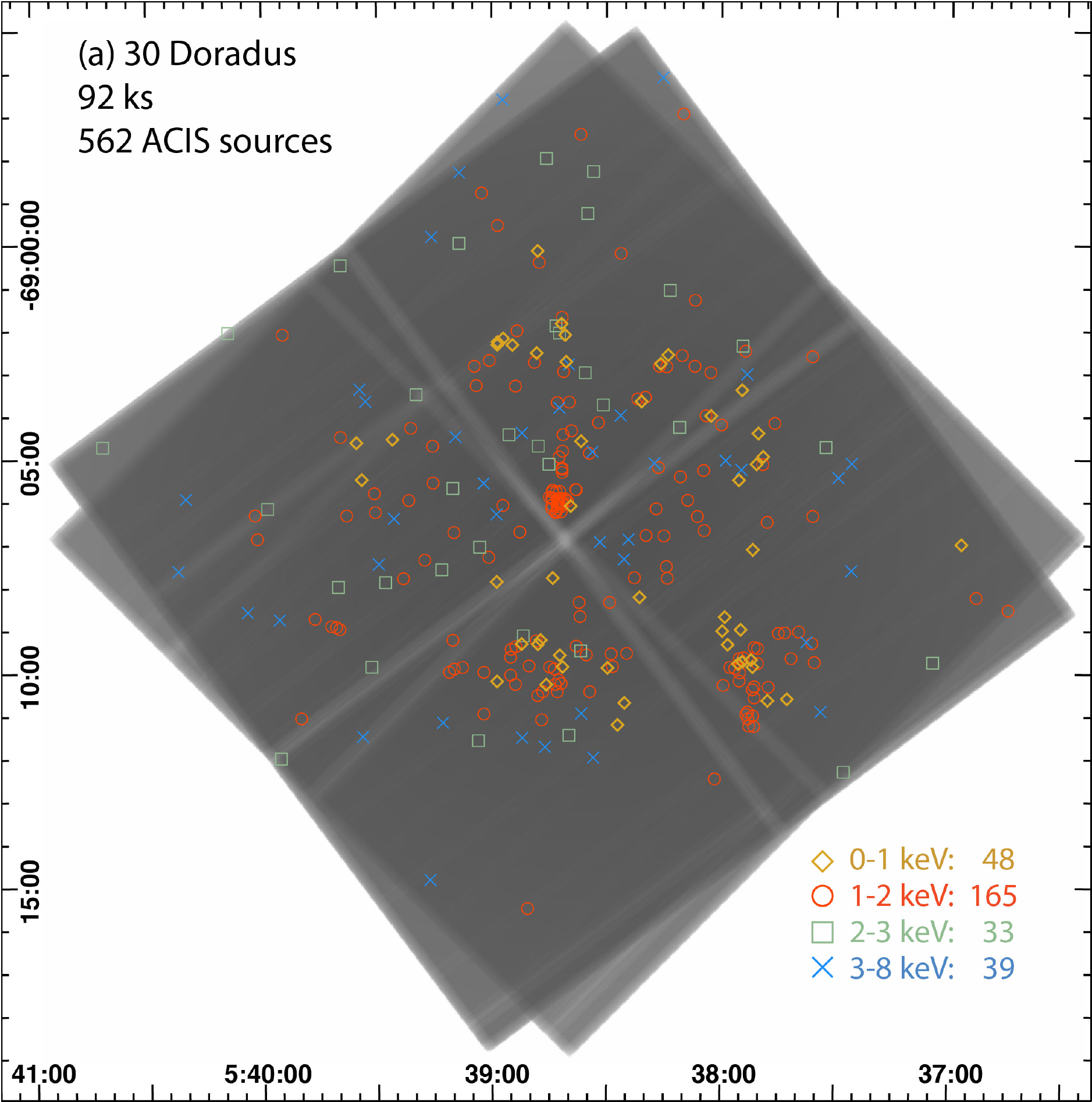}
\includegraphics[width=0.48\textwidth]{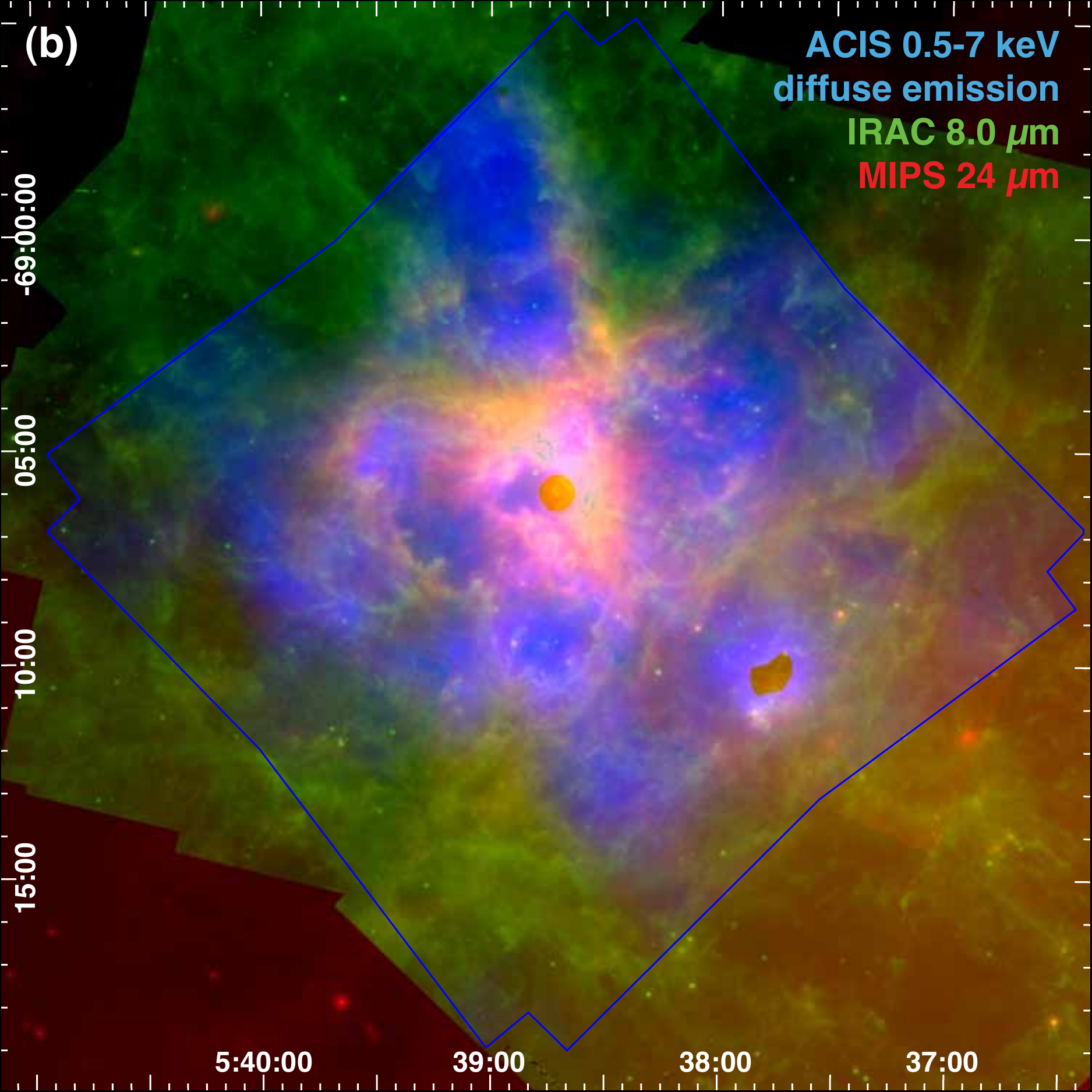}
\caption{30 Doradus (The Tarantula Nebula).
(a) ACIS exposure map with brighter ($\geq$5 net counts) ACIS point sources overlaid, with symbols and colors denoting median energy for each source.
(b) ACIS diffuse emission (full-band, 0.5--7~keV) in blue, \Spitzer/IRAC 8~$\mu$m emission in green, and \Spitzer/MIPS 24~$\mu$m emission in red.  The central cluster R136 and PSR~J0537-6910 (and its pulsar wind nebula) have been masked in the ACIS image.
\label{30dor.fig}}
\end{figure}

The 2006 \Chandra observations reach just the top of the MSFR X-ray luminosity function, detecting a few WR and early-O stars in R136 (Figure~\ref*{30dor.fig}a); many of these exhibit hard X-ray emission consistent with colliding-wind binaries \citep{Townsley06b} and/or magnetic massive stars \citep{Petit13}, as we have now come to expect, having examined the other MOXC MSFRs in these last few pages.  Details of the X-ray sources in R136 and across the wider ACIS field will be presented in an upcoming paper (Townsley et al.\ in prep.), thus no zoomed images are included here.

The $\sim$2.5-Myr-old MSFR NGC~2060 lies $\sim$6$\arcmin$ southwest of R136.  In X-rays, NGC~2060 is dominated by the young SNR N157B containing the 16-ms PSR~J0537-6910, the most energetic pulsar known \citep{Marshall98}, and its prominent cometary pulsar wind nebula \citep{Chen06}, a TeV gamma ray source \citep[probably due to inverse-Compton scattering of the surrounding bright IR radiation field;][]{HESS12}.  The pulsar is slightly piled-up, even this far off-axis.  SNRs must pervade 30~Dor but are difficult to detect individually due to age and environment \citep{Chu90}.  

30~Dor exhibits at least five plasma-filled superbubbles with $\sim$100-pc scales \citep{Wang91}, products of strong OB winds and multiple supernovae (Figure~\ref*{30dor.fig}b).  \Chandra has already shown \citep{Townsley06a, Townsley11c, Lopez11, Pellegrini11} that these are spatially complex X-ray structures with a range of X-ray plasma temperatures, ionization timescales, absorptions, and luminosities.  Our upcoming paper on the 2006 dataset (Townsley et al.\ in prep.) will give coarse maps of plasma properties from X-ray spectral fitting.  New \Chandra observations of 30~Dor were recently approved for 2014; we expect these deeper data to resolve several thousand X-ray point sources and to give much finer spatial detail on the diffuse X-ray emission.

\section{SUMMARY \label{sec:summary}}

The Massive Star-Forming Regions Omnibus X-ray Catalog (MOXC) collates 20,623 X-ray point sources in 12 (mostly Galactic) MSFRs, obtained from observations with the ACIS-I camera on \Chandra between 2000 March and 2013 January.  Photometric and other properties are provided in a consistent manner for all point sources in a single large electronic table, to facilitate the comparison of these MSFRs and their MYSCs.  The first seven targets are part of the MYStIX project \citep{Feigelson13}; the remaining MSFRs are generally more massive and more distant, and include 30~Doradus in the Large Magellanic Cloud.  

These \Chandra datasets are characterized by a wide range of detection sensitivities and spatial coverage.  We have employed our own custom software, {\em ACIS Extract}, to accomodate these observational challenges; this software is supported, well-documented, and available to the community.\footnote{ The {\em ACIS Extract} software package and User's Guide are available at \url{http://www.astro.psu.edu/xray/acis/acis_analysis.html}. }  For MYStIX, the X-ray catalogs from MOXC have been combined with other \Chandra data analyzed with the same methods \citep{Kuhn13a}, along with near-IR data \citep{King13,Naylor13} and \Spitzer\ mid-IR data \citep{Kuhn13b,Povich13}; this multiwavelength dataset has been subjected to a Bayesian source classification scheme, resulting in a list of ``MYStIX Probable Complex Members'' \citep{Broos13} that will be used for science analysis.

The MSFRs considered here exhibit many kinds of young stellar cluster masses and morphologies, from monolithic MYSCs to clusters-of-clusters to a variety of distributed clumps.  The spatial structure of the MYStIX targets will be studied in detail in a series of papers by Kuhn et al.
Here we noted in particular the small clumps of pre-MS stars that sometimes surround otherwise-isolated massive stars (e.g.\ in NGC~6357 and W51A).

Many young, embedded massive stars in the MOXC sample exhibit hard X-ray emission, probably indicating the presence of colliding-wind binaries and/or significant magnetic fields.  Much of the X-ray luminosity of these sources is likely hidden from us, in a brighter, soft X-ray plasma component that is completely absorbed by the large intervening column.  Conversely, the \UCHIIR ionizing sources that are not detected in these \Chandra observations are perhaps dominated by this soft component, implying that they are single massive stars and/or that they lack strong magnetic fields.  

Based on simple morphological arguments, especially when compared to ISM structures dramatically illustrated by \Spitzer\ and {\em WISE} data, we have shown that all MOXC MSFRs have measurable diffuse X-ray emission.  This illustrates the birth of the hot ISM through massive star winds and cavity supernovae and provides definitive evidence that such processes are at work in all MSFRs.  This soft diffuse X-ray emission is observable even around distant, very young (hence highly embedded) MSFRs---as long as they have already formed their massive stars---presumably because the natal GMCs that contain these MSFRs are highly inhomogeneous, allowing the hot plasma to escape through fissures in this clumpy medium \citep{Rogers13}.

The MOXC paper is both the last of the MYStIX data papers and the first of our ``beyond-MYStIX'' \Chandra studies of more distant, energetic, and massive MSFRs.  A wide range of MYStIX science studies is now underway, using the catalogs presented here and in the other MYStIX data papers mentioned above.  In addition, much remains to be done on just the ACIS data; we will next turn our attention to X-ray spectral fitting of both the point sources and the diffuse components of these MSFRs.  Of particular interest will be the fascinating variety of X-ray emission displayed by the large number of massive stars in the MOXC sample.


\appendix
\section{X-RAY CHARACTERIZATIONS OF OTHER MYStIX TARGETS}\label{othertargets.sec}

For completeness and to facilitate comparison, this appendix gives the same X-ray characterizations for other targets in the MYStIX sample as those shown in Section~\ref{sec:targets} for MOXC targets.  \Chandra point source catalogs for these other MYStIX targets can be found in \citet{Kuhn13a} and \citet{Kuhn10}.  Brief descriptions of these MSFRs are given in the MYStIX introductory paper \citep{Feigelson13}.  As we found above for the MOXC targets, these MSFRs also show diffuse X-ray emission that is strikingly complementary to IR structures, filling bubbles and exhibiting shadowing and displacement by shells and pillars of cold material.  Massive star feedback clearly manifests itself through these ubiquitous X-ray plasmas, revealing the birth of the hot ISM in MSFRs.


\begin{figure}[htb]
\centering
\includegraphics[width=0.48\textwidth]{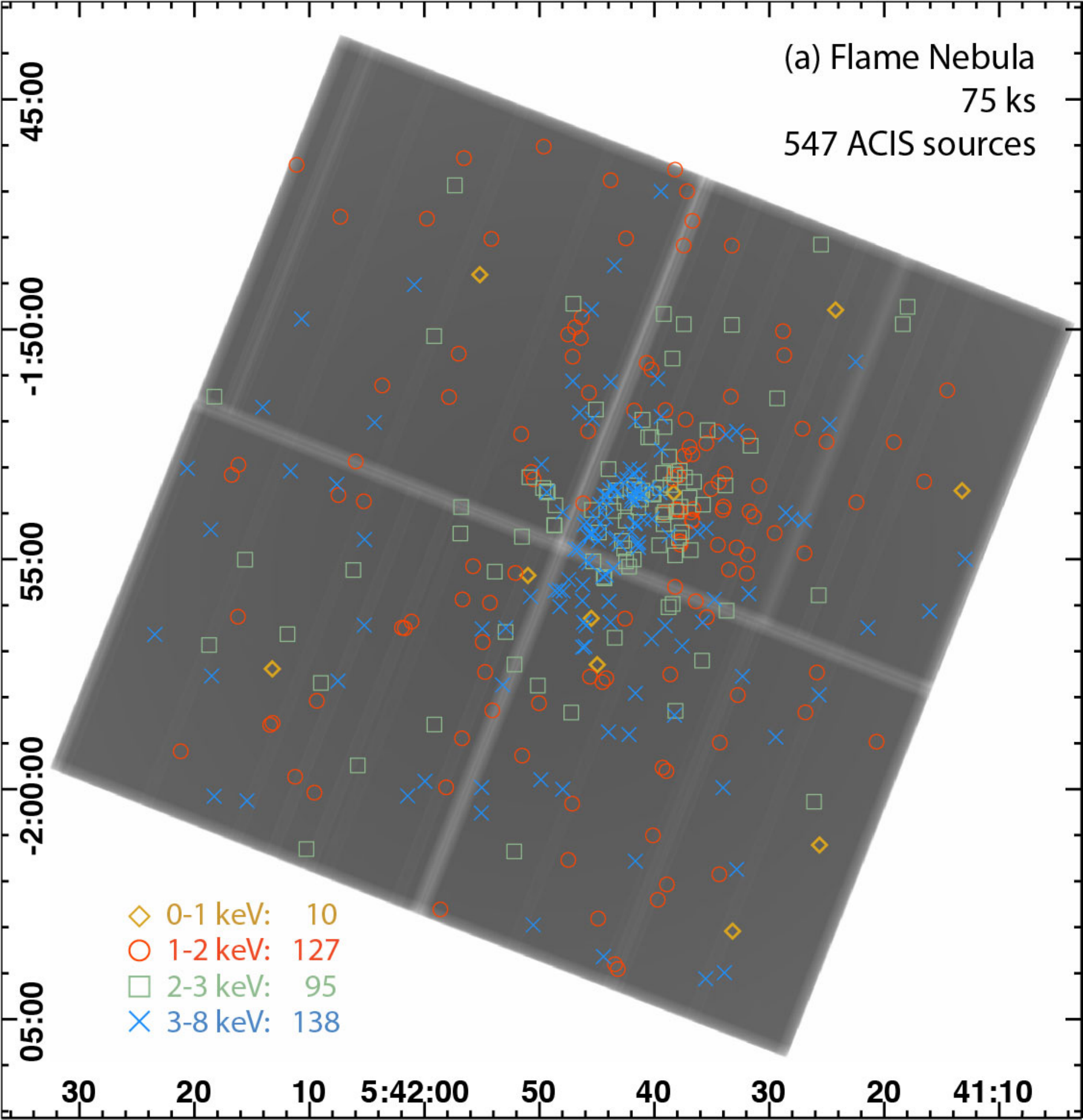}
\includegraphics[width=0.48\textwidth]{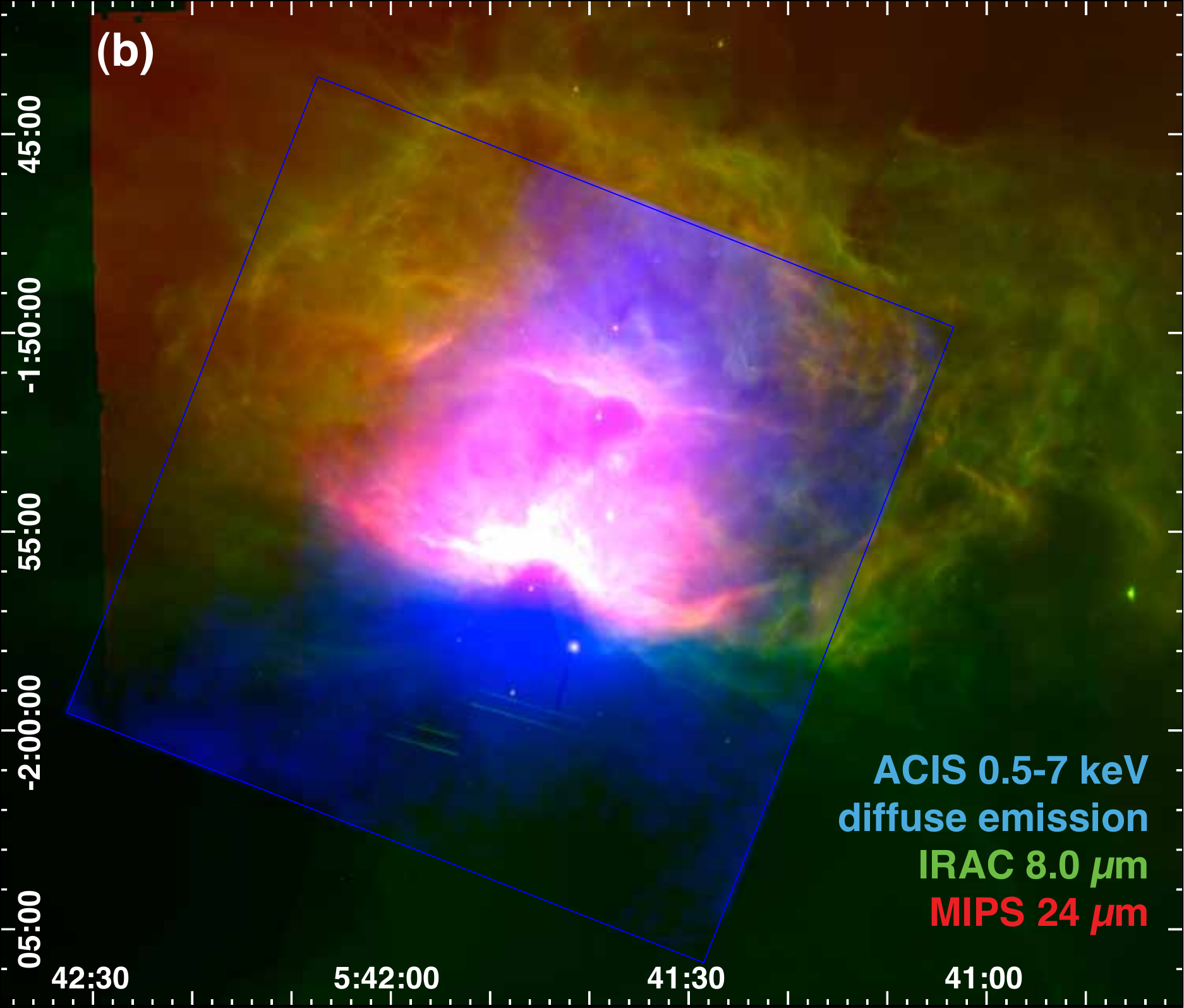}
\caption{The Flame Nebula (NGC~2024).
(a) ACIS exposure map with brighter ($\geq$5 net counts) ACIS point sources overlaid, with symbols and colors denoting median energy for each source.
(b) ACIS diffuse emission (full-band, 0.5--7~keV) in blue, {\em Spitzer}/IRAC 8~$\mu$m emission in green, and {\em Spitzer}/MIPS 24~$\mu$m emission in red.   
\label{flame.fig}}
\end{figure}


\begin{figure}[htb]
\centering
\includegraphics[width=0.48\textwidth]{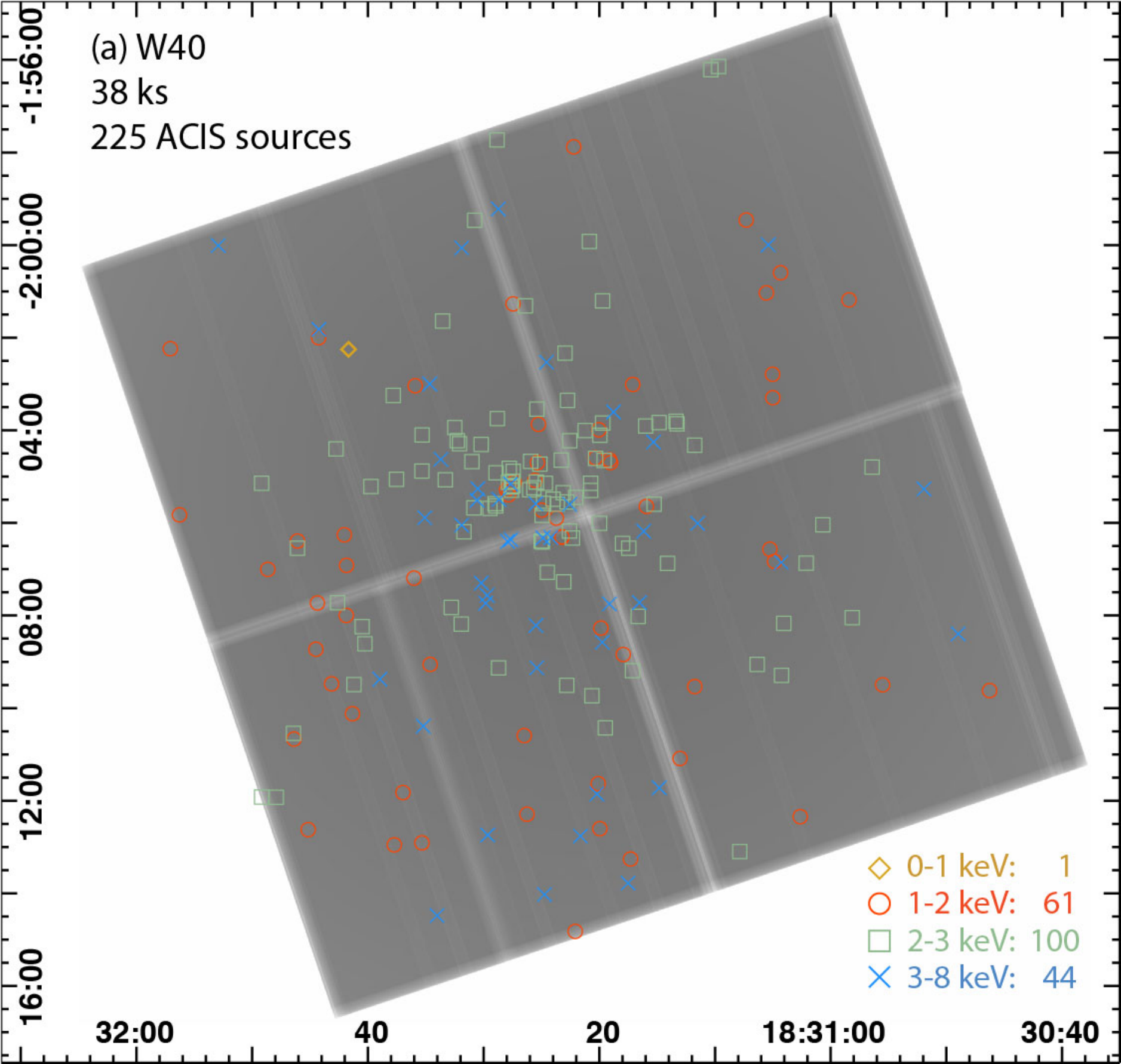}
\includegraphics[width=0.48\textwidth]{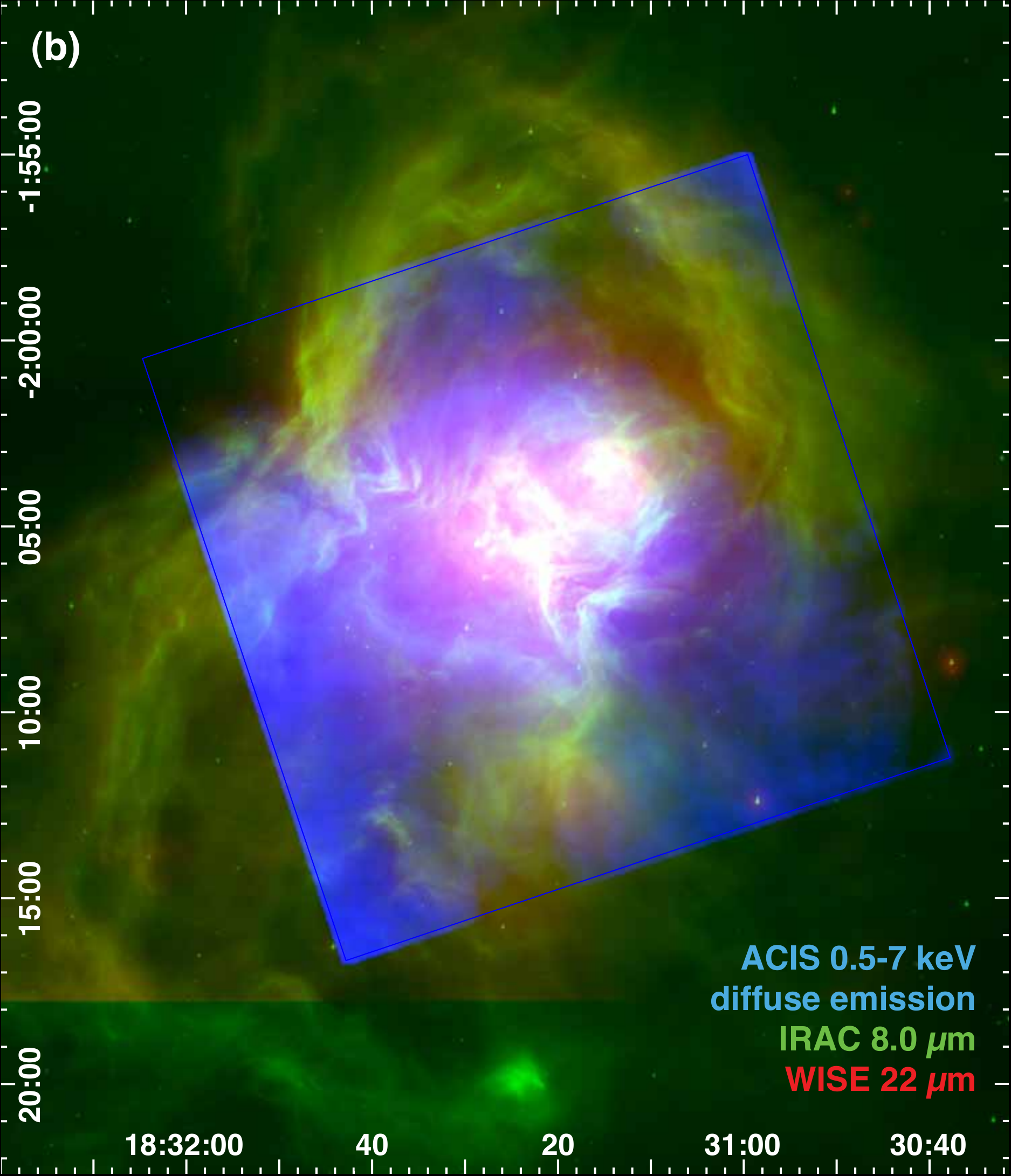}
\caption{W40.
(a) ACIS exposure map with brighter ($\geq$5 net counts) ACIS point sources overlaid, with symbols and colors denoting median energy for each source.
(b) ACIS diffuse emission (full-band, 0.5--7~keV) in blue, {\em Spitzer}/IRAC 8~$\mu$m emission in green, and {\em WISE} Band 4 (22~$\mu$m) emission in red.   
\label{w40.fig}}
\end{figure}


\begin{figure}[htb]
\centering
\includegraphics[width=0.48\textwidth]{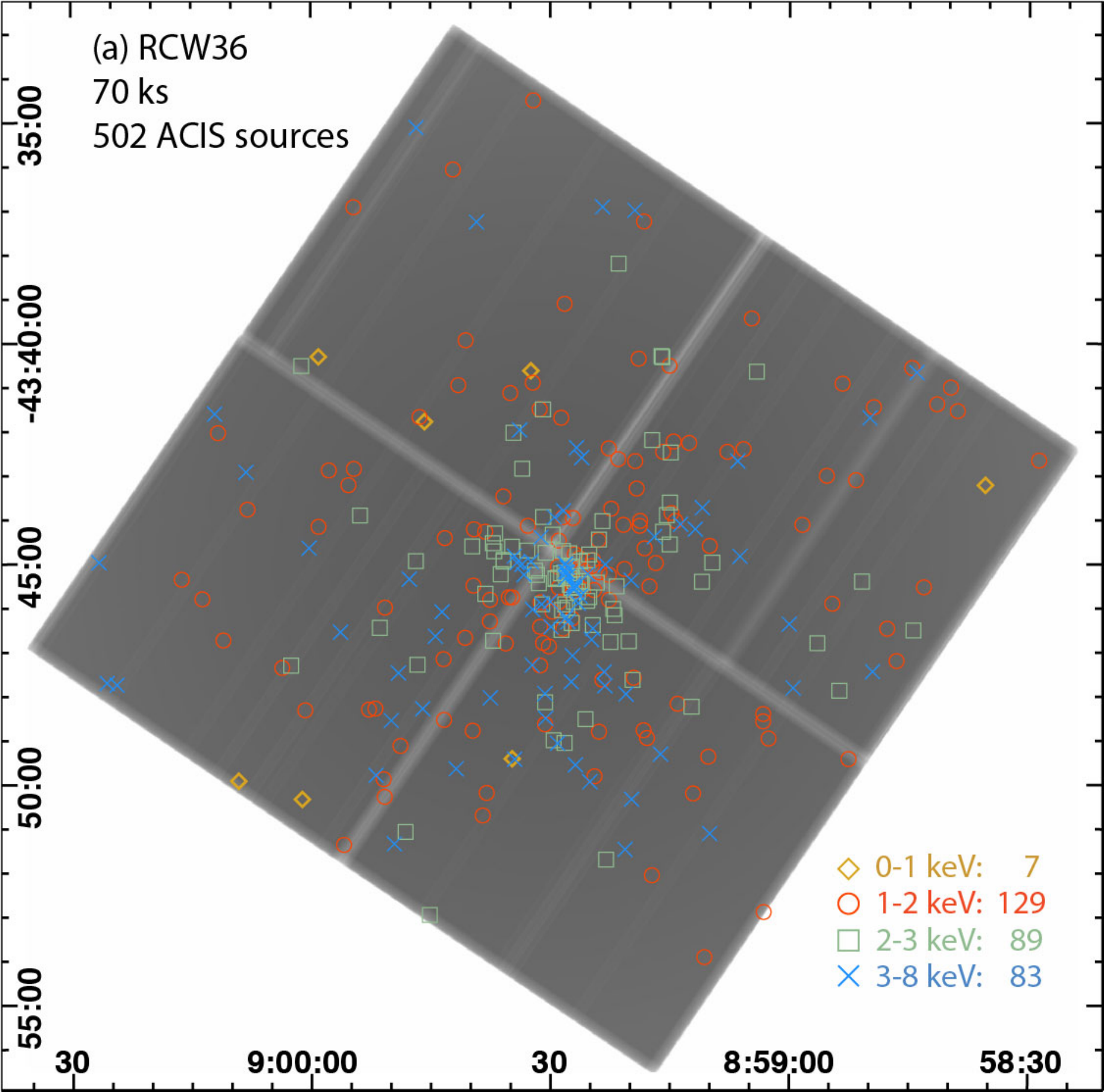}
\includegraphics[width=0.48\textwidth]{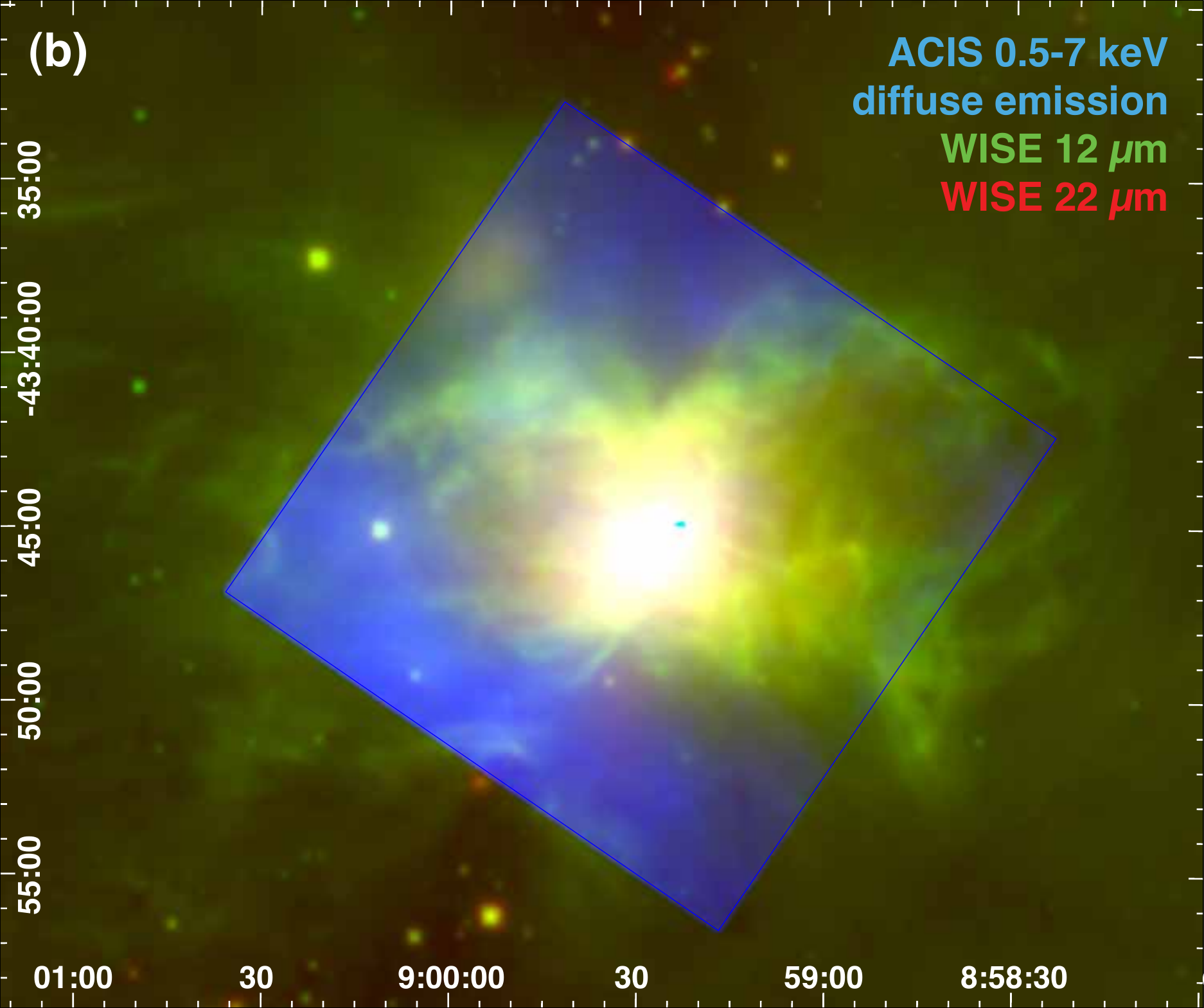}
\caption{RCW~36.
(a) ACIS exposure map with brighter ($\geq$5 net counts) ACIS point sources overlaid, with symbols and colors denoting median energy for each source.
(b) ACIS diffuse emission (full-band, 0.5--7~keV) in blue, {\em WISE} Band 3 (12~$\mu$m) emission in green, and {\em WISE} Band 4 (22~$\mu$m) emission in red.    
\label{rcw36.fig}}
\end{figure}


\begin{figure}[htb]
\centering
\includegraphics[width=0.48\textwidth]{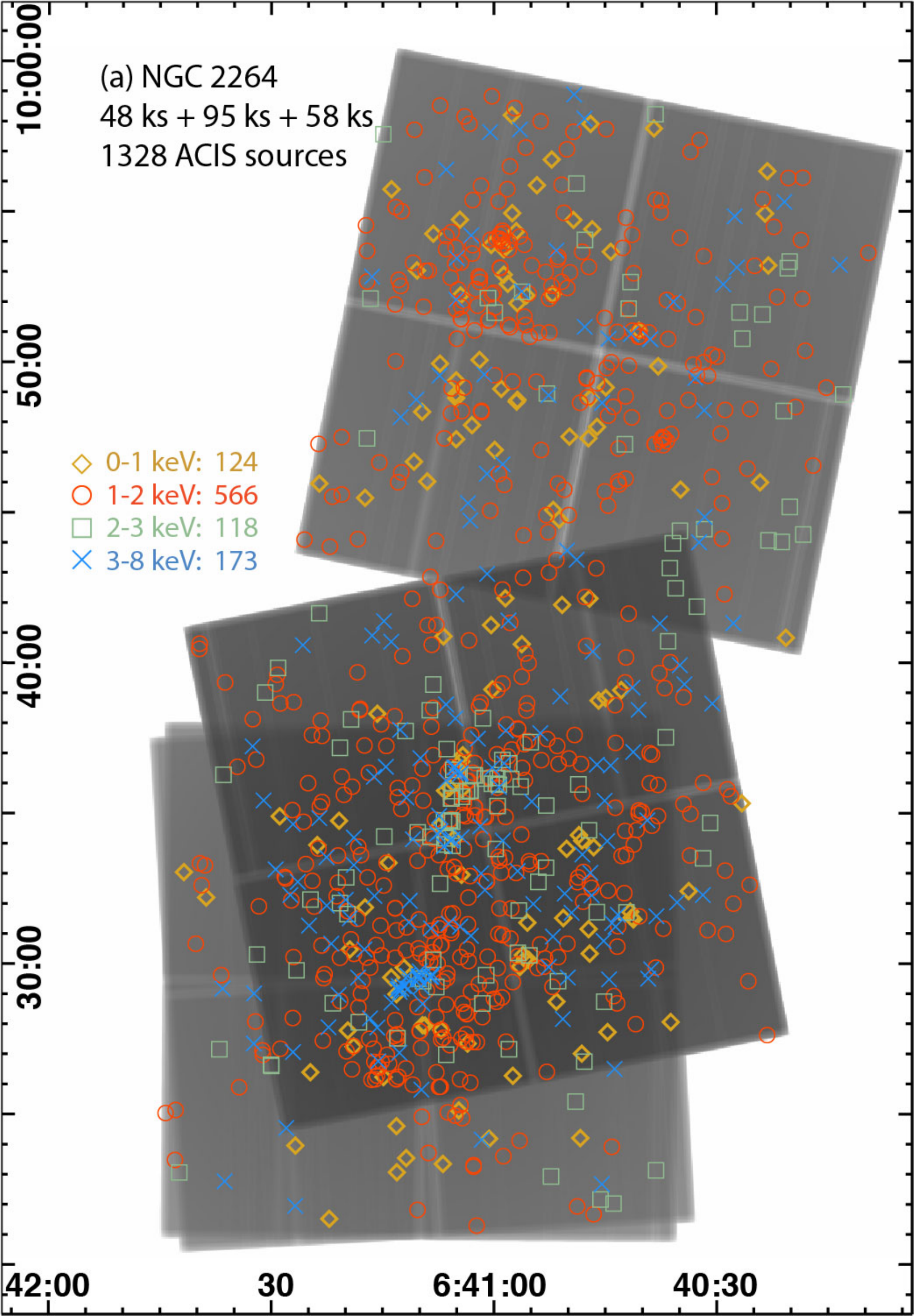}
\includegraphics[width=0.48\textwidth]{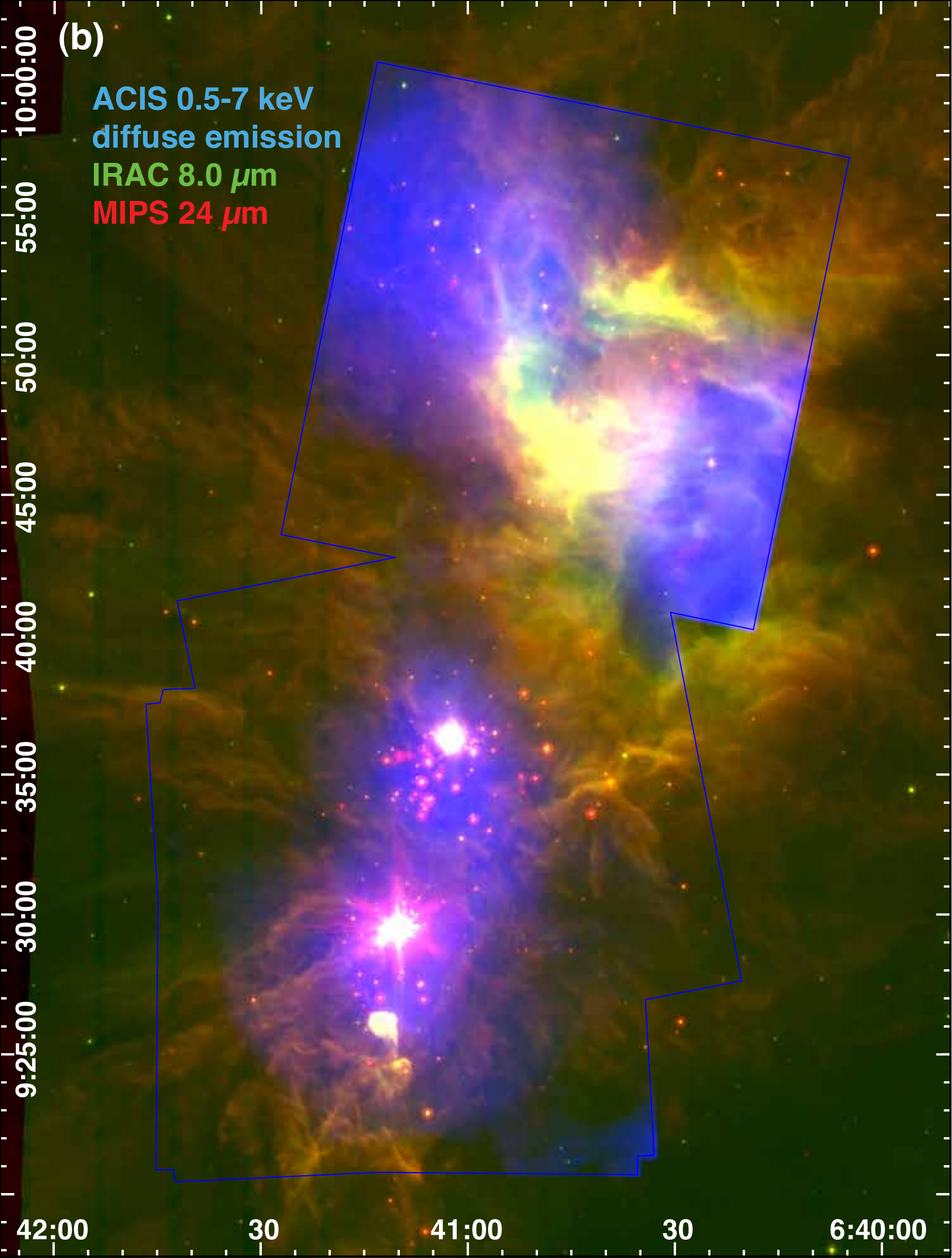}
\caption{NGC~2264.
(a) ACIS exposure map with brighter ($\geq$5 net counts) ACIS point sources overlaid, with symbols and colors denoting median energy for each source.
(b) ACIS diffuse emission (full-band, 0.5--7~keV) in blue, {\em Spitzer}/IRAC 8~$\mu$m emission in green, and {\em Spitzer}/MIPS 24~$\mu$m emission in red.   
\label{ngc2264.fig}}
\end{figure}


\begin{figure}[htb]
\centering
\includegraphics[width=0.48\textwidth]{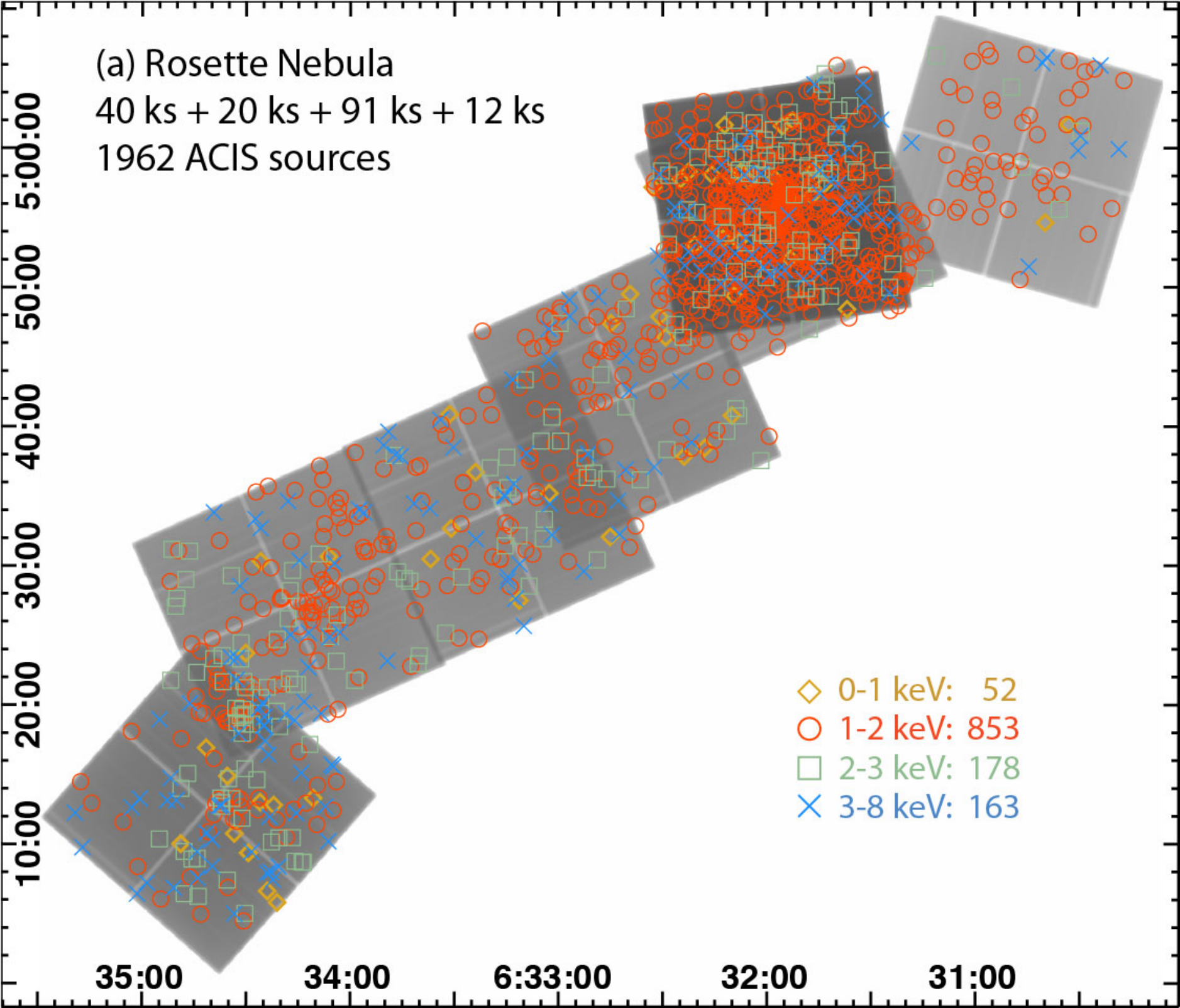}
\includegraphics[width=0.48\textwidth]{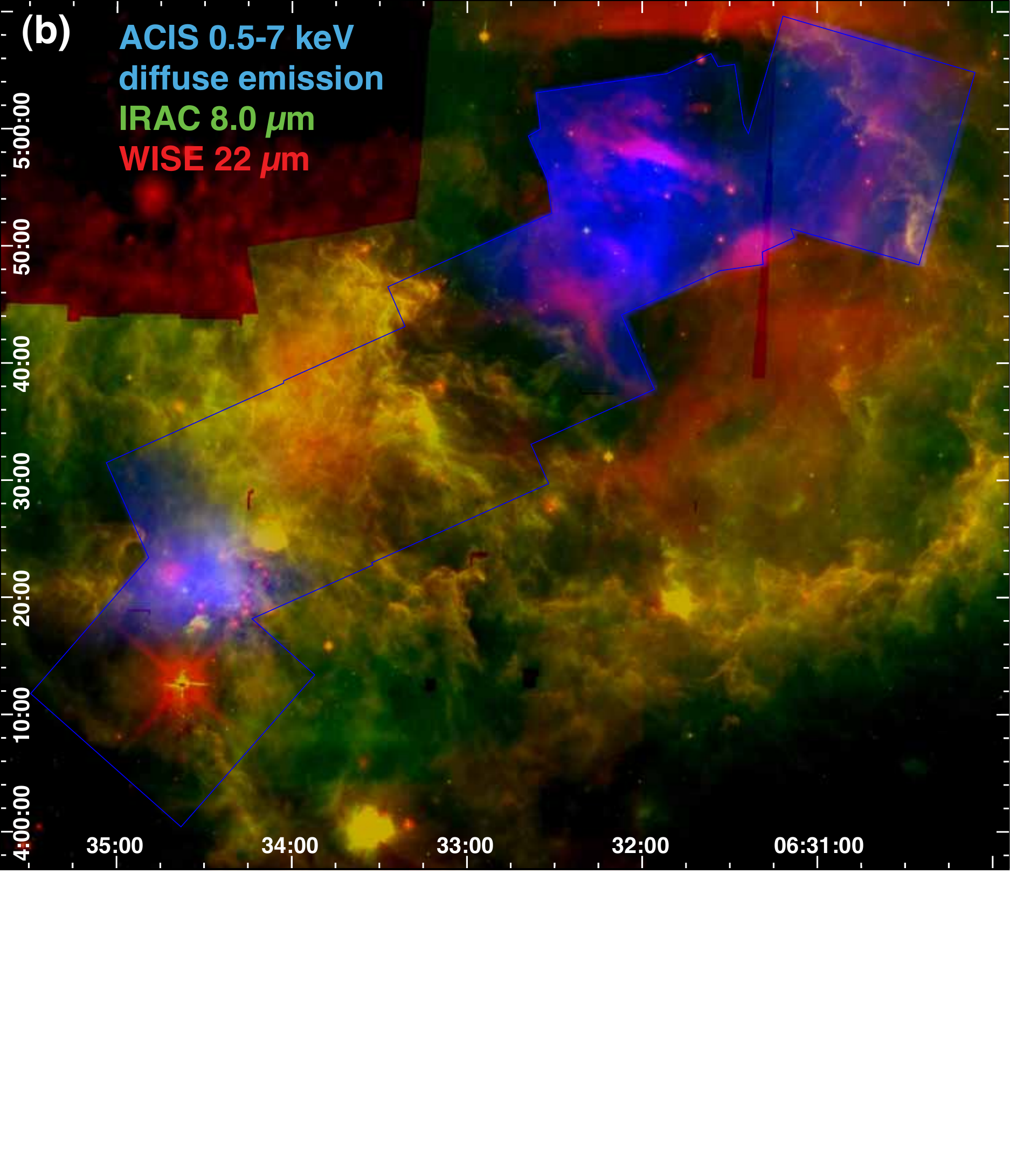}
\caption{The Rosette Nebula (NGC~2244) and Rosette Molecular Cloud.
(a) ACIS exposure map with brighter ($\geq$5 net counts) ACIS point sources overlaid, with symbols and colors denoting median energy for each source.
(b) ACIS diffuse emission (full-band, 0.5--7~keV) in blue, {\em Spitzer}/IRAC 8~$\mu$m emission in green, and {\em WISE} Band 4 (22~$\mu$m) emission in red.   
\label{rosette.fig}}
\end{figure}


\begin{figure}[htb]
\centering
\includegraphics[width=0.48\textwidth]{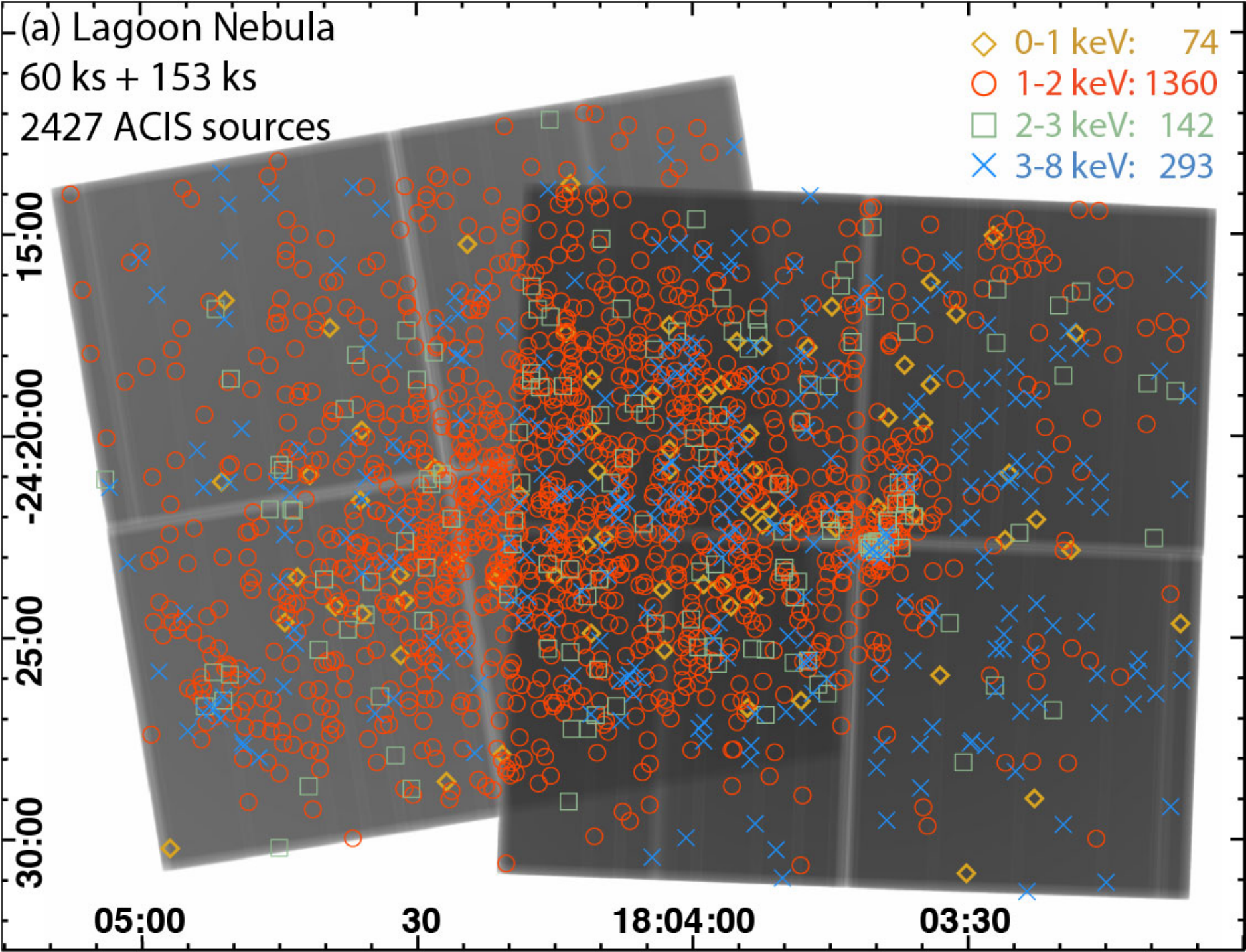}
\includegraphics[width=0.48\textwidth]{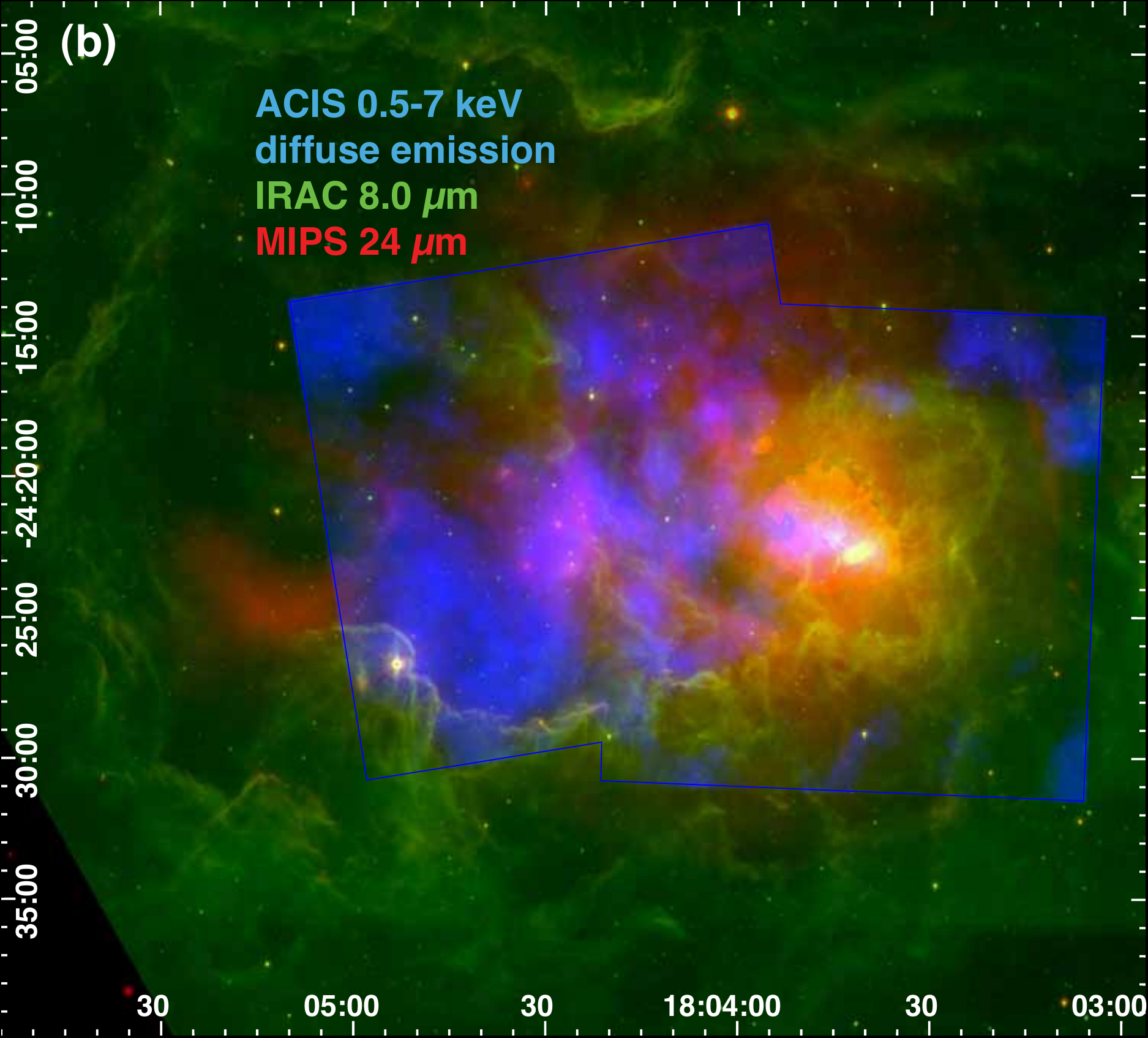}
\caption{M8 (The Lagoon Nebula, NGC~6530).
(a) ACIS exposure map with brighter ($\geq$5 net counts) ACIS point sources overlaid, with symbols and colors denoting median energy for each source.
(b) ACIS diffuse emission (full-band, 0.5--7~keV) in blue, {\em Spitzer}/IRAC 8~$\mu$m emission in green, and {\em Spitzer}/MIPS 24~$\mu$m emission in red.   
\label{lagoon.fig}}
\end{figure}


\begin{figure}[htb]
\centering
\includegraphics[width=0.48\textwidth]{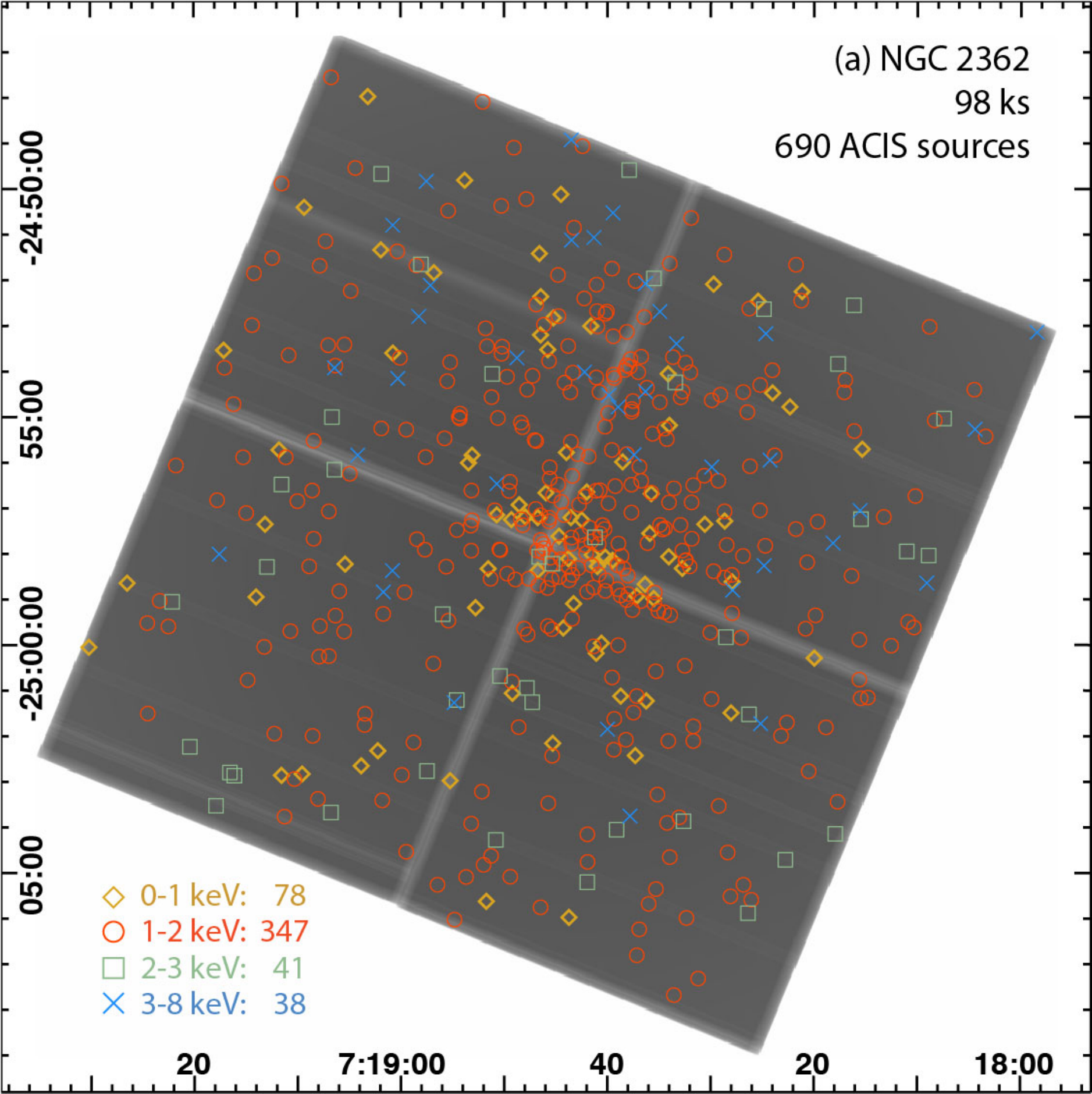}
\includegraphics[width=0.48\textwidth]{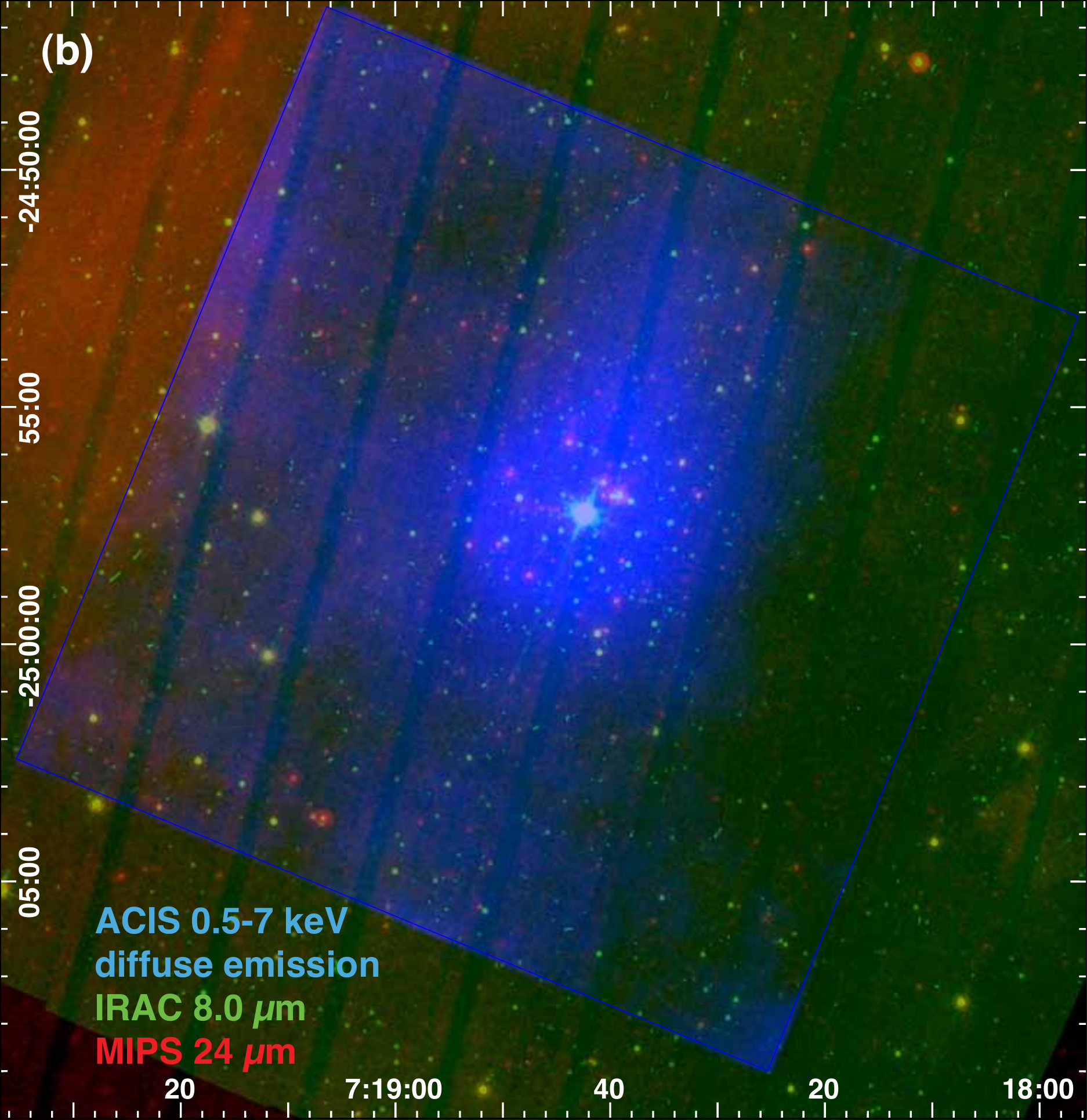}
\caption{NGC~2362.
(a) ACIS exposure map with brighter ($\geq$5 net counts) ACIS point sources overlaid, with symbols and colors denoting median energy for each source.
(b) ACIS diffuse emission (full-band, 0.5--7~keV) in blue, {\em Spitzer}/IRAC 8~$\mu$m emission in green, and {\em Spitzer}/MIPS 24~$\mu$m emission in red.   
\label{ngc2362.fig}}
\end{figure}


\begin{figure}[htb]
\centering
\includegraphics[width=0.48\textwidth]{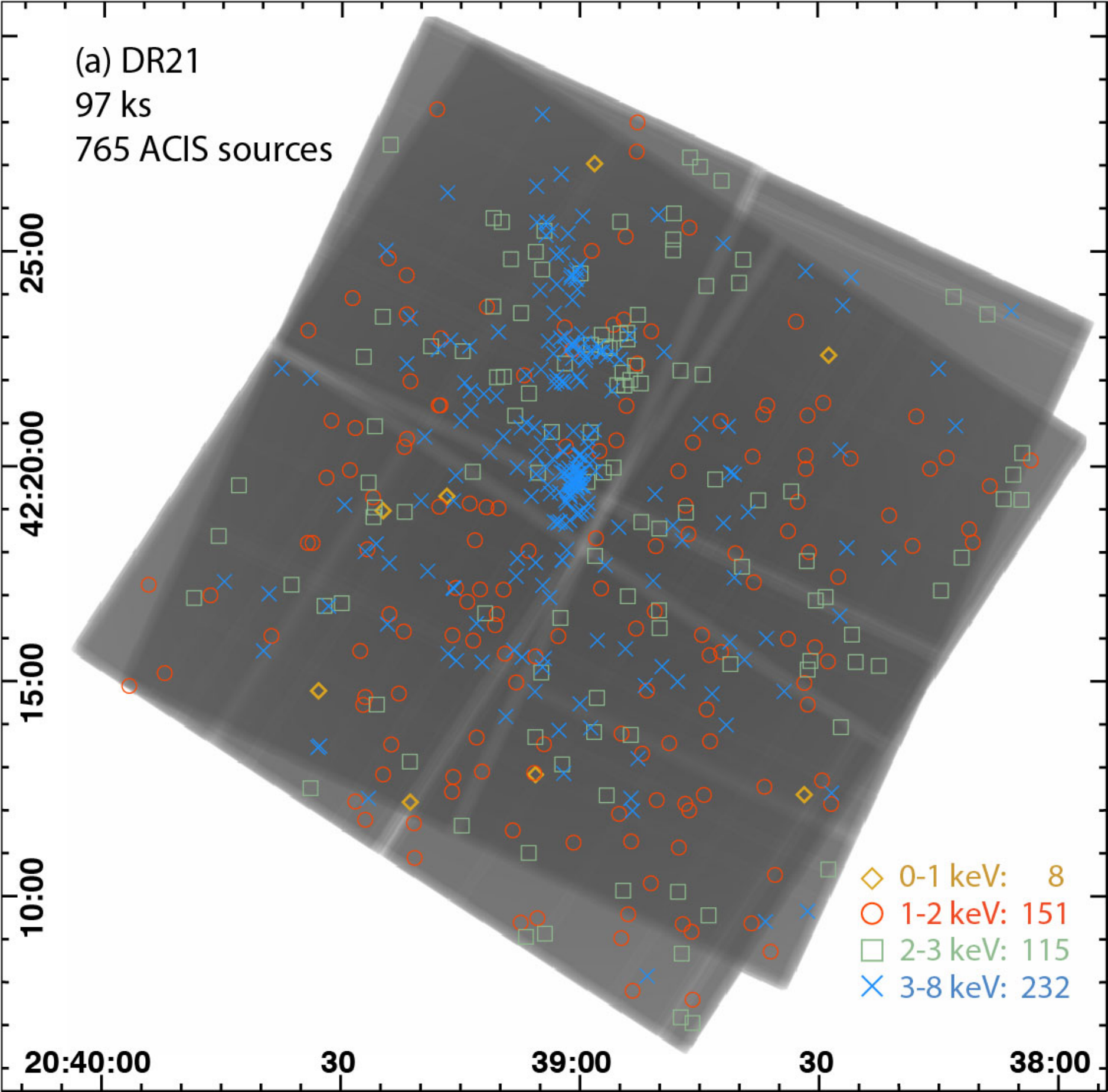}
\includegraphics[width=0.48\textwidth]{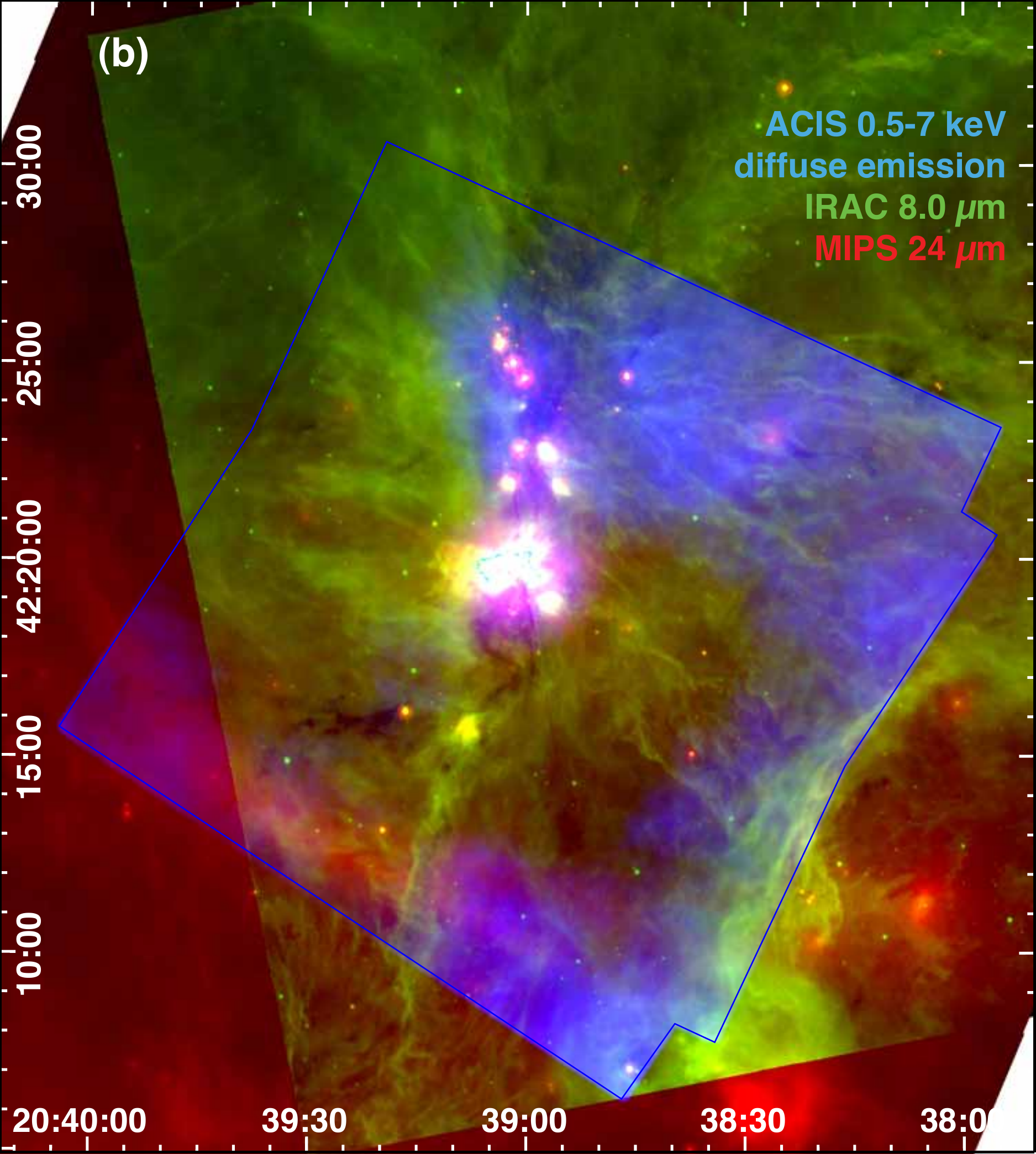}
\caption{DR~21.
(a) ACIS exposure map with brighter ($\geq$5 net counts) ACIS point sources overlaid, with symbols and colors denoting median energy for each source.
(b) ACIS diffuse emission (full-band, 0.5--7~keV) in blue, {\em Spitzer}/IRAC 8~$\mu$m emission in green, and {\em Spitzer}/MIPS 24~$\mu$m emission in red.   
\label{dr21.fig}}
\end{figure}


\begin{figure}[htb]
\centering
\includegraphics[width=0.48\textwidth]{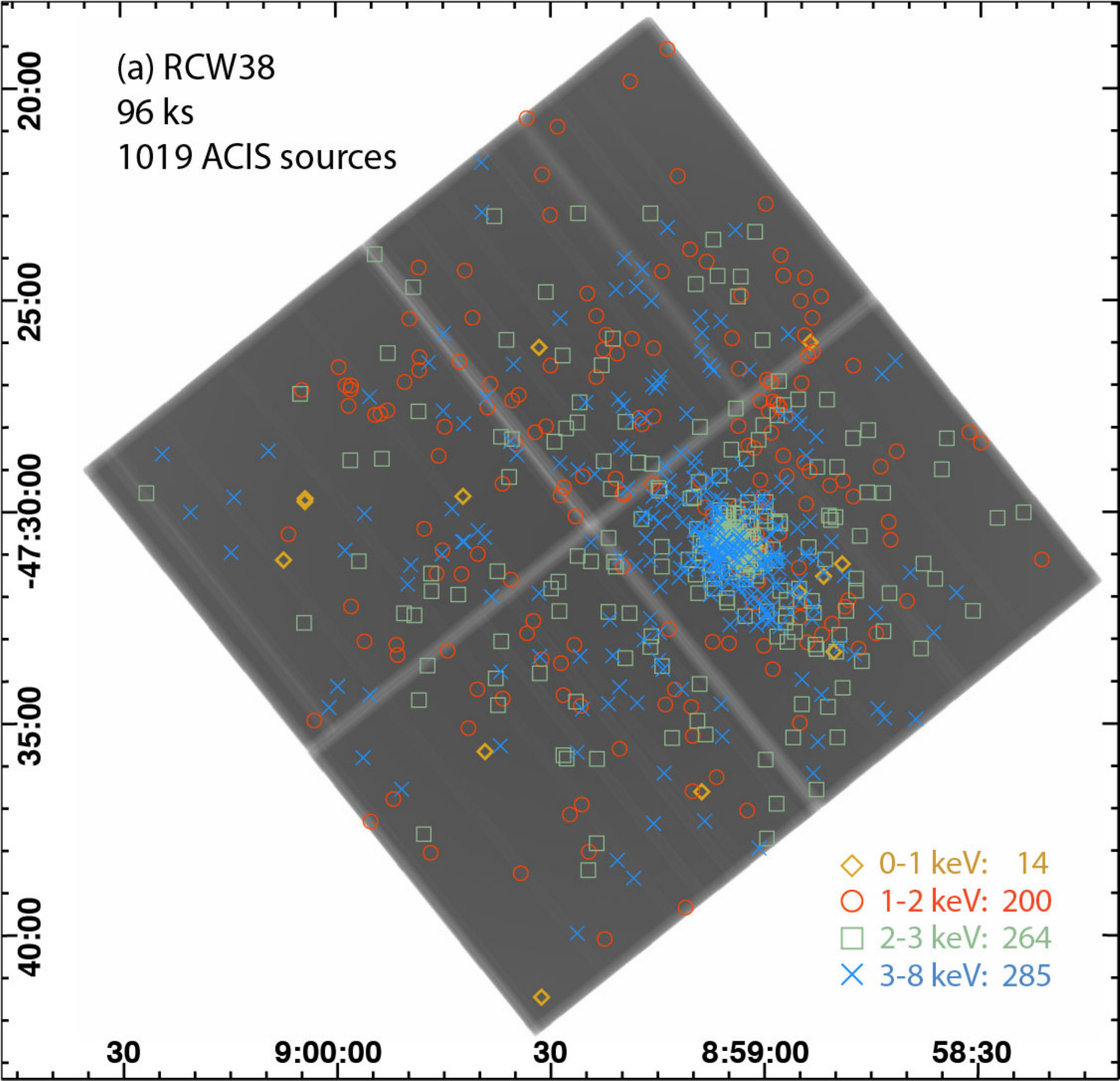}
\includegraphics[width=0.48\textwidth]{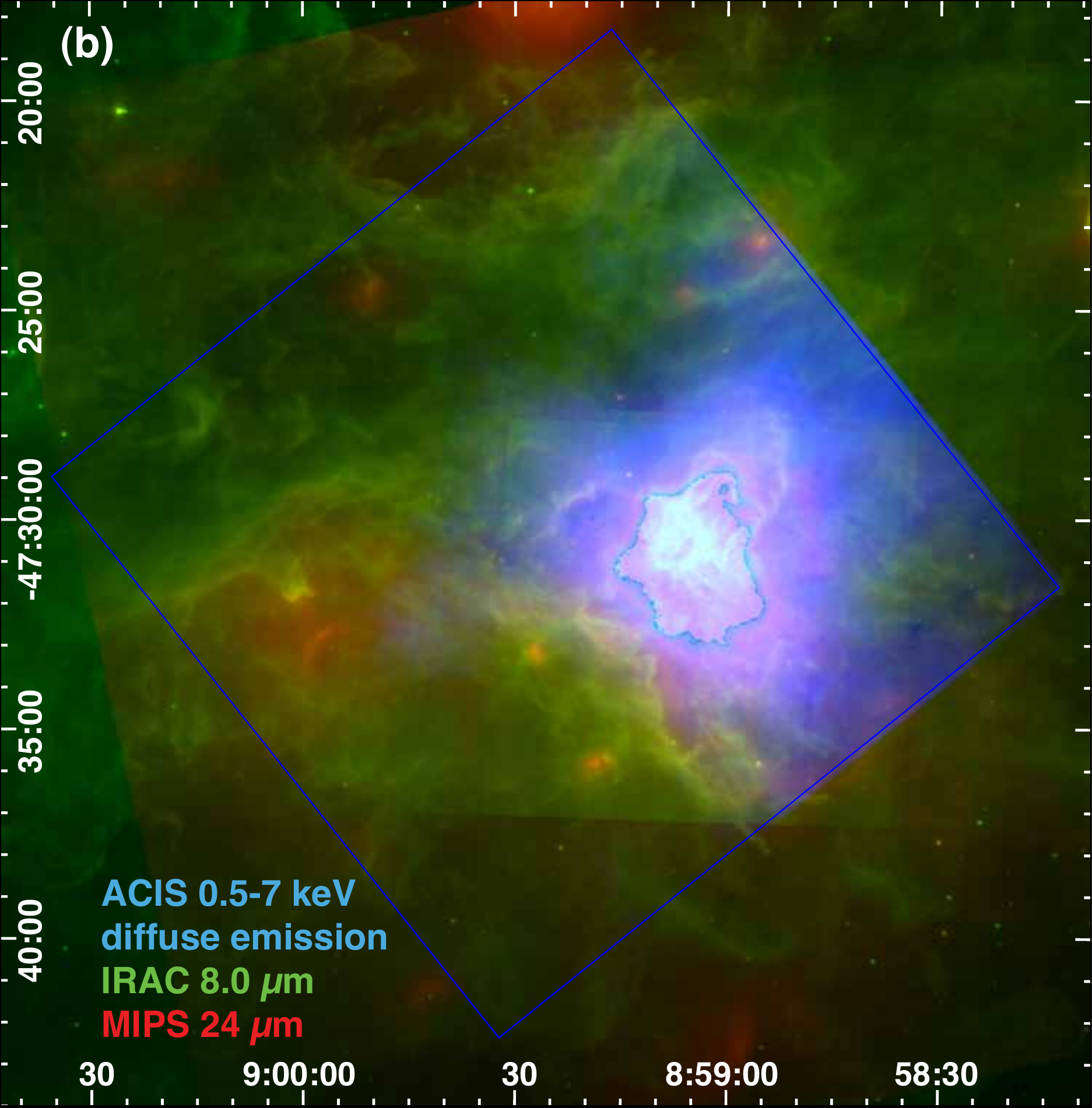}
\caption{RCW~38.
(a) ACIS exposure map with brighter ($\geq$5 net counts) ACIS point sources overlaid, with symbols and colors denoting median energy for each source.
(b) ACIS diffuse emission (full-band, 0.5--7~keV) in blue, {\em Spitzer}/IRAC 8~$\mu$m emission in green, and {\em Spitzer}/MIPS 24~$\mu$m emission in red.   
\label{rcw38.fig}}
\end{figure}


\begin{figure}[htb]
\centering
\includegraphics[width=0.48\textwidth]{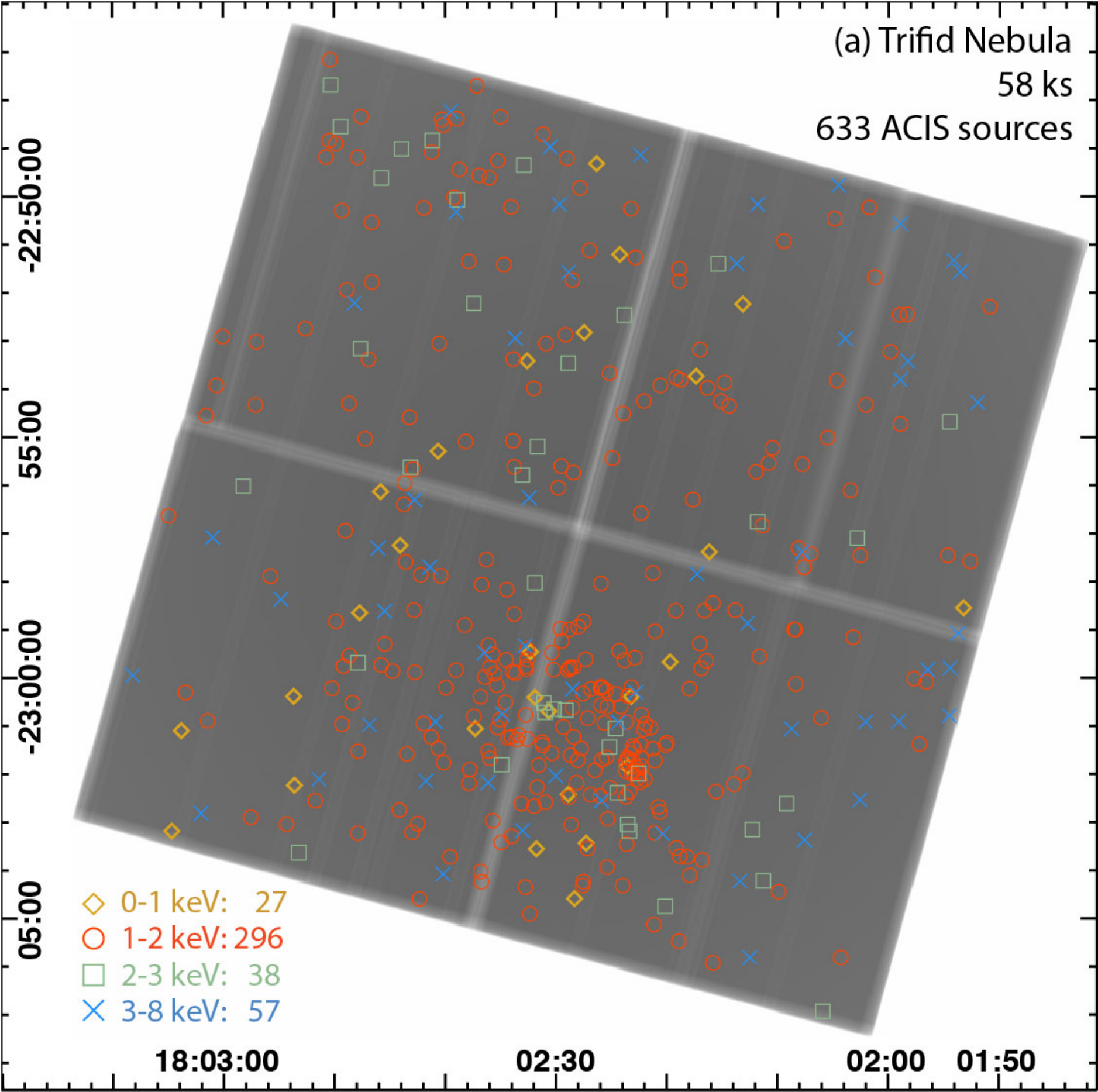}
\includegraphics[width=0.48\textwidth]{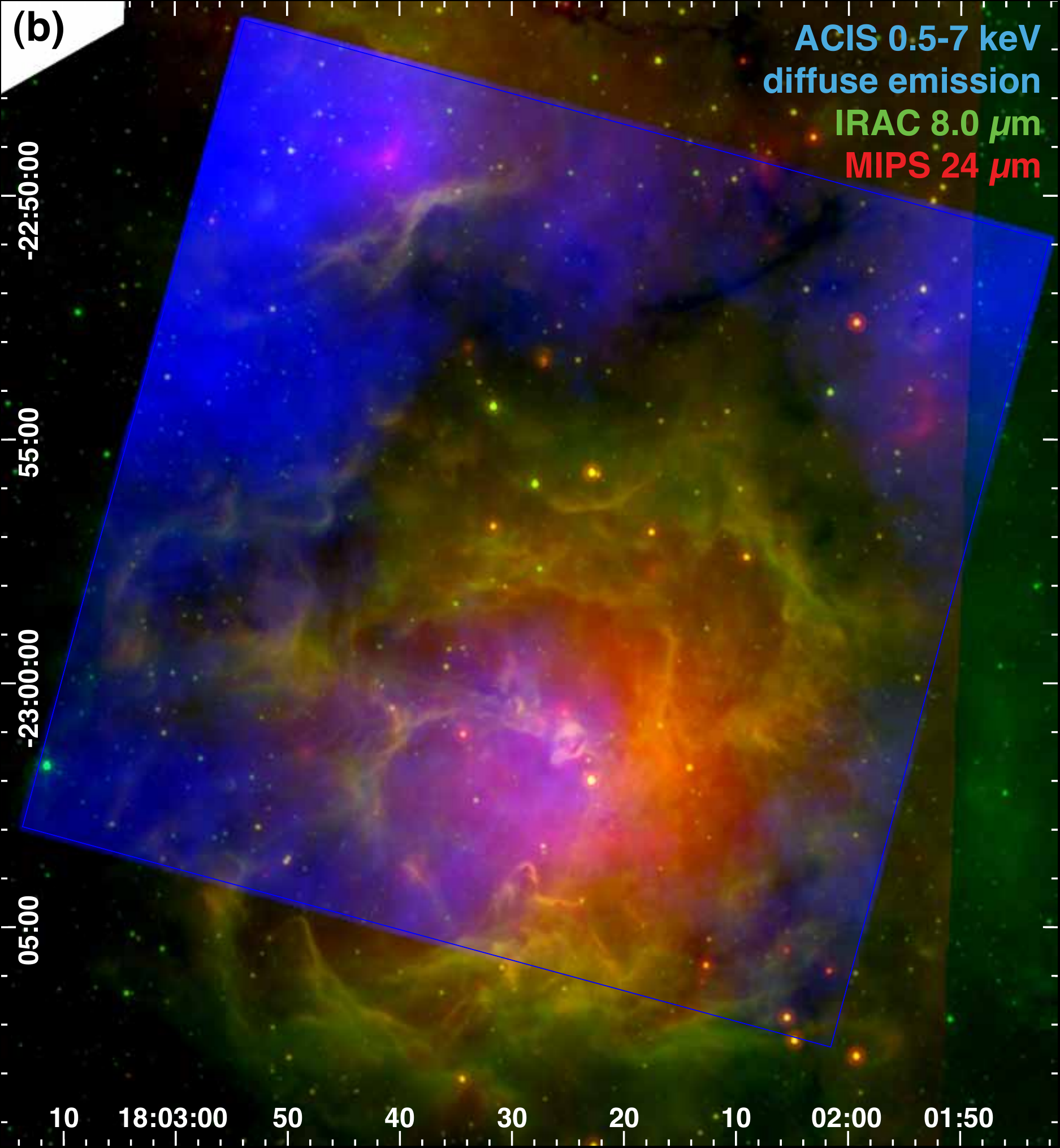}
\caption{M20 (The Trifid Nebula, NGC~6514).
(a) ACIS exposure map with brighter ($\geq$5 net counts) ACIS point sources overlaid, with symbols and colors denoting median energy for each source.
(b) ACIS diffuse emission (full-band, 0.5--7~keV) in blue, {\em Spitzer}/IRAC 8~$\mu$m emission in green, and {\em Spitzer}/MIPS 24~$\mu$m emission in red.   
\label{trifid.fig}}
\end{figure}


\begin{figure}[htb]
\centering
\includegraphics[width=0.48\textwidth]{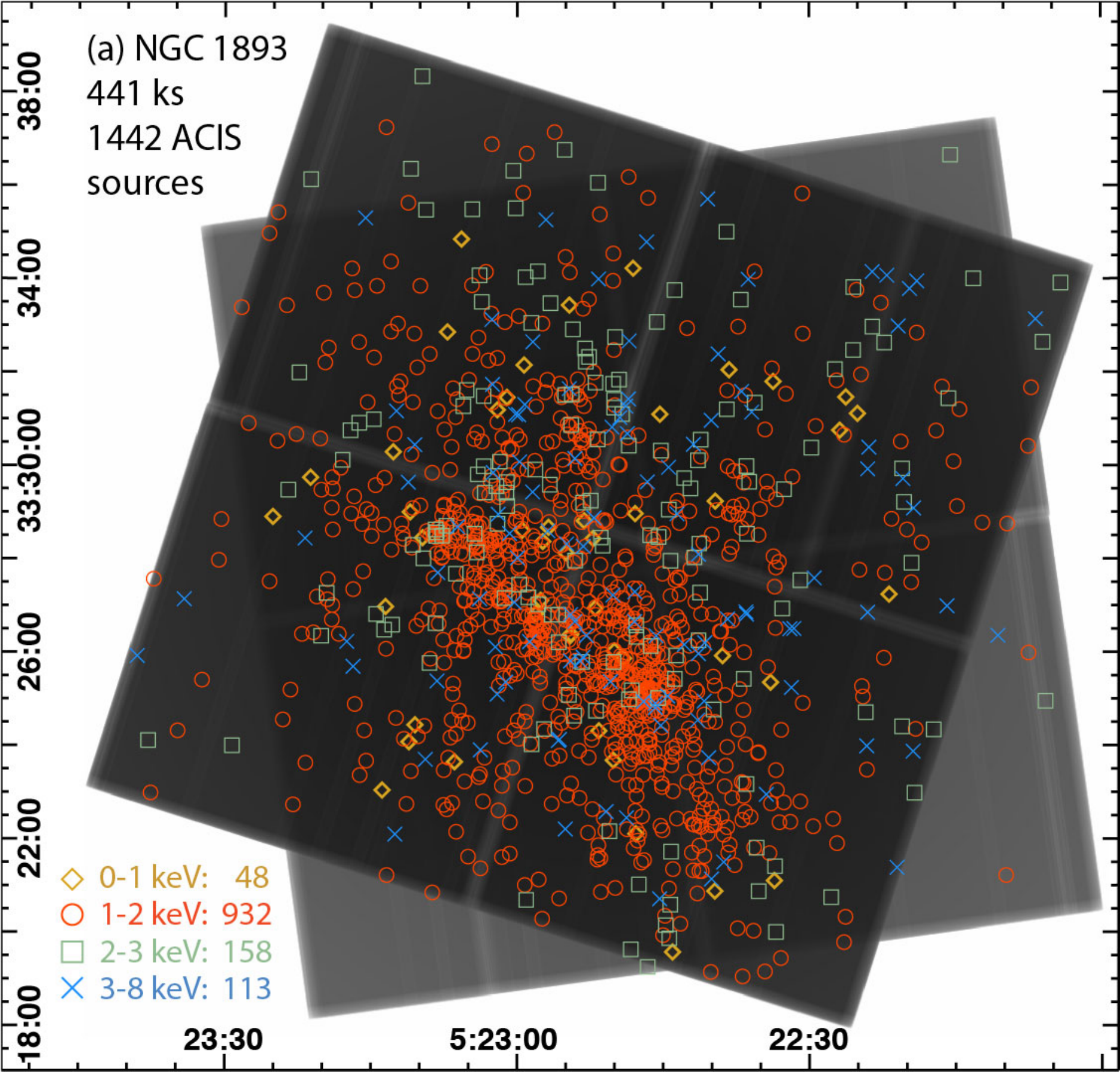}
\includegraphics[width=0.48\textwidth]{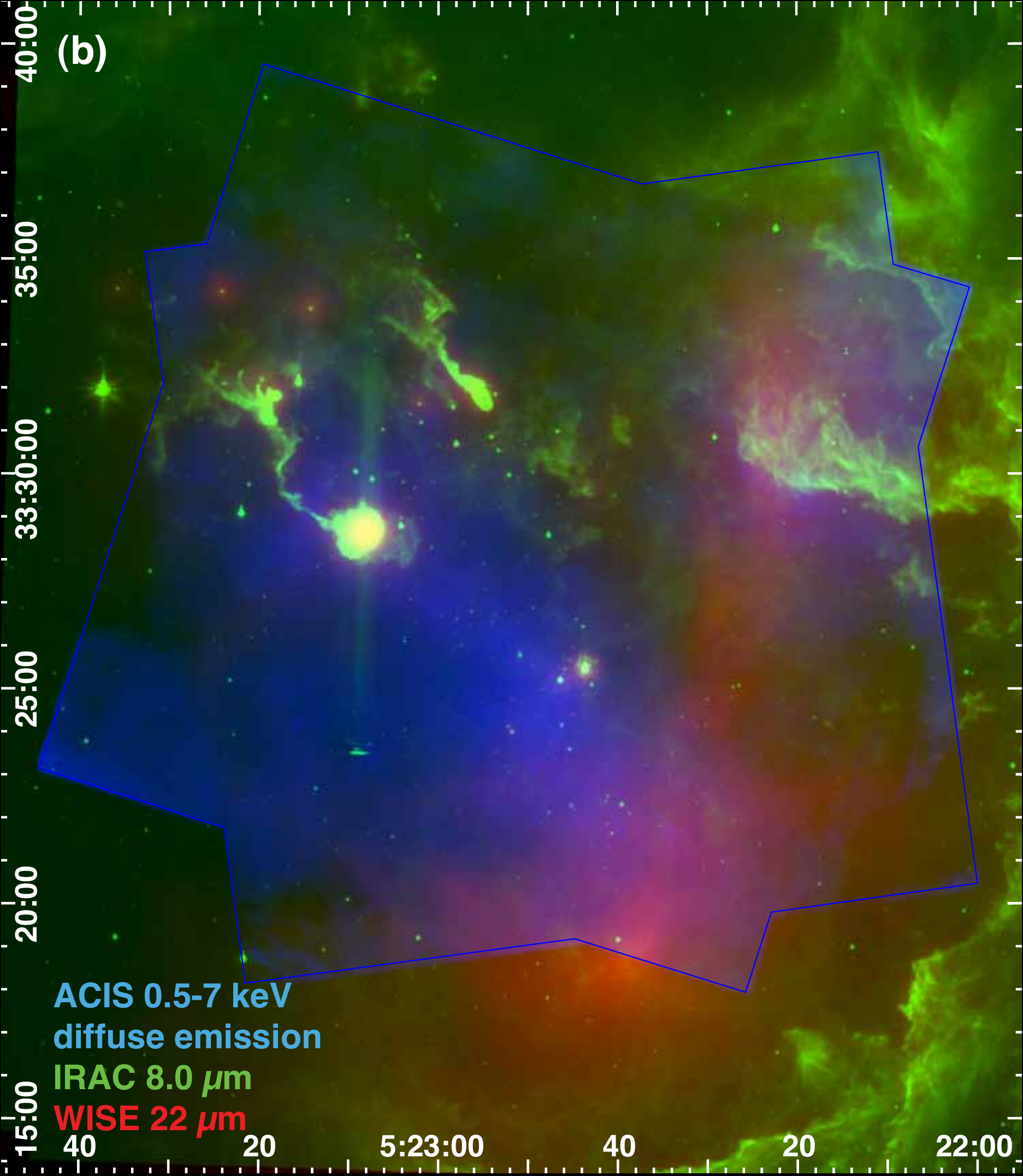}
\caption{NGC~1893.
(a) ACIS exposure map with brighter ($\geq$5 net counts) ACIS point sources overlaid, with symbols and colors denoting median energy for each source.
(b) ACIS diffuse emission (full-band, 0.5--7~keV) in blue, {\em Spitzer}/IRAC 8~$\mu$m emission in green, and {\em WISE} Band 4 (22~$\mu$m) emission in red.   
\label{ngc1893.fig}}
\end{figure}

\clearpage


\acknowledgments
We appreciate the time and effort donated by our anonymous referee to comment on this paper.  
This work was supported by {\em Chandra X-ray Observatory} general observer (GO) grants 
GO4-5007X,   
GO5-6080X,   
GO5-6143X,   
GO6-7006X,   
GO6-7134X,   
GO8-9006X,   
GO9-0003X,   
AR9-0001X,   
GO0-11026X,  
GO1-12013X,  
and GO2-13013X  
(PI:  L.\ Townsley),
GO0-11030X (Science PI:  M.\ Povich), 
and GO7-8025X (Science PI:  J.\ Bouwman), 
by MYStIX project grants NNX09AC74G (NASA; PI:  E.\ Feigelson) and AST-0908038 (NSF; co-PI's:  E.\ Feigelson and L.\ Townsley),
and by the Penn State ACIS Instrument Team Contract SV4-74018, issued by the \Chandra X-ray Center, which is operated by the Smithsonian Astrophysical Observatory for and on behalf of NASA under contract NAS8-03060.
The Guaranteed Time Observations (GTO) included here were selected by the ACIS Instrument Principal Investigator, Gordon P.\ Garmire, of the Huntingdon Institute for X-ray Astronomy, LLC, which is under contract to the Smithsonian Astrophysical Observatory; Contract SV2-82024.
This research used data products from the \Chandra Data Archive, software provided by the \Chandra X-ray Center (CXC) in the application package {\it CIAO}, and {\it SAOImage DS9} software developed by the Smithsonian Astrophysical Observatory.  This research also used data products from the {\em Spitzer Space Telescope}, operated by the Jet Propulsion Laboratory (California Institute of Technology) (JPL/CalTech) under a contract with NASA, and data products from the {\em Wide-field Infrared Survey Explorer} ({\em WISE}), which is a joint project of the University of California, Los Angeles and JPL/CalTech, funded by NASA.  This research used NASA's Astrophysics Data System Bibliographic Services, and the SIMBAD database and VizieR catalog access tool provided by CDS, Strasbourg, France.

{\em Facilities:} \facility{CXO (ACIS)}, \facility{Spitzer (IRAC, MIPS)}, \facility{WISE}.


\end{document}

%% file: target_table.tex
\noindent
\setlength{\tabcolsep}{0.05in}
\begin{deluxetable}{lcccccrccrl}
\tablecaption{MOXC Targets \label{targets.tbl}}
\tabletypesize{\tiny}

\tablehead{
\colhead{MOXC \rule{0mm}{3ex}} & 
\colhead{Galactic} &  
\colhead{Celestial J2000} & 
\colhead{Distance\tablenotemark{a}} & 
\colhead{Scale\tablenotemark{b}} & 
\colhead{$\langle A_V \rangle$\tablenotemark{c}} & 
\colhead{Nominal} & 
\colhead{$\log L_{tc}$} & 
\colhead{$M_{50\%}$}   & 
\colhead{X-ray} &
\colhead{Distance} \\
\colhead{Target}  & 
\colhead{($l,b$)}  &                                                                         
\colhead{($RA,Dec$)} & 
\colhead{(kpc)}    & 
\colhead{(arcmin/pc)} & 
\colhead{(mag)} & 
\colhead{Exp (ks)} &
\colhead{(erg/s)} & 
\colhead{($M_{\sun}$)} & 
\colhead{Srcs (\#)} &
\colhead{Reference\tablenotemark{a}} \\
\numberthecolumn & \numberthecolumn & \numberthecolumn & \numberthecolumn & \numberthecolumn & \numberthecolumn & \numberthecolumn & \numberthecolumn & \numberthecolumn & \numberthecolumn & \numberthecolumn
\setcounter{column_number}{1}
}
\startdata
\multicolumn{2}{l}{\bf \mystix} \\
 \object{NGC 6334   }&       (351.42,\phn$+$0.64) & 17 20 54.0  $-$35 47 04 &\phn1.7\phn\phn     & 2.02 &\phn4.1      &  40 & 29.83 & 0.4 & 1510    &  \\
 \object{NGC 6357   }&       (353.11,\phn$+$0.65) & 17 25 34.2  $-$34 23 12 &\phn1.7\phn\phn     & 2.02 &\phn5\phd\phn&  40 & 29.93 & 0.5 & 3108    &  \\
 \object{M16        }&    (\phn16.95,\phn$+$0.81) & 18 18 44.7  $-$13 47 56 &\phn1.75\phn        & 1.96 &\phn2.6      &  79 & 29.48 & 0.3 & 2830    &  \\ 
 \object{M17        }&    (\phn15.06,\phn$-$0.69) & 18 20 29.9  $-$16 10 45 &\phn2.0\phn\phn     & 1.72 &\phn8\phd\phn& 325 & 29.27 & 0.2 & 2999    &  \\
 \object{W3         }&       (133.95,\phn$+$1.06) & 02 27 04.1  $+$61 52 22 &\phn2.04\phn        & 1.69 & 20\phd\phn  &  78 & 30.21 & 0.8 & 2094    &  \\ 
 \object{W4         }&       (134.72,\phn$+$0.92) & 02 32 42.5  $+$61 27 22 &\phn2.04\phn        & 1.69 &\phn2.5      &  78 & 29.64 & 0.3 &  647    &  \\ 
 \object{NGC 3576   }&       (291.22,\phn$-$0.50) & 11 12 04.5  $-$61 05 43 &\phn2.8\phn\phn     & 1.23 &\phn8\phd\phn&  52 & 30.33 & 0.9 & 1522    &  \\ 
\hline
\multicolumn{2}{l}{\bf Beyond \mystix\ \rule{0mm}{3ex}} \\
 \object{G333.6-0.2 }&       (333.60,\phn$-$0.21) & 16 22 09.2  $-$50 06 04 &\phn2.6\phn\phn     & 1.32 & 15\phd\phn  &  60 & 30.41 & 1.0    &  653 & \citet{Figueredo05} \\
 \object{W51A       }&    (\phn49.49,\phn$-$0.37) & 19 23 40.0  $+$14 31 06 &\phn5.1\phn\phn     & 0.67 & 20\phd\phn  &  70 & 31.04 & bright &  641 & \citet{Xu09} \\
 \object{G29.96-0.02}&    (\phn29.96,\phn$-$0.02) & 18 46 04.0  $-$02 39 20 &\phn6.2\phn\phn     & 0.55 & 21\phd\phn  &  27 & 31.63 & bright &  172 & \citet{Russeil11} \\
 \object{NGC 3603   }&       (291.62,\phn$-$0.52) & 11 15 07.2  $-$61 15 35 &\phn7.0\phn\phn     & 0.49 &\phn4.5      & 490 & 30.05 & 0.6    & 3885 & \citet{Harayama08} \\
 \object{30 Doradus }&      (279.46,\phn$-$31.67) & 05 38 42.4  $-$69 06 02 & 50.0\phn\phn       & 0.07 &\phn0.5      &  92 & 32.18 & bright &  562 & \citet{Pietrzynski13} \\
\enddata
\tablenotetext{a}{Distances for MYStIX targets are taken from \citet{Feigelson13}; primary references are given there.}
\tablenotetext{b}{Image scale assuming the distance given in Col.\ (4).}
\tablenotetext{c}{Approximate average absorption to the target, estimated from a variety of literature sources.  Most MSFRs have highly variable and spatially complex obscuration, so this value 
should be used only as a rough indicator.}
\end{deluxetable} 


%
%
%
%
%
%
%
%

%% file: observing_log.tex
\begin{deluxetable}{crrcrccrcll}
\centering 
\tabletypesize{\tiny} \tablewidth{0pt}

\tablecaption{ Log of {\em Chandra} ACIS-I Observations 
 \label{tbl:obslog}}

\tablehead{
\colhead{Target} & 
\colhead{ObsID} & 
\colhead{Sequence} & 
\colhead{Start Time} & 
\colhead{Exposure\tablenotemark{a}} & 
\multicolumn{2}{c}{Aimpoint\tablenotemark{b}} & 
\colhead{Roll} & 
\colhead{ACIS Mode\tablenotemark{c}} &  
\colhead{PI}  &
\colhead{CALDB\tablenotemark{d}}  \\
\cline{6-7}

\colhead{} & 
\colhead{} & 
\colhead{} & 
\colhead{(UT)} & 
\colhead{(s)} & 
\colhead{$\alpha_{\rm J2000}$} & 
\colhead{$\delta_{\rm J2000}$} & 
\colhead{(\arcdeg)} & 
\colhead{} \\
\numberthecolumn &\numberthecolumn & \numberthecolumn & \numberthecolumn & \numberthecolumn & \numberthecolumn & \numberthecolumn & \numberthecolumn & \numberthecolumn & \numberthecolumn & \numberthecolumn 
\setcounter{column_number}{1}
}

\startdata
  \multicolumn{1}{l}{\bf NGC 6334} \\ 
        NGC 6334 REGION 1 & \dataset[ADS/Sa.CXO#obs/02574]{  2574} &  200182 & 2002-08-31T12:49 &   39648 & 17:20:54.00 & -35:47:03.9 &  269 &     TIMED FAINT & Yuichiro Ezoe  & 3.2.3 \\ %
        NGC 6334 REGION 2 & \dataset[ADS/Sa.CXO#obs/02573]{  2573} &  200181 & 2002-09-02T00:12 &   23835 & 17:20:01.00 & -35:56:07.0 &  268 &     TIMED FAINT & Yuichiro Ezoe  & 3.2.3 \\ %
          IGR J17204-3554 & \dataset[ADS/Sa.CXO#obs/08975]{  8975} &  400826 & 2009-01-26T23:19 &    1015 & 17:20:25.00 & -35:53:31.2 &   98 &     TIMED FAINT & Mariano Mendez & 3.5.1  \\ %
\\
  \multicolumn{1}{l}{\bf NGC 6357} \\ 
         NGC 6357 Field I & \dataset[ADS/Sa.CXO#obs/04477]{  4477} &  200246 & 2004-07-09T13:46 &   37689 & 17:24:43.39 & -34:11:56.0 &  288 &    TIMED VFAINT & Gordon Garmire & 3.2.1 \\ %
               G353.2+0.7 & \dataset[ADS/Sa.CXO#obs/10988]{ 10988} &  200619 & 2010-05-07T06:54 &   39651 & 17:26:01.69 & -34:15:15.4 &   72 &    TIMED VFAINT & Leisa Townsley & 4.1.7 \\ %
               G353.1+0.6 & \dataset[ADS/Sa.CXO#obs/10987]{ 10987} &  200618 & 2010-07-17T21:55 &   40526 & 17:25:34.19 & -34:23:11.6 &  283 &    TIMED VFAINT & Leisa Townsley & 4.3.0 \\ %
             G353.08+0.36 & \dataset[ADS/Sa.CXO#obs/13267]{ 13267} &  200737 & 2012-07-06T09:15 &   56227 & 17:26:38.50 & -34:34:23.8 &  293 &    TIMED VFAINT & Gordon Garmire & 4.5.0 \\ %
             G352.90+1.02 & \dataset[ADS/Sa.CXO#obs/13622]{ 13622} &  200776 & 2013-01-29T04:30 &   39458 & 17:23:30.70 & -34:21:10.5 &   96 &    TIMED VFAINT & Leisa Townsley & 4.5.5 \\ %
             G353.08+1.24 & \dataset[ADS/Sa.CXO#obs/13623]{ 13623} &  200777 & 2013-01-31T12:19 &   39458 & 17:23:06.90 & -34:04:36.5 &   95 &    TIMED VFAINT & Leisa Townsley & 4.5.5 \\ %
\\
  \multicolumn{1}{l}{\bf Eagle Nebula (M16)} \\ 
                     M 16 & \dataset[ADS/Sa.CXO#obs/00978]{   978} &  200085 & 2001-07-30T18:55 &   77126 & 18:18:44.71 & -13:47:56.5 &  257 &    TIMED VFAINT & Jeffrey Linsky  & 3.2.4 \\ %
                 NGC 6611 & \dataset[ADS/Sa.CXO#obs/08931]{  8931} &  200512 & 2008-05-28T20:52 &   79136 & 18:19:12.00 & -13:33:00.0 &  110 &     TIMED FAINT & Mario Guarcello & 3.4.4 \\ %
                 NGC 6611 & \dataset[ADS/Sa.CXO#obs/08932]{  8932} &  200513 & 2008-06-02T16:36 &   30154 & 18:19:35.99 & -13:47:23.9 &  116 &     TIMED FAINT & Mario Guarcello & 3.4.4 \\ %
                 NGC 6611 & \dataset[ADS/Sa.CXO#obs/09865]{  9865} &  200513 & 2008-06-04T11:13 &   16502 & 18:19:35.99 & -13:47:23.9 &  116 &     TIMED FAINT & Mario Guarcello & 3.4.4 \\ %
                 NGC 6611 & \dataset[ADS/Sa.CXO#obs/09864]{  9864} &  200513 & 2008-06-07T23:15 &   23802 & 18:19:35.99 & -13:47:23.9 &  116 &     TIMED FAINT & Mario Guarcello & 3.4.4 \\ %
                 NGC 6611 & \dataset[ADS/Sa.CXO#obs/09872]{  9872} &  200513 & 2008-06-09T10:53 &    9065 & 18:19:35.99 & -13:47:23.9 &  116 &     TIMED FAINT & Mario Guarcello & 3.4.4 \\ %
\\
  \multicolumn{1}{l}{\bf M17} \\ 
                      M17 & \dataset[ADS/Sa.CXO#obs/00972]{   972} &  200079 & 2002-03-02T17:05 &   39436 & 18:20:29.89 & -16:10:45.5 &   89 &     TIMED FAINT & Gordon Garmire & 3.2.3 \\ %
           M17 Pointing I & \dataset[ADS/Sa.CXO#obs/06420]{  6420} &  200395 & 2006-08-01T02:06 &  149439 & 18:20:29.89 & -16:10:44.9 &  261 &    TIMED VFAINT & Leisa Townsley & 4.4.9 \\ %
           M17 Pointing I & \dataset[ADS/Sa.CXO#obs/06421]{  6421} &  200395 & 2006-08-07T00:15 &   39435 & 18:20:29.89 & -16:10:44.9 &  261 &    TIMED VFAINT & Leisa Townsley & 4.4.9 \\ %
          M17 Pointing II & \dataset[ADS/Sa.CXO#obs/06422]{  6422} &  200396 & 2006-08-21T01:20 &   25530 & 18:21:33.49 & -16:11:55.8 &  267 &    TIMED VFAINT & Leisa Townsley & 4.4.9 \\ %
          M17 Pointing II & \dataset[ADS/Sa.CXO#obs/07391]{  7391} &  200396 & 2006-08-27T12:10 &   59090 & 18:21:33.49 & -16:11:55.8 &  267 &    TIMED VFAINT & Leisa Townsley & 4.4.9 \\ %
                      M17 & \dataset[ADS/Sa.CXO#obs/06403]{  6403} &  200380 & 2006-11-06T22:22 &   34716 & 18:20:29.89 & -16:10:44.9 &  278 &    TIMED VFAINT & Gordon Garmire & 4.4.9 \\ %
                      M17 & \dataset[ADS/Sa.CXO#obs/08460]{  8460} &  200380 & 2006-11-08T04:20 &   29762 & 18:20:29.89 & -16:10:44.9 &  278 &    TIMED VFAINT & Gordon Garmire & 4.4.9 \\ %
                      M17 & \dataset[ADS/Sa.CXO#obs/08461]{  8461} &  200380 & 2006-11-11T00:13 &   32550 & 18:20:29.89 & -16:10:44.9 &  278 &    TIMED VFAINT & Gordon Garmire & 4.4.9 \\ %
               NGC 6618PG & \dataset[ADS/Sa.CXO#obs/10993]{ 10993} &  200624 & 2010-05-27T02:14 &   39540 & 18:21:11.29 & -15:58:53.0 &  104 &    TIMED VFAINT & Matthew Povich & 4.1.7 \\ %
\\
  \multicolumn{1}{l}{\bf W3} \\ 
                      W3B & \dataset[ADS/Sa.CXO#obs/00611]{   611} &  200036 & 2000-03-23T11:59 &   18467 & 02:25:38.09 & +62:05:52.5 &  323 &     TIMED FAINT & Edward Churchwell & 3.3.0 \\ %
                      W3B & \dataset[ADS/Sa.CXO#obs/00446]{   446} &  200036 & 2000-04-03T02:57 &   20062 & 02:25:38.09 & +62:05:52.5 &  332 &     TIMED FAINT & Edward Churchwell & 3.3.0 \\ %
             W3 Main IRS5 & \dataset[ADS/Sa.CXO#obs/05890]{  5890} &  900390 & 2005-01-04T08:52 &   39632 & 02:25:40.60 & +62:05:52.4 &  262 &    TIMED VFAINT & Leisa Townsley    & 3.2.1 \\ %
                   W3(OH) & \dataset[ADS/Sa.CXO#obs/05889]{  5889} &  900389 & 2005-04-23T03:34 &   71243 & 02:27:04.09 & +61:52:22.0 &  352 &    TIMED VFAINT & Leisa Townsley    & 3.2.1 \\ %
                 W3 North & \dataset[ADS/Sa.CXO#obs/06335]{  6335} &  900391 & 2005-07-02T01:37 &   20997 & 02:26:50.80 & +62:15:51.9 &   78 &    TIMED VFAINT & Leisa Townsley    & 3.2.1 \\ %
                 W3 North & \dataset[ADS/Sa.CXO#obs/05891]{  5891} &  900391 & 2005-07-17T14:52 &   46064 & 02:26:50.80 & +62:15:51.9 &   91 &    TIMED VFAINT & Leisa Townsley    & 3.2.1 \\ %
                 W3 North & \dataset[ADS/Sa.CXO#obs/06348]{  6348} &  900391 & 2005-11-21T19:57 &   12002 & 02:26:50.80 & +62:15:51.9 &  208 &    TIMED VFAINT & Leisa Townsley    & 4.4.8 \\ %
                  IC 1795 & \dataset[ADS/Sa.CXO#obs/07356]{  7356} &  200419 & 2007-12-04T13:17 &   49411 & 02:26:34.39 & +62:00:42.9 &  225 &    TIMED VFAINT & Jeroen Bouwman    & 3.4.1 \\ %
\\
  \multicolumn{1}{l}{\bf W4} \\ 
                  IC 1805 & \dataset[ADS/Sa.CXO#obs/07033]{  7033} &  900449 & 2006-11-25T23:09 &   78034 & 02:32:42.49 & +61:27:21.6 &  212 &    TIMED VFAINT & Leisa Townsley & 3.2.4 \\ %
\\                                                                                                                                        
  \multicolumn{1}{l}{\bf NGC 3576} \\ 
                 NGC 3576 & \dataset[ADS/Sa.CXO#obs/04496]{  4496} &  200265 & 2005-07-21T14:04 &   40109 & 11:11:53.80 & -61:18:24.9 &  222 &    TIMED VFAINT & Leisa Townsley & 3.2.1 \\ %
                 NGC 3576 & \dataset[ADS/Sa.CXO#obs/06349]{  6349} &  200265 & 2005-07-23T05:59 &   11660 & 11:11:53.80 & -61:18:24.9 &  222 &    TIMED VFAINT & Leisa Townsley & 3.2.1 \\ %
                 HD 97484 & \dataset[ADS/Sa.CXO#obs/08905]{  8905} &  200486 & 2007-11-10T13:36 &   58390 & 11:12:04.50 & -61:05:43.0 &  111 &    TIMED VFAINT & Leisa Townsley & 3.4.1 \\ %
\\
  \multicolumn{1}{l}{\bf G333.6-0.2 } \\ 
               G333.6-0.2 & \dataset[ADS/Sa.CXO#obs/09911]{  9911} &  200528 & 2009-06-14T12:19 &   60096 & 16:22:09.19 & -50:06:03.4 &  327 &    TIMED VFAINT & Leisa Townsley & 3.5.3 \\ %
\\
  \multicolumn{1}{l}{\bf W51A       } \\ 
                      W51 & \dataset[ADS/Sa.CXO#obs/02524]{  2524} &  200132 & 2002-06-17T03:38 &   50998 & 19:23:40.00 & +14:31:05.9 &  147 &     TIMED FAINT & Gordon Garmire & 3.2.3 \\ %
                      W51 & \dataset[ADS/Sa.CXO#obs/03711]{  3711} &  200132 & 2002-06-18T10:45 &   19215 & 19:23:40.00 & +14:31:05.9 &  147 &     TIMED FAINT & Gordon Garmire & 3.2.3 \\ %
\\
  \multicolumn{1}{l}{\bf G29.96-0.02} \\ 
              G29.96-0.02 & \dataset[ADS/Sa.CXO#obs/10681]{ 10681} &  200566 & 2009-02-25T15:59 &   27522 & 18:46:04.00 & -02:39:20.0 &   81 &    TIMED VFAINT & Gordon Garmire & 3.5.1 \\ %
\\
  \multicolumn{1}{l}{\bf NGC 3603   } \\ 
                 NGC 3603 & \dataset[ADS/Sa.CXO#obs/00633]{   633} &  200058 & 2000-05-01T23:29 &   46599 & 11:15:07.20 & -61:15:35.2 &  298 &     TIMED FAINT & Michael Corcoran & 3.3.0 \\ %
                 NGC 3603 & \dataset[ADS/Sa.CXO#obs/12328]{ 12328} &  200666 & 2010-10-07T07:27 &  165185 & 11:15:07.20 & -61:15:35.2 &  151 &    TIMED VFAINT & Leisa Townsley   & 4.3.1 \\ %
                 NGC 3603 & \dataset[ADS/Sa.CXO#obs/12329]{ 12329} &  200666 & 2010-10-15T12:57 &  145162 & 11:15:07.20 & -61:15:35.2 &  142 &    TIMED VFAINT & Leisa Townsley   & 4.3.1 \\ %
                 NGC 3603 & \dataset[ADS/Sa.CXO#obs/12330]{ 12330} &  200666 & 2010-10-18T06:56 &   86103 & 11:15:07.20 & -61:15:35.2 &  139 &    TIMED VFAINT & Leisa Townsley   & 4.3.1 \\ %
                 NGC 3603 & \dataset[ADS/Sa.CXO#obs/13162]{ 13162} &  200666 & 2010-10-23T17:21 &   51568 & 11:15:07.20 & -61:15:35.2 &  132 &    TIMED VFAINT & Leisa Townsley   & 4.3.1 \\ %
\\
  \multicolumn{1}{l}{\bf 30 Doradus     } \\ 
               30 Doradus & \dataset[ADS/Sa.CXO#obs/05906]{  5906} &  600425 & 2006-01-21T19:04 &   12318 & 05:38:42.40 & -69:06:02.0 &  323 &    TIMED VFAINT & Leisa Townsley & 3.2.1 \\ %
               30 Doradus & \dataset[ADS/Sa.CXO#obs/07263]{  7263} &  600425 & 2006-01-22T16:51 &   42529 & 05:38:42.40 & -69:06:02.0 &  323 &    TIMED VFAINT & Leisa Townsley & 3.2.1 \\ %
               30 Doradus & \dataset[ADS/Sa.CXO#obs/07264]{  7264} &  600425 & 2006-01-30T15:06 &   37594 & 05:38:42.40 & -69:06:02.0 &  314 &    TIMED VFAINT & Leisa Townsley & 3.2.1 \\ %
               
\enddata

\tablenotetext{a}{Exposure times are the net usable times after various filtering steps are applied in the data reduction process. 
For the following ObsIDs, we discarded exposure time as noted to remove periods of high instrumental background:
2573 (16 ks),
13267 (0.4 ks),
6422 (9 ks),
7391 (0.7 ks),
8461 (0.6 ks),
5891 (0.5 ks),
7033 (1 ks),
4496 (0.4 ks),
6349 (9 ks),
2524 (1 ks),
and 633 (3 ks).
The time variability of the ACIS background is discussed in \S6.16.3 of the \anchorparen{http://asc.harvard.edu/proposer/POG/}{\Chandra Proposers' Observatory Guide} and in the ACIS Background Memos at \url{http://asc.harvard.edu/cal/Acis/Cal_prods/bkgrnd/current/}.
} 

\tablenotetext{b}{The aimpoints (given in celestial coordinates) are obtained from the satellite aspect solution before astrometric correction is applied.  
Units of right ascension ($\alpha$) are hours, minutes, and seconds; units of declination ($\delta$) are degrees, arcminutes, and arcseconds.}

\tablenotetext{c}{ACIS observing modes are described in \S6.12 of the \Chandra Proposers' Observatory Guide.}

\tablenotetext{d}{The \Chandra Calibration Database version used for event calibration.}

\end{deluxetable}

%% file: xray_column_labels.tex
\begin{deluxetable}{lll}
\tablecaption{MOXC X-ray Sources and Properties \label{xray_properties.tbl}
}
\tablewidth{7in}
\tabletypesize{\scriptsize}

\tablehead{   
\colhead{Column Label} & \colhead{Units} & \colhead{Description} \\   
\numberthecolumn & \numberthecolumn & \numberthecolumn  
\setcounter{column_number}{1}
}
\startdata
 RegionName                 & \nodata              & name of the MSFR \\ 
 Name                       & \nodata              & \parbox[t]{3.5in}{X-ray source name in IAU format; prefix is CXOU~J} \\
 Label\tnm{a}               & \nodata              & X-ray source name used within the project       \\
 RAdeg                      & deg                  & right ascension (J2000)                                         \\
 DEdeg                      & deg                  & declination (J2000)                                             \\
 PosErr                     & arcsec               & 1-$\sigma$ error circle around (RAdeg,DEdeg)                               \\
 PosType                    & \nodata              & algorithm used to estimate position \citep[][\S7.1]{Broos10}  \\
                &                      &                                                                 \\
 ProbNoSrc\_min             & \nodata              & smallest of ProbNoSrc\_t, ProbNoSrc\_s, ProbNoSrc\_h      \\
 ProbNoSrc\_t               & \nodata              & {\em p}-value\tnm{b} for no-source hypothesis \citep[][\S4.3]{Broos10} \\
 ProbNoSrc\_s               & \nodata              & {\em p}-value for no-source hypothesis                                \\
 ProbNoSrc\_h               & \nodata              & {\em p}-value for no-source hypothesis                                \\
                &                      &                                                                 \\
 ProbKS\_single\tnm{c}      & \nodata              & \parbox[t]{3.5in}{smallest {\em p}-value for the one-sample Kolmogorov--Smirnov statistic under the no-variability null hypothesis within a single-observation}\\
 ProbKS\_merge\tnm{c}       & \nodata              & \parbox[t]{3.5in}{smallest {\em p}-value for the one-sample Kolmogorov--Smirnov statistic under the no-variability null hypothesis over merged observations}      \\
                \\
 ExposureTimeNominal        & s                    & total exposure time in merged observations                     \\
 ExposureFraction\tnm{d}    & \nodata              & fraction of ExposureTimeNominal that source was observed\\
 RateIn3x3Cell\tnm{e}       &  count /frame        & 0.5:8 keV, in 3$\times$3 CCD pixel cell  \\
 NumObservations            & \nodata              & total number of observations extracted                                \\
 NumMerged                  & \nodata              & \parbox[t]{3.5in}{number of observations merged to estimate photometry properties}\\
 MergeBias                  & \nodata              & fraction of exposure discarded in merge\\
                            &                      &                                                                 \\
 Theta\_Lo                  & arcmin               & smallest off-axis angle for merged observations                 \\
 Theta                      & arcmin               &  average off-axis angle for merged observations                 \\
 Theta\_Hi                  & arcmin               &  largest off-axis angle for merged observations                 \\
                            \\
 PsfFraction                & \nodata              & average PSF fraction (at 1.5 keV) for merged observations \\
 SrcArea                    & (0.492 arcsec)$^2$   & average aperture area for merged observations                   \\
 AfterglowFraction\tnm{f}   & \nodata              & suspected afterglow fraction                   \\
                &                      &                                                                 \\
 SrcCounts\_t               & count                & observed counts in merged apertures                             \\
 SrcCounts\_s               & count                & observed counts in merged apertures                             \\
 SrcCounts\_h               & count                & observed counts in merged apertures                             \\
                &                      &                                                                 \\
 BkgScaling                 & \nodata              & scaling of the background extraction \citep[][\S5.4]{Broos10}                 \\
                &                      &                                                                 \\
 BkgCounts\_t               & count                & observed counts in merged background regions                    \\
 BkgCounts\_s               & count                & observed counts in merged background regions                    \\
 BkgCounts\_h               & count                & observed counts in merged background regions                    \\
                &                      &                                                                 \\
 NetCounts\_t               & count                & net counts in merged apertures                                  \\
 NetCounts\_s               & count                & net counts in merged apertures                                  \\
 NetCounts\_h               & count                & net counts in merged apertures                                  \\
                            \\
 NetCounts\_Lo\_t\tnm{g}    & count                & 1-$\sigma$ lower bound on NetCounts\_t               \\
 NetCounts\_Hi\_t           & count                & 1-$\sigma$ upper bound on NetCounts\_t               \\
 
 NetCounts\_Lo\_s           & count                & 1-$\sigma$ lower bound on NetCounts\_s               \\
 NetCounts\_Hi\_s           & count                & 1-$\sigma$ upper bound on NetCounts\_s               \\
 
 NetCounts\_Lo\_h           & count                & 1-$\sigma$ lower bound on NetCounts\_h               \\
 NetCounts\_Hi\_h           & count                & 1-$\sigma$ upper bound on NetCounts\_h             \\  
                            &                      &                                                                 \\
 MeanEffectiveArea\_t\tnm{h}& cm$^2$~count~photon$^{-1}$ & mean ARF value                                                  \\
 MeanEffectiveArea\_s       & cm$^2$~count~photon$^{-1}$ & mean ARF value                                                  \\
 MeanEffectiveArea\_h       & cm$^2$~count~photon$^{-1}$ & mean ARF value                                                  \\
                &                      &                                                                 \\
 MedianEnergy\_t\tnm{i}     & keV                  & median energy, observed spectrum                                \\
 MedianEnergy\_s            & keV                  & median energy, observed spectrum                                \\
 MedianEnergy\_h            & keV                  & median energy, observed spectrum                                \\
                &                      &                                                                 \\
 PhotonFlux\_t\tnm{j}       & photon /cm**2 /s     & apparent photon flux \\
 PhotonFlux\_s              & photon /cm**2 /s     & apparent photon flux \\
 PhotonFlux\_h              & photon /cm**2 /s     & apparent photon flux \\
                \\
 EnergyFlux\_t              & erg~cm$^{-2}$~s$^{-1}$ & max(EnergyFlux\_s,0) + max(EnergyFlux\_h,0)      \\
 EnergyFlux\_s\tnm{k}       & erg~cm$^{-2}$~s$^{-1}$ & apparent energy flux \\
 EnergyFlux\_h\tnm{k}       & erg~cm$^{-2}$~s$^{-1}$ & apparent energy flux
\enddata

                                                                                                                                              
\tablecomments{
These X-ray column labels were previously published by the CCCP \citep{Broos11a} and are produced by the
{\em ACIS Extract} (AE) software package \citep{Broos10,AE2012}.
The AE software and User's Guide are available at \anchor{http://www.astro.psu.edu/xray/acis/acis_analysis.html}{\url{http://www.astro.psu.edu/xray/acis/acis_analysis.html}}.
}
\tablecomments{
The suffixes ``\_t'', ``\_s'', and ``\_h'' on names of photometric quantities designate the {\em total} (0.5--8~keV), {\em soft} (0.5--2~keV), and {\em hard} (2--8~keV) energy bands. 
}
\tablecomments{
Source significance quantities (ProbNoSrc\_t, ProbNoSrc\_s, ProbNoSrc\_h, ProbNoSrc\_min) are computed using a subset of each source's extractions chosen to maximize significance \citep[][\S6.2]{Broos10}.
Source position quantities (RAdeg, DEdeg, PosErr) are computed using a subset of each source's extractions chosen to minimize the position uncertainty \citep[][\S6.2 and 7.1]{Broos10}.   
All other quantities are computed using a subset of each source's extractions chosen to balance the conflicting goals of minimizing photometric uncertainty and of avoiding photometric bias \citep[][\S6.2 and 7]{Broos10}. 
}

\tablenotetext{a}{Source labels identify a \Chandra pointing; they do not convey membership in astrophysical clusters.}

\tablenotetext{b}{In statistical hypothesis testing, the {\em p}-value is the probability of obtaining a test statistic at least as extreme as the one that was actually observed, when the null hypothesis is true.}

\tablenotetext{c}{See \citet[][\S7.6]{Broos10} for a description of the variability metrics, and caveats regarding possible spurious indications of variability using the ProbKS\_merge metric. }

\tablenotetext{d}{Due to dithering over inactive portions of the focal plane, a \Chandra source often is not observed during some fraction of the nominal exposure time.  (See \url{http://cxc.harvard.edu/ciao/why/dither.html}.)  The reported quantity is FRACEXPO, produced by the \CIAO\ tool {\em mkarf}.}

\tablenotetext{e}{ACIS suffers from a non-linearity at high count rates known as {\em photon pile-up}, described in Section~\ref{sec:pileup} below. 
RateIn3x3Cell is an estimate of the observed count rate falling on an event detection cell of size 3$\times$3 \ACIS\ pixels, centered on the source position.
When RateIn3x3Cell $> 0.05$ (count/frame), the reported source properties may be biased by pile-up effects.
See Table~\ref{pile-up_risk.tbl} for a list of MOXC sources with significant pile-up.
}

\tablenotetext{f}{Some background events arising from an effect known as ``afterglow'' (\url{http://cxc.harvard.edu/ciao/why/afterglow.html}) may contaminate source extractions, despite careful procedures to identify and remove them during data preparation \citep[][\S3]{Broos10}.
After extraction, we attempt to identify afterglow events using the AE tool {\em ae\_afterglow\_report}, and report the fraction of extracted events attributed to afterglow; see the \anchorparen{http://www.astro.psu.edu/xray/acis/acis_analysis.html}{{\it ACIS Extract} manual}.} 

\tablenotetext{g}{Confidence intervals (68\%) for NetCounts quantities are estimated by the \CIAO\ tool {\em aprates} (\url{http://asc.harvard.edu/ciao/ahelp/aprates.html}).}

\tablenotetext{h}{The ancillary response file (ARF) in \ACIS\ data analysis represents both the effective area of the observatory and the fraction of the observation for which data were actually collected for the source (ExposureFraction).}

\tablenotetext{i}{MedianEnergy is the median energy of extracted events, corrected for background \citep[][\S7.3]{Broos10}. } 

\tablenotetext{j}{PhotonFlux = (NetCounts / MeanEffectiveArea / ExposureTimeNominal) \citep[][\S7.4]{Broos10}. }

\tablenotetext{k}{EnergyFlux = $1.602 \times 10^{-9} {\rm (erg/keV)} \times $(NetCounts/ExposureTimeNominal/MeanEffectiveArea) $\times$ MedianEnergy \citep[][\S2.2]{Getman10}. }

\end{deluxetable}

%% file: piled_table.tex
\begin{deluxetable}{llllccccc}
\tablecaption{Sources Exhibiting Photon Pile-up \label{pile-up_risk.tbl}}
\tablewidth{0pt}
\tabletypesize{\scriptsize}

\tablehead{
\colhead{MSFR} & \colhead{Name} & \colhead{Label} & \colhead{Identifier} & \colhead{ObsID} & \colhead{$\theta$} & \colhead{PsfFraction}& \colhead{Correction}  \\  
                                                                           &&&&& \colhead{(\arcmin)} \\
\numberthecolumn & \numberthecolumn & \numberthecolumn & \numberthecolumn & \numberthecolumn & \numberthecolumn & \numberthecolumn & \numberthecolumn  
\setcounter{column_number}{1}
}
\startdata
NGC 6334      & 172001.73-355816.2 &  p2\_417  & 2MASS J17200173-3558162 & 2573 & 1.9 & 0.91 & 1.032 \\ 
              & 172031.79-355111.5 &  p1\_339  & 2MASS J17203178-3551111 & 2574 & 5.9 & 0.90 & 1.027 \\ 
              &                    &           &                         & 8975 & 2.5 & 0.89 & 1.062 \\ 
\hline
NGC 6357      & 172259.76-340439.6 &  p6\_135  & 2MASS J17225977-3404395 &13623 & 1.4 & 0.91 & 1.081 \\ 
              & 172443.49-341156.9 &  p1\_591  & Pismis 24-1 SW + NEab   & 4477 & 0.3 & 0.89 & 1.132 \\ 
              & 172443.95-341145.6 &  p1\_644  & 2MASS J17244396-3411458 & 4477 & 0.5 & 0.89 & 1.068 \\ 
              & 172444.72-341202.6 &  p1\_713  & Pismis 24-17            & 4477 & 0.3 & 0.89 & 1.066 \\ 
              & 172508.85-341112.4 &  p1\_1372 & WR 93 (HD 157504)       & 4477 & 5.4 & 0.90 & 1.054 \\ 
              & 172534.23-342311.7 &  p3\_718  & [N78] 49                &10987 & 0.5 & 0.89 & 1.060 \\ 
\hline
Eagle Nebula  & 181836.42-134802.4 &  p1\_644  & HD 168076               & 0978 & 1.9 & 0.90 & 1.081 \\ 
              & 181837.04-134529.4 &  p1\_719  & NGC 6611 213            & 0978 & 3.2 & 0.70 & 1.048 \\ 
\hline
M17           & 182025.34-161021.9 &  lp\_955  & UGPS J182025.34-161021.9& 6421 & 1.1 & 0.90 & 1.072 \\ 
              & 182026.60-161055.6 &  lp\_1094 & UGPS J182026.59-161055.6&  972 & 1.0 & 0.89 & 1.177 \\ 
              & 182029.81-161045.5 &  lp\_1433 & CEN 1b                  & 6403 & 0.3 & 0.90 & 1.132 \\ 
              &                    &           &                         & 6420 & 0.3 & 0.90 & 1.360 \\ 
              &                    &           &                         & 6421 & 0.3 & 0.89 & 1.145 \\ 
              &                    &           &                         & 8460 & 0.3 & 0.89 & 1.140 \\ 
              &                    &           &                         & 8461 & 0.3 & 0.89 & 1.147 \\ 
              &                    &           &                         &  972 & 0.3 & 0.90 & 1.148 \\ 
              & 182029.89-161044.4 &  lp\_1452 & CEN 1a                  & 6403 & 0.3 & 0.89 & 1.311 \\ 
              &                    &           &                         & 6420 & 0.3 & 0.89 & 1.335 \\ 
              &                    &           &                         & 6421 & 0.3 & 0.90 & 1.307 \\ 
              &                    &           &                         & 8460 & 0.3 & 0.89 & 1.336 \\ 
              &                    &           &                         & 8461 & 0.3 & 0.90 & 1.314 \\ 
              &                    &           &                         &  972 & 0.3 & 0.90 & 1.384 \\ 
              & 182030.63-161028.4 &  lp\_1543 & UGPS J182030.63-161028.4& 6403 & 0.6 & 0.90 & 1.068 \\ 
              & 182034.49-161011.8 &  lp\_1846 & Cl* NGC 6618 Sch 1      & 6403 & 1.4 & 0.90 & 1.059 \\ 
              &                    &           &                         & 6420 & 1.5 & 0.90 & 1.050 \\ 
              &                    &           &                         & 8460 & 1.4 & 0.90 & 1.031 \\ 
              &                    &           &                         & 8461 & 1.4 & 0.90 & 1.060 \\ 
              &                    &           &                         &  972 & 1.0 & 0.90 &\nodata\\ 
              & 182039.04-160836.9 &  lp\_1980 & UGPS J182039.04-160836.9& 6403 & 3.3 & 0.89 & 1.033 \\ 
              & 182132.25-161028.4 &  lp\_2497 & UGPS J182132.25-161028.5& 6422 & 1.7 & 0.90 & 1.032 \\ 
\hline
W4            & 023242.54+612721.7 &  p1\_479  & HD 15558                & 7033 & 0.3 & 0.90 & 1.101 \\ 
\hline
NGC 3576      & 111124.01-611722.8 &  p1\_162  & 2MASS J11112401-6117227 & 6349 & 3.5 & 0.89 & 1.054 \\ 
              & 111153.31-611845.9 &  p1\_605  &                         & 4496 & 0.3 & 0.71 & 1.128 \\ 
              &                    &           &                         & 6349 & 0.3 & 0.71 & 1.141 \\ 
              & 111204.50-610543.0 &  p2\_701  & EM Car (HD 97484)       & 8905 & 0.5 & 0.67 & 1.086 \\ 
\hline
W51A          & 192333.75+142953.7 &  p1\_194  & UGPS J192333.74+142953.7& 2524 & 2.0 & 0.90 & 1.085 \\ 
              &                    &           &                         & 3711 & 2.0 & 0.90 & 1.048 \\ 
              & 192350.15+143302.6 &  p1\_776  & UGPS J192350.15+143302.6& 2524 & 3.1 & 0.89 & 1.058 \\ 
\hline
NGC 3603      & 111459.49-611433.8 &  p1\_1164 & MTT 68                  & 0633 & 1.6 & 0.90 & 1.252 \\ 
              &                    &           &                         &12328 & 1.0 & 0.89 & 1.145 \\ 
              &                    &           &                         &12329 & 1.0 & 0.90 & 1.193 \\ 
              &                    &           &                         &12330 & 1.0 & 0.90 & 1.227 \\ 
              &                    &           &                         &13162 & 0.9 & 0.89 & 1.381 \\ 
              & 111507.30-611538.4 &  p1\_2668 & NGC 3603-A1             & 0633 & 0.2 & 0.75 & 1.277 \\ 
              &                    &           &                         &12328 & 0.5 & 0.67 & 1.252 \\ 
              &                    &           &                         &12329 & 0.5 & 0.65 & 1.264 \\ 
              &                    &           &                         &12330 & 0.5 & 0.67 & 1.262 \\ 
              &                    &           &                         &13162 & 0.5 & 0.62 & 1.213 \\ 
              & 111507.40-611538.6 &  p1\_2704 & NGC 3603-B              & 0633 & 0.2 & 0.46 & 1.290 \\ 
              &                    &           &                         &12328 & 0.5 & 0.50 & 1.194 \\ 
              &                    &           &                         &12329 & 0.5 & 0.55 & 1.187 \\ 
              &                    &           &                         &12330 & 0.5 & 0.49 & 1.285 \\ 
              &                    &           &                         &13162 & 0.5 & 0.57 & 1.248 \\ 
              & 111507.58-611537.9 &  p1\_2777 & NGC 3603-C              & 0633 & 0.3 & 0.89 & 1.855 \\ 
              &                    &           &                         &12328 & 0.5 & 0.89 & 1.091 \\ 
              &                    &           &                         &12329 & 0.5 & 0.89 & 1.189 \\ 
              &                    &           &                         &12330 & 0.5 & 0.89 & 1.132 \\ 
              &                    &           &                         &13162 & 0.5 & 0.89 & 1.383 \\ 
              & 111509.35-611602.0 &  p1\_3368 & Cl* NGC 3603 Sher 47    & 0633 & 0.3 & 0.90 & 1.118 \\ 
              &                    &           &                         &12328 & 0.9 & 0.89 & 1.140 \\ 
              &                    &           &                         &12329 & 0.9 & 0.89 & 1.123 \\ 
              &                    &           &                         &12330 & 1.0 & 0.89 & 1.177 \\ 
              &                    &           &                         &13162 & 1.0 & 0.90 & 1.131 \\ 
              & 111521.31-611504.3 &  p1\_4736 & 2MASS J11152132-6115043 & 0633 & 1.9 & 0.91 & 1.076 \\ 
              &                    &           &                         &12328 & 2.2 & 0.91 & 1.053 \\ 
              &                    &           &                         &12329 & 2.2 & 0.91 & 1.060 \\ 
              &                    &           &                         &12330 & 2.2 & 0.91 & 1.045 \\ 
              &                    &           &                         &13162 & 2.1 & 0.91 & 1.041 \\ 
\hline
30 Doradus    & 053747.43-691020.0 & p1\_255   & PSR J0537-6910          & 5906 & 6.5 & 0.89 & 1.074  \\ 
              &                    &           &                         & 7263 & 6.5 & 0.91 & 1.073  \\ 
              &                    &           &                         & 7264 & 6.4 & 0.89 & 1.057  \\ 
              & 053844.26-690605.9 & p1\_995   & Mk34                    & 5906 & 0.2 & 0.89 & 1.095  \\ 
              &                    &           &                         & 7263 & 0.2 & 0.89 & 1.110  \\ 
              &                    &           &                         & 7264 & 0.2 & 0.89 & 1.111  \\ 
              & 053841.61-690513.3 & p1\_752   & R140a1a2                & 5906 & 1.1 & 0.43 & 1.079  \\ 
              &                    &           &                         & 7263 & 1.1 & 0.63 & 1.070  \\ 
              &                    &           &                         & 7264 & 1.1 & 0.63 & 1.067     
\enddata
\tablecomments{
Col.\ (1): Name of the MSFR.
\\Col.\ (2): X-ray source name in IAU format; prefix is CXOU~J ({\em Name} in Table~\ref{xray_properties.tbl}).
\\Col.\ (3): X-ray source name used within the project ({\em Label} in Table~\ref{xray_properties.tbl}).
\\Col.\ (4): Source name from VizieR or SIMBAD.
\\Col.\ (5): \Chandra Observation Identification.
\\Col.\ (6): Off-axis angle ({\em Theta} in Table~\ref{xray_properties.tbl}).
\\Col.\ (7): Fraction of the PSF (at 1.497 keV) enclosed within the extraction region ({\em PsfFraction} in Table~\ref{xray_properties.tbl}). A reduced PSF fraction (significantly below 90\%)  indicates that the source is in a crowded region. 
\\Col.\ (8): Estimated ratio of pile-up-free to observed (piled) count rates in the 0.5--8~keV energy band.
}
\end{deluxetable}